\documentclass[review]{elsarticle}

\usepackage{lineno}
\usepackage{hyperref,mathtools,amsmath,upgreek}
\usepackage{bm}
\usepackage{mathtools}
\usepackage{subfig}
\usepackage{graphicx}
\usepackage{hyperref}
\usepackage{xcolor}
\usepackage[margin=2.5cm]{geometry}
\usepackage{subfig}
\usepackage{color,soul}

\journal{Computer Physics Communications}

\biboptions{sort&compress}

\DeclarePairedDelimiter\abs{\lvert}{\rvert}%
\DeclarePairedDelimiter\norm{\lVert}{\rVert}%
\makeatletter
\let\oldabs\abs
\def\abs{\@ifstar{\oldabs}{\oldabs*}}
\let\oldnorm\norm
\def\norm{\@ifstar{\oldnorm}{\oldnorm*}}
\makeatother

\newcommand{\subsubfloat}[2]{%
  \begin{tabular}{@{}c@{}}#1\\#2\end{tabular}%
}

\newcommand{\mykB}{k_\mathrm{B}}

\begin{document}

\begin{frontmatter}

\title{Comparison of effective and stable Langevin dynamics integrators}

\author[Simone_address]{Bogdan Tanygin\corref{mycorrespondingauthor}}
\cortext[mycorrespondingauthor]{Corresponding author}
\ead{bogdan@tanygin-holding.com}

\author[Simone_address]{Simone Melchionna}

\address[Simone_address]{IAC-CNR, Istituto per le Applicazioni del Calcolo ``M. Picone'', Via dei Taurini 19, 00185 Rome, Italy.}

\begin{abstract}

Langevin and Brownian simulations play a prominent role in computational research, and state of the art integration algorithms provide trajectories with different stability ranges and accuracy in reproducing statistical averages. The practical usability of integrators is an important aspect to allow choosing large time steps while ensuring numerical stability and overall computational efficiency. In this work, different use cases and practical features are selected in order to perform a cumulative comparison of integrators with a focus on evaluating the derived velocity and position autocorrelation functions, a comparison that is often disregarded in the literature. A standard industrial open-source software methodology is suggested to compare systematically the different algorithms.

\end{abstract}


\end{frontmatter}

\section{Introduction}

Complex fluids, such as polymeric and colloidal suspensions, are of high interest for physics, chemistry and life sciences~\cite{Schlick2010,Frenkel2002,jones2002soft} with profound applications in biotechnologies, nanotechnology, healthcare, and pharmacology~\cite{Jordan1999a,Mashaghi2013}. For instance, biological fluids at the nano and microscale as much as structural transformations of macromolecules ~\cite{Schneider1999,Beard2000} are key to understanding organic life and searching for new drugs and vaccines. These states of matter can be  simulated by using a wide range of particle and field-based numerical techniques, derived from quantum, classical or kinetic equations of motion ~\cite{coffey2012langevin,Fisher1964,Kolafa1994,Economou2001,bernaschi2019mesoscopic}. Practically effective and feasible calculations can be performed by means of computer simulation via the techniques of molecular dynamics, Monte Carlo and derived ones~\cite{satoh2010introduction,allen2017computer,Schlick2010, fyta2006multiscale}. Molecular dynamics stands as the reference technique for simulating continuous trajectories of  systems ranging from biomaterials~\cite{Schneider1999,Beard2000} to galaxies~\cite{burtscher2011efficient}.

As the simulated systems exhibit the hallmark of scale separation, computational strategies rely on the reduction of fast degrees of freedom in favor of retaining the slow, important ones. A similar consideration apply to representing systems in contact with a heat reservoir that does not need to be represented in full detail. The implicit solvent model~\cite{roux1999implicit} and the usage of effective thermostats to sample states from the canonical ensemble~\cite{Schlick2010} consider the motion of particles within a stochastic environment formed by other particles represented in effective terms. As a result, the molecular trajectory is approximated by Langevin dynamics (LD)~\cite{Schlick2010,mcquarrie2000statistical} featuring frictional and random forces, the latter encoded by a delta-correlated stationary Gaussian process~\cite{Pottier2014}, while the Brownian dynamics refers to dynamics in presence of explicit effective hydrodynamic forces ~\cite{mcquarrie2000statistical}. The equations of motion of Langevin or Brownian dynamics are not phenomenological but derived through rigorous multiscale theoretical analysis.

The trajectory of a single particle undergoing LD can be analyzed  by treating the stochastic force as a generalized function and statistical averages are derived in closed-form. Numerically speaking, the particle dynamics is computed by a time-discrete approximation of the LD equations and by replacing the Dirac delta function by suitable discrete approximations, as been proposed in the literature~\cite{satoh2010introduction,allen2017computer,Schlick2010}.

The main motivation of the present research is to evaluate and compare a large class of LD propagators since such comparison is only present in the literature for specific use cases and integrators~\cite{leimkuhler2013robust,Schlick2010}.
For example, the important integrator proposed by Grønbech-Jensen and Farago (GJF)~\cite{gronbech2013simple} is missing. Also, previous benchmarks focused on specific scientific applications which can limit their generalization to other contexts. The present work targets additional use cases and should be viewed as an extension of previous work rather than a comprehensive and ultimate review.

The additional crucial aspect of the present report is the suggested methodology. Each integrator has been implemented on top of the well-maintained open-source molecular dynamics package ESPResSo~\cite{Limbach2006,Arnold2013,TheESPResSoprojectCopyrightC2016}. Each method passed automated Python tests scripted by independent researchers providing objectivity, reusability and future maintenance of its implementation. The  results reported in the following are based on modification of some of these tests and, importantly, the same Python scripts are used to validate each integrator without \textit{ad hoc} adaptation. The code for testing is available in GitHub and referred here (see Table~\ref{table:integrators-def}). Clean and repeatable run time environment is organized using Docker technology~\cite{dockerRef} and GitHub tags of the source code including the simulation scripts~\cite{scriptsCPC2024}. The present analysis is straightforward and appears to minimize code errors in different programming languages and environments. Hence, we hereby encourage other researchers to follow such open-source approach to avoid publishing algorithms as pseudo-codes or based on proprietary or hidden frameworks, unmaintained personal software codes, and/or out of reach to other investigators.

In general, we have defined a ranking criteria method, including accuracy~\cite{higham2002accuracy} for all possible parameters (including infinitesimal, medium, or unlimited large time steps), stationarity of the most relevant thermodynamic and diffusive properties and responses to perturbations (e.g. the harmonic  or uniform external field). The theoretical derivation of integrators has been reviewed by highlighting, for example, the symplectic similarity of different integrators. We believe that  these criteria rank the methods properly in order to find  effective stochastic integrators which can be  applied to different interdisciplinary problems with minimal adaptation. 

The present comparison has its own limitations since our implementations might be suboptimal in terms of efficiency or  results may differ from those reported in the original publications in terms of stability and numerical error. The mismatch may arise if averaging procedure do not reproduce in full the published forms or, for instance, by different usage of the Kahan summation algorithm to reduce error by keeping a separate running compensation variable calculated at each step~\cite{Higham1993TheAO}. 

The present study focuses systems with diagonal mass matrix and diagonal friction tensor and under external harmonic and constant forces and with further focus to a single degree of freedom. A report concerning the general case of \textit{rotational} LD will be published separately. Our provisional results reveal that the ranking of different algorithms cannot be put on the same level because stochastic rotational motion discretization requires solving more sophisticated analytical and technical challenges.

The article is organized as follows: section \ref{integrators} outlines the selected algorithms including references to their implementation.  In addition to all applicable tests, the integrators were tested with different physical requirements in section \ref{validation}. The final scoring of the methods is summarized and discussed in section \ref{results_dis}. Practical recommendations regarding the benefits and constraints of usage of the selected integrators conclude the paper.

\section{Langevin integrators} \label{integrators}
\subsection{Theoretical foundation}

Langevin dynamics integrators can be formulated based on the generalized mechanical framework. Let us indicate cartesian coordinates with $\mathbf{q}$ and $\mathbf{p}$, where bold character indicates a vector referring to position and momentum, respectively, and for each degree of freedom. Assuming a given initial time $t=t_1$, any integration scheme is described by the propagator $\mathcal{P}\left(t_1,t_2\right)$~\cite{Tuckerman1992} that propagates the state as:
\begin{eqnarray}
	\mathcal{P} \left(t_1,t_2\right) \cdot \left( \mathbf{p}, \mathbf{q} \right)\biggr\rvert_{t=t_1} = \left( \mathbf{p}, \mathbf{q} \right)\biggr\rvert_{t=t_2}, \label{eq:propagation}
\end{eqnarray}
where the propagator acts in terms of successive time steps as
\begin{eqnarray}
    \mathcal{P}\left(t_1,t_2\right)\cdot
    =
    \lim_{\Delta \rightarrow 0} \prod_{k=0}^{n-1}e^{i\mathcal{L}\left(\mathbf{q}, \mathbf{p}, t_1+k\Delta \right)\Delta}  \label{eq:propagationDetailed}
\end{eqnarray}
where $\Delta = \left(t_2-t_1\right)/n$ is the elementary timestep. Note that the notation used here follows the one  suggested in~\cite{Martys1999} in order to highlight the time-reversible and symplectic nature of integrators of deterministic, time-symmetric mechanics (hereinafter named $t$--symmetrical ones). The operator~\eqref{eq:propagationDetailed}  should be conventionally applied to the system mechanical state~\eqref{eq:propagation} in a right-to-left direction. The above formula is based on the replacement of continuous time derivative, referred to as  the $t_{\partial}$--approximation, with a discrete one.

Assuming that the system is conservative and samples the microcanonical constant--$\left(N,V,E\right)$ ensemble, with a conservative potential $U\left( \mathbf{q},t \right)$ and no solenoidal vector type of either electromagnetic or similar interaction fields, the Liouville operator is given by:
\begin{eqnarray}
    i\mathcal{L}\left( \mathbf{p},\mathbf{q},t \right)\cdot
    =
    \left\lbrace \cdots,\mathcal{H}\left( \mathbf{p},\mathbf{q},t \right) \right\rbrace
    =
    \frac{\partial \mathcal{H}\left( \mathbf{p},\mathbf{q},t \right)}{\partial \mathbf{p}} \frac{\partial}{\partial \mathbf{q}} - \frac{\partial \mathcal{H}\left( \mathbf{p},\mathbf{q},t \right)}{\partial \mathbf{q}} \frac{\partial}{\partial \mathbf{p}}, \label{eq:Liouville}
\end{eqnarray}
where $\mathcal{H}\left( \mathbf{q},\mathbf{p},t \right) = K\left( \mathbf{q},\mathbf{p} \right) + U\left( \mathbf{q},t \right) $ is the Hamiltonian including kinetic and potential energy and where the temporal dependence may indicate environmental parametric changes. In cartesian coordinates the kinetic energy then becomes:
\begin{eqnarray}
  K = K\left( \mathbf{p} \right) \label{eq:independent-deg-freedom}
\end{eqnarray}
corresponding to pure translational motion.
An arbitrary-shaped rigid body rotational motion and more complex mechanical models will be considered in an ensuing publications. Assuming the implicit solvent model, the Liouville operator is given by:
\begin{eqnarray}
    i\mathcal{L}\left( \mathbf{p},\mathbf{q},t \right)\cdot
    &=&
     \frac{\partial \mathcal{H}\left( \mathbf{p},\mathbf{q},t \right)}{\partial \mathbf{p}} \frac{\partial}{\partial \mathbf{q}} + \pmb{\Phi}\left(\mathbf{p},\textbf{q},t\right) \frac{\partial}{\partial \mathbf{p}} \label{eq:LiouvilleMod}
\\
    \pmb{\Phi}\left(\mathbf{p},\textbf{q},t\right)
    &=&
    - \frac{\partial U\left( \mathbf{q},t \right)}{\partial \mathbf{q}} - \frac{\partial \mathcal{R}\left(\mathbf{p},\textbf{q} \right)}{\partial \dot{\mathbf{q}}} + \mathbf{f}\left(t\right), \label{eq:GenForce}
\end{eqnarray}
where the total generalized force (hereinafter, a force) $\pmb{\Phi}\left(\mathbf{p},\mathbf{q},t\right)$ includes the conservative term $\mathbf{F}\left(\mathbf{q},t \right)$, the friction term (expressed here via the general Rayleigh dissipation
function \cite{minguzzi2015rayleigh}) and the randomly fluctuating force, respectively. The last two components are interconnected by the fluctuation-dissipation theorem. The noise term $\mathbf{f}\left(t\right)$ is modelled by a delta-correlated stationary process~\cite{Pottier2014}:
\begin{eqnarray}
	\langle \mathbf{f}\left(t\right) \rangle &=& 0 \label{eq:noise1}
 \\
    \langle \textbf{f}\left(t\right)
    \otimes
    \textbf{f}\left(t'\right) \rangle
    &=&
    2k_{\mathrm{B}} T \textbf{Z}\textbf{I} \delta\left( t-t'\right), \label{eq:noise2}	
\end{eqnarray}
where $\textbf{Z}$ is a Stokesian friction tensor, $\textbf{I}$ is an identity matrix, and $\otimes$ is a tensor product. The corresponding~\eqref{eq:propagationDetailed} propagator $\mathcal{P}\left(t_1,t_2\right)$ defines a stochastic differential equation of motion that reduces to the Langevin equation:
\begin{eqnarray}
    \dot{\textbf{p}}
    &=&
    \textbf{F}\left(\mathbf{q},t \right)
    - \textbf{Z} \dot{\textbf{q}}
    + \textbf{f}\left(t\right) \label{eq:langevin}
  \\
    \dot{\mathbf{q}}
    &=&
    \frac
    {\partial \mathcal{H}\left( \mathbf{p},\mathbf{q},t \right)}
    {\partial \mathbf{p}} \label{eq:eom}
  \\
  \mathbf{p} &=& \mathbf{M} \dot{\mathbf{q}}, \label{eq:mass}
\end{eqnarray}
where $\mathbf{M}$ is the mass matrix. Here we consider only the diagonal mass matrix and friction tensors:
\begin{eqnarray}
    \mathrm{diag} \left( \textbf{m} \right)
    \stackrel{\mathrm{def}}{=}
    \textbf{M} \label{eq:mass-matrix}
\\
    \mathrm{diag} \left( \pmb{\gamma} \right)
    \stackrel{\mathrm{def}}{=}
    \textbf{Z}
    \label{eq:friction-matrix}
\end{eqnarray}

By averaging over the stochastic trajectories from the Langevin equation, a probability density function obeys the Klein-Kramers equation~\cite{forbert2000fourth}. We will keep the Langevin equation-based formalism to model each individual particle's behavior in this study.

As this study focuses on a diagonal kinetic energy term ~\eqref{eq:independent-deg-freedom}, for the sake of simplicity all the following formalism refers to a single degree of freedom expressed in cartesian coordinates $\left(q,p\right)$, unless stated otherwise. A generalization to tensorial form is straightforward and well described in the corresponding original references of each integrator. 

The first large family of integrators is defined by applying conventional discretization methods to Eq.~\eqref{eq:langevin}---~\eqref{eq:eom}, as in the case of the Trotter factorization~\cite{Trotter1959,Tuckerman1992,ricci2003algorithms,thalmann2007trotter} of  Eq.~\eqref{eq:propagationDetailed} with encoded time-reversibility and by employing different factorization levels. The time-reversible case corresponds to the most prominent examples of symplectic integrators: Verlet, velocity Verlet, and leapfrog schemes~\cite{Schlick2010}.
Analytical integration of  Eq.~\eqref{eq:langevin} or the selected factorization factors over a single time step results in a closed-form expression which can be propagated in sequence over multiple time steps.  Many modern LD integrators are based on a combination of this approach. Higher orders of expansion of~\eqref{eq:propagationDetailed} by the operator factorization or other means leads to more sophisticated integrators with better accuracy. The fourth-order methods were introduced~\cite{suzuki1995new, chin1997symplectic, forbert2000fourth}.

\subsection{Selected integrators} \label{selected_int}

Following a thorough review of the specialized literature and based on the experience gathered on the implementation of various LD integrators, we selected  a number of LD integrators. Taking into account our experience and that from other research groups, including those working on open-source Molecular Dynamics, the selected set of integrators represents quite well those widely-adopted by the numerical community.

A well-known integrator was introduced by Brooks,  Brünger, and Karplus,  known as BBK~\cite{brunger1984stochastic}.  It is similar to the Verlet integrator~\cite{Verlet1967b}, also known as the Störmer method or velocity Verlet integrator depending on a specific representation.  An alternative name for BBK is the generalized Verlet~\cite{Schlick2010} which means that the original Verlet method is supplemented by a discretized Dirac delta (stochastic force) and friction terms (a momentum-dependent force):
\begin{eqnarray}
  p_{k+1/2}&=&
     \left[
       F_k
       - \gamma \frac{p_k}{m}
       +f_k
     \right]
       \frac{\Delta}{2} \label{eq:bbk1}
\\
  q_{k+1}&=&
    q_k + \frac{p_{k+1/2}\Delta}{m} \label{eq:bbk2}
\\
  p_{k+1}&=&
    \left[
      F_{k+1}
      -\gamma \frac{p_{k+1}}{m}
      +f_{k+1}
    \right]
    \frac{\Delta}{2} \label{eq:bbk3}\\
  f_{k+1} &=&
        \sqrt{
          \frac
          {2k_{\mathrm{B}}T\gamma}
          {\Delta}
        }
        \xi_{k+1} \label{eq:bbkrnd}
\end{eqnarray}
Here and hereinafter, subscripts define a value at the discrete timestep $t = \left(t_1+k\Delta\right)$; $\xi_k$ is a pseudo-random number with a unitary standard deviation and probability distribution defined below. This integrator propagates the state $\left( \textbf{p},\textbf{q} \right)$ from time $t_1+k\Delta$ to $t_1+\left(k + 1 \right)\Delta$. The method was named the generalized Verlet because it is supplemented by fluctuation-dissipation terms that is moved out of the conservative Hamiltonian system and time reversibility is not applicable anymore. However, its  symplectic form derived from the Trotter factorization reduces to Hamiltonian mechanics in the case of non-dissipative dynamics.

The main challenge here is the implicit momentum component in eq.~\eqref{eq:bbk3}.  The BBK integrator benefits from the linear algebraic form of ~\eqref{eq:eom},~\eqref{eq:bbk1}--\eqref{eq:bbk3}, where explicit solution for $p_{k+1}$ can be easily derived~\cite{Schlick2010} leading to the original form of the BBK integrator~\cite{brunger1984stochastic}.
However,  there are many variants of Verlet-like LD integrator ~\cite{TheESPResSoprojectCopyrightC2016,Martys1999,groot1997dissipative}, where $ p_{k+1}$ is simply replaced by an approximate value $\tilde{p}_{k+1}$, most frequently by $p_{k}$ on the right side of~\eqref{eq:bbk3}. This approach for translational and rotational motion was suggested in ~\cite{Martys1999} with original sources of~\cite{groot1997dissipative} which contains the explanation of this selection as an empirical (technical) choice of the free parameter $\lambda = 0.5$ defining this integrator via the approximation: $ \tilde{p}_{k+1} \approx p_{k} + \lambda  \Delta \left(F_k- \gamma p_k/m + f_k\right)$.
Hereinafter, any cartesian coordinates-related approximations will be referred to as a $q_{\lambda}$--approximation. There is no special name for this class of integrators.  It is often referred to as the regular velocity Verlet algorithm because the frictional part of the force is simply added before the velocity is updated. This introduces a certain terminological confusion.  We will follow the naming of the latter sources and  refer this method to as $\lambda05$-velocity Verlet or, shorter,  $\lambda05$-VV. Finally, we will consider both cartesian coordinates in the BBK method and formally refer to this integrator as vBBK.

In absence of conservative force, both BBK and $\lambda05$-VV  have upper stability limits in terms of the timestep $\Delta$ ~\cite{wang2003analysis,Schlick2010} which can be expressed as the following $t_{\partial}$--approximation:
\begin{eqnarray}
    \Delta \ll \left. m \right/ \gamma
    \stackrel{\mathrm{def}}{=}
    \Delta_0
    \label{eq:bbk-temporal-limit}
\end{eqnarray}

The Grønbech-Jensen and Farago method~\cite{gronbech2013simple} is based on the one-trapezoidal $q_{\Delta}$-- and force $F_{\Delta}$--approximation of integrals of \eqref{eq:langevin}--\eqref{eq:eom} within the $\Delta$ producing errors up to $O\left[ \left( \Delta / \Delta_0 \right)^2 \right]$:
\begin{eqnarray}
    q_{k+1}-q_k
    &\approx&
    \frac{\Delta}{2m}
    \left(p_{k+1}+p_{k}\right) \label{eq:gjfmodel1}
\\
    \int_{t_1+k \Delta}^{t_1+\left(k + 1 \right)\Delta} F\left[\textbf{q}\left(t\right),t\right] \,dt
    &\approx& \frac
      { F_k
       +F_{k+1}}
      {2} \Delta \label{eq:gjfmodel2}
\end{eqnarray}
leading to the following scheme:
\begin{eqnarray}
    q_{k+1} &=& q_{k}+
      \frac{b\Delta}{m}
      \left[
        p_k+
        \frac{\Delta}{2}
        \left(
          F_k
          + f_k
        \right)
      \right] \label{eq:gjf-eom-pos} \\
    p_{k+1} &=& a p_k
      + \frac{\Delta}{2}
        \left(
          a F_k
          + F_{k+1}
        \right)
      + b f_k \Delta \label{eq:gjf-eom-vel} \\
    a &\stackrel{\mathrm{def}}{=}& \frac
      {1 - \frac{\beta \Delta}{2}}
      {1 + \frac{\beta \Delta}{2}} \label{eq:gjf-adef} \\
    b &\stackrel{\mathrm{def}}{=}& \frac
        {1}
        {1 + \frac{\beta \Delta}{2}} \label{eq:gjf-bdef} \\
    \beta &\stackrel{\mathrm{def}}{=}&
    \left. \gamma \right/ m \label{eq:beta-ref}
\end{eqnarray}
Here, the big-$O$ and the little-$o$ notations apply to a dimensionless representation of equations as defined below. The convergence of GJF to the analytical solution for $k \xrightarrow{} \infty$ was proven~\cite{gronbech2013simple} to be independent on $\Delta$ as long as $a_i < 1$, which is a lighter $t$--approximation of the type~\eqref{eq:bbk-temporal-limit}. The main approximations underlying the GJF scheme is described in~\eqref{eq:gjfmodel1}.

The same integrals can be derived in closed-form if a different $F_{\mathrm{C}}$--approximation is taken:

\begin{eqnarray}
    F\left[\textbf{q}\left(t\right),t\right] \rvert_{t \in \left(t_1+k \Delta, t_1+\left(k + 1 \right)\Delta \right)}
    \approx
    F_{k} = \mathrm{const}  \label{eq:ebmodel}
\end{eqnarray}
This constant force approximation allows integration of \eqref{eq:langevin}--\eqref{eq:eom} as a solution of the inhomogeneous differential equation that leads to the Ermak-Buckholz (EB)~\cite{ermak1980numerical} method:
\begin{eqnarray}
    p_{k+1} &=& 
      p_{k} e^{-\beta \Delta}
      + \frac{E F_{k}}{\beta}
      + \tilde{p}_{k} \label{eq:eb-p} \\
    q_{k+1} &=&
      q_{k}
      + \frac{E p_k}{\gamma}
      + \frac{F_{k}}{\gamma}
        \left[
          \Delta
        - \frac{E}{\beta}
        \right]
      + \tilde{q}_{k} \label{eq:eb-q} \\
      E &\stackrel{\mathrm{def}}{=}&
        1 - e^{-\beta \Delta} \label{eq:E-def}
\end{eqnarray}

Here, the noise components $\tilde{p}_k$ and $\tilde{q}_k$ are \emph{correlated} random steps in phase space:
\begin{eqnarray}
    \left< \tilde{p}_k \right> &=& 0 \label{eq:eb-aver-p} 
\\
    \left< \tilde{q}_k \right> &=& 0 \label{eq:eb-aver-q} 
\\
    \left< \tilde{p}_{k} \tilde{q}_{k} \right> &=&
        \frac
            {k_{\mathrm{B}}T E^2}
            {\beta} \label{eq:eb-aver-pq}
\\
    \left< \left( \tilde{p}_{k} \right)^2 \right> &=&
        k_{\mathrm{B}}T m
        \left( 1 - e^{-2\beta \Delta} \right) \label{eq:eb-aver-p2}
\\
    \left< \left( \tilde{q}_{k} \right)^2 \right> &=&
        \frac
            {k_{\mathrm{B}}T}
            {m \beta^2}
        \left(
            2 \beta \Delta
            - 3
            + 4 e^{- \beta \Delta}
            - e^{- 2 \beta \Delta}
        \right) \label{eq:eb-aver-q2}
\end{eqnarray}

A detailed algorithm for selecting the random steps is described in~\cite{ermak1980numerical} and mirrored in our implementation~\cite{EBimpl}. Depending on the order of selection of the random steps two variations of EB method are possible: the velocity-EB and the position-EB (vEB, pEB accordingly).

The Brownian Dynamics (BD) scheme ~\cite{Schlick2010} was originally suggested by Ermak and McCammon (EM)~\cite{ermak1978brownian}. It is the overdamped limit of the EB~\eqref{eq:ebmodel}---~\eqref{eq:eb-aver-q2} which is another kind of $t_{\Delta}$--approximation with:
\begin{eqnarray}
    \Delta \gg \Delta_0
    \label{eq:em-model}
\end{eqnarray}
It leads to  vanishing of all  terms of order $o\left( \left. \Delta \right/ \Delta_0 \right)$ in~\eqref{eq:eb-p}---\eqref{eq:eb-aver-q2} transforming everything back to the classical formulae for diffusion~\cite{einstein1905movement} and viscous drift.

Our previously published method~\cite{tanygin2019langevin} (``AI integrator'') corresponds to EB with $q$--approximation of zero cross-correlation~\eqref{eq:eb-aver-pq}. Another similar integrator is the Long-Timestep Inertial Dynamics (LTID)~\cite{beard2000inertial} corresponding to the $F_{\mathrm{C}}$--approximation and propagation structure of the EB method~\eqref{eq:ebmodel}--\eqref{eq:E-def} with additional $f_{\mathrm{C}}$--approximation: instead of correlated $\tilde{p}_k$ and $\tilde{q}_k$ terms, the average over $\Delta$ random force of the BBK and GJF type~\eqref{eq:bbkrnd} is added to the conservative force as if it was constant during the elementary timestep as well. The LTID integrator has its limit form which was referred to as the inertial Brownian Dynamics (IBD)~\cite{beard2000inertial} if other $o\left( \left. \Delta \right/ \Delta_0 \right)$--terms are retained but terms like $e^{- \beta \Delta}$ are dropped. It is \textit{a priori} known and validated by our tests that the AI, LTID and IBD schemes have performances that fall between those of  EB and  EM. Therefore, we will consider only LTID for the sake of simplicity.

The EB method is an integrator of special interest because each state is integrated almost without approximations and the timestep can be as large as possible. Its only approximation is a constant conservative force within the single time step~\eqref{eq:ebmodel}. It is possible to exclude or minimize such approximation with a proper two-point interpolation of the conservative force. Depending on a particular physical problem, different basis functions for interpolation can be selected. One special form is the integrator named ``Langevin impulse'' (LI)~\cite{skeel2002impulse}:
\begin{eqnarray}
    F\left[\textbf{q}\left(t\right),t\right]
    &\approx&
      \left(
        F_{k}
       - F_{k-1}
      \right)
      \chi\left(  t - t_{k}  \right)
      \label{eq:limodel} \\
    \chi\left(t\right) &=&
      \frac
      {1 - e^{-\beta \left(  t - t_{k}  \right)}}
      {e^{\beta\Delta} - 1} \label{eq:limodel-chi}
\end{eqnarray}

The method of van Gunsteren and Berendsen~\cite{van1982algorithms} (vGB82) uses a different interpolation function that assumes that forces change linearly in time:
\begin{eqnarray}
    \chi_i\left(t\right) =
      \frac
      {t - t_{k}}
      {\Delta}
        \label{eq:vgb82-chi}
\end{eqnarray}

It is obvious that methods involving a closed-form solution of the Langevin equation (e.g. EM, LTID, EB, LI and vGB82) benefit from using large time steps as long as their respective force approximation suits. Most of them (except EM) reduce to the Hamiltonian equations of motion~\eqref{eq:langevin}---\eqref{eq:eom} at $\Delta \xrightarrow{} 0$. However, a part of these methods does not have a quasi-symplectic structure~\cite{melchionna2007design} and time-reversibility in the limiting case of deterministic dynamics. An interesting integrator combining unrestricted time steps and temporally symmetrical structure is the so-called BAOAB integrator~\cite{leimkuhler2013robust}:
\begin{eqnarray}
    p_{k+1/3} &=&
      p_{k}
    + F_k
      \frac{\Delta}{2} \label{eq:baoab-eom-1}
\\
    q_{k+1/2} &=&
    q_k
  + p_{k+1/3}
    \frac{\Delta}{2m} \label{eq:baoab-eom-2}
\\
    p_{k+2/3} &=&
      p_{k+1/3}
      e^{-\beta \Delta}
    + \xi_{k}
      \sqrt{
      k_{\mathrm{B}}T
      \left(
        1
      - e^{-2 \beta \Delta}
      \right)
      m
      } \label{eq:baoab-eom-3}
\\
    q_{k+1} &=&
    q_{k+1/2}
  + p_{k+2/3}
    \frac{\Delta}{2m} \label{eq:baoab-eom-4}
\\
    p_{k+1} &=&
      p_{k+2/3}
    + F_{k+1}
      \frac{\Delta}{2} \label{eq:baoab-eom-5}
\end{eqnarray}

The original paper compares several alternatives (including BBK, vGB82, LI) for Molecular Dynamics configurational sampling and concludes that BAOAB is the most accurate one. We will consider this integrator for our selected use cases.

An additional factor is the number of conservative force recalculations per the simulation step. It is important for single-step or a small number of steps usage of the integrators for systems with forces that require high computational efforts (e.g., long-range interactions). In a sequential run of multiple timesteps, all the selected integrators require one conservative force calculation per step because $F_{k}$ can be reused from the end of the previous step.

\begin{table}[h!]
\centering
\begin{tabular}{||p{25mm} p{20mm} p{25mm} p{30mm} p{20mm}||} 
 \hline
 Method & $t$--symmetry limit & Approximations & Conservative \newline force calculations \newline per single step &  Source code \\ [0.5ex] 
 \hline\hline
 BBK~\cite{brunger1984stochastic} & Yes & $t_{\partial}$ & 2 & \cite{vBBKimpl}
 \\ 
 $\lambda05$--VV~\cite{Martys1999} & Yes & $t_{\partial},q_{\lambda}$ & 2 & \cite{lambda05impl}
 \\
 GJF~\cite{gronbech2013simple} & Yes & $t_{\partial},q_{\Delta},F_{\Delta}$ & 2 & \cite{GJFimpl}
 \\
 vEB/pEB~\cite{ermak1980numerical} & No & $t_{\partial},F_{\mathrm{C}}$ & 1 & \cite{EBimpl}
 \\
 EM (BD)~\cite{ermak1978brownian} & No & $t_{\partial},t_{\Delta},F_{\mathrm{C}}$ & 1 & \cite{EMimpl}
 \\
 LTID~\cite{beard2000inertial} & No & $t_{\partial},F_{\mathrm{C}},f_{\mathrm{C}}$ & 1 & \cite{LTIDimpl}
 \\
 vGB82~\cite{van1982algorithms} & Yes & $t_{\partial}$ & 2 & \cite{vGB82impl}
 \\
 LI~\cite{skeel2002impulse} & Yes & $t_{\partial}$ & 2 & \cite{LIimpl}
 \\
 BAOAB~\cite{leimkuhler2013robust} & Yes & $t_{\partial}$ & 2 & \cite{BAOABimpl}
 \\ [1ex]
 \hline
\end{tabular}
\caption{List of selected Langevin integrators. The implied approximations are described in the text}
\label{table:integrators-def}
\end{table}

\section{Results} \label{validation}
All implementation and  results reported here are based on  dimensionless units defined via the conversion~\cite{ermak1980numerical}:
\begin{eqnarray}
  t^{*} &=& t / \Delta_0 \label{eq:t_dim}
\\
  p^{*} &=& p / \sqrt{m_0 k_{\mathrm{B}}T} \label{eq:p_dim},
\end{eqnarray}
where the characteristic mass $m_0$ is equal to that of a given particle because we will be considering identical particles only; all dimensionless units will be denoted by $^*$. The friction tensor is built with equal diagonal-only components $\gamma_0$ as well. All parameters are taken as unitary unless stated otherwise. All other units can be easily derived from~\eqref{eq:t_dim}--\eqref{eq:p_dim}. The equations defining the various integrators keep their structure and do simplify. The main benefit of using dimensionless units~\cite{ermak1980numerical} is that Langevin equation is parameter-free in the free particle case~\eqref{eq:langevin-dimless}, i.e. viscosity and temperature are hidden in the units definition while interactions introduce extra parameters, and can be cast as
\begin{eqnarray}
  \dot{\textbf{p}}^*
    =
    - \textbf{p}^*
    + \textbf{f}^*\left(t^*\right) \label{eq:langevin-dimless}
\end{eqnarray}
Our tests have been performed for the integrators reported in Table~\ref{table:integrators-def}. 
In order to test the integrators with different time steps, the following range of values have been selected: $\Delta^* \in \left[ 0.01,1.5 \right]$ where, in principle certain integrators may be stable only in a smaller portion of this range. The indicative results will be shown for the selected  time steps and for $\lambda05$--VV, GJF, LI, and BAOAB. The full range of analysis will be provided in the figures for precision parameters obtained from the detailed results. 
The pseudo-random number $\xi_k$ can receive different seed values to initiate the random sequence. 
We selected a large enough number of particles (and 3 times more degrees of freedom) and a number of simulation steps to collect statistically meaningful results. The standard error has been calculated depending on the seed values and other simulation parameters variation.

In the following plots, a line chart cut means an end of the integrator stability area for this test. The EM integrator is not designed to provide momentum at all. Hence, corresponding dependencies will be skipped in the following results.

\subsection{Thermal properties} \label{thermal-val}

A first important validation relates to thermal consistency ~\cite{Martys1999,groot1997dissipative}. A first test is given by an ideal gas made of $N=$ 1000 identical point-like particles and no external potential applied. It is  expected to obtain a temperature on each degree of freedom $\langle T^* \rangle \approx 1 $ and stationary over time. We will validate this behavior by using the linear regression analysis of $\langle \Delta T^* \rangle = \langle T^* \rangle - 1$ for 100 000  time steps covering the  interval $\left[ t_1^*,t_2^* \right]$:
\begin{eqnarray}
  \langle \Delta T^* \rangle \left( \tau^* \right)
  &\approx&
  \epsilon'
  + \frac
    {\epsilon'' \tau^*}
    {t_2^* - t_1^*}
  \label{eq:thermal-eps}
\\
  \tau^* &=& t^* - t_1^* \label{eq:tau-def}
\end{eqnarray}
The two variables $\epsilon'$ and $\epsilon''$ evaluate deviations from the expected temperature and deviations from stationarity. Hence, they can be considered as precision (quality) parameters for the thermal test~(Fig.~\ref{fig:Tdrift}). 

\subsubsection{Ferrofluid case} \label{thermal-ff-val}
We then validate the thermal performances of integrators for a non-ideal system composed by particles interacting with long-range interactions, a complex enough test. The ferrofluid particles interact with dipole-dipole forces where the $i$-th magnetic particle contribution to the vector potential is given by:
\begin{eqnarray}
  \textbf{A}^{*}_i \left( \textbf{q} \right)
             &=& \frac
             {\textbf{m}^{*}_i \times \pmb{\rho}^{*}_i}
             {\left( \rho^{*}_i \right)^3} \label{eq:ffpotential} \\
  \pmb{\rho}^{*}_i
  &\stackrel{\mathrm{def}}{=}&
    \textbf{q}^{*} - \textbf{q}_i^{*}
  \label{eq:ffrhodef}
\end{eqnarray}
where the magnetic moment $\textbf{m}^{*}_i$ is selected as constant, unitary, and identical for all particles. The particles repulsion is modelled by the soft-sphere interaction with the order $n=2$~\cite{lange2009comparison}:
\begin{eqnarray}
    U^{\mathrm{SS*}}_i
    =
    \frac
    {E^{*}}{|\pmb{\rho}^{*}_i|^n}
    \biggr\rvert_{\rho_i^{*} < \rho_0^{*}}
    \label{eq:ss_potential} \\
    U^{\mathrm{SS*}}_i = 0
    \biggr\rvert_{\rho_i^{*} \geq \rho_0^{*}}
\end{eqnarray}
The cut-off distance $\rho_0^{*}=10$ was selected for performance and stability reasons. The repulsion potential energy magnitude $E^{*}=500$ has been selected to account the dipole-dipole interaction better. The in silico temperature deviation $ \langle \Delta T^* \rangle \left( \tau^* \right)$ from the correct unitary value has been calculated for each moment of time (Fig.~\ref{fig:Tdrift-dd}). The simulation was performed for the only case of time step $\Delta^{*} = 0.1$.

\subsection{Diffusion}
We consider the translational diffusivity of particles with the same setup of section~\ref{thermal-val}. The accuracy of a given integrator in reproducing the diffusive properties is key to compare particle trajectories with theoretical predictions and experiments ~\cite{gronbech2013simple}.
The mean square displacement at time $t_1^*$ is given by the expression \cite{ermak1980numerical}\begin{eqnarray}
  \delta q^{*2} \left( \tau^* \right) =
\left. \left< \left[
  \textbf{q}^* \left( t^* \right) - \textbf{q}^* \left( t_1^* \right)
  \right]^2 \right> \right/ \left( 3 N \right) &=&
            2 \tau^*
            - 3
            + 4 e^{- \tau^*}
            - e^{- 2 \tau^*} \label{eq:diff-validation}
\end{eqnarray}
Here and hereinafter, the subscript $_\mathrm{sim}$ differentiates numerical values unless expressed explicitly. The expression is a version of the diffusion equation~\cite{einstein1905movement} in the case of a particle at rest  at time $\tau^{*}=0$ that is assigned a given initial velocity and in absence of thermal noise (in the so-called ''kick" experiment). We intend to reproduce the evolution for the time interval  $\tau^{*}=[0, 10]$. Testing this  time range was performed for a small time step $\Delta^*$=0.01. The precision parameter is taken to be:
\begin{eqnarray}
  \frac{1}{t_2^* - t_1^*}
  \int_{0}^{t_2^* - t_1^*}
    \left( \delta r* \right)^2
  d\tau^*
  &=&
  \epsilon'
  \label{eq:diff-eps1}
\\
  \left( \delta r* \right)^2
  &=&
  \frac
  {\left( \delta q^* \right)^2
  -\left( \delta q^*_{\mathrm{sim}} \right)^2}
  {\left( \delta q^* \right)^2}. \label{eq:diff-dr2}
\end{eqnarray}

For large time steps and later times ($\tau^{*} > 10$) the deviation of positional variance $\left( \delta r* \right)^2 \left(\tau^{*}\right)$ and its regression analysis (with the same $\epsilon'$ and $\epsilon''$ parameters as before~\eqref{eq:thermal-eps}) have been calculated for the different integrators~(Fig. \ref{fig:diffusion}-\ref{fig:diffusion-eps}). They represent the basic shift and stationarity of the average particle position variance against expected values~\eqref{eq:diff-validation}.

\subsection{Autocorrelation}
\begin{figure}
  \centering

  \subsubfloat{\includegraphics[width=0.32\columnwidth]{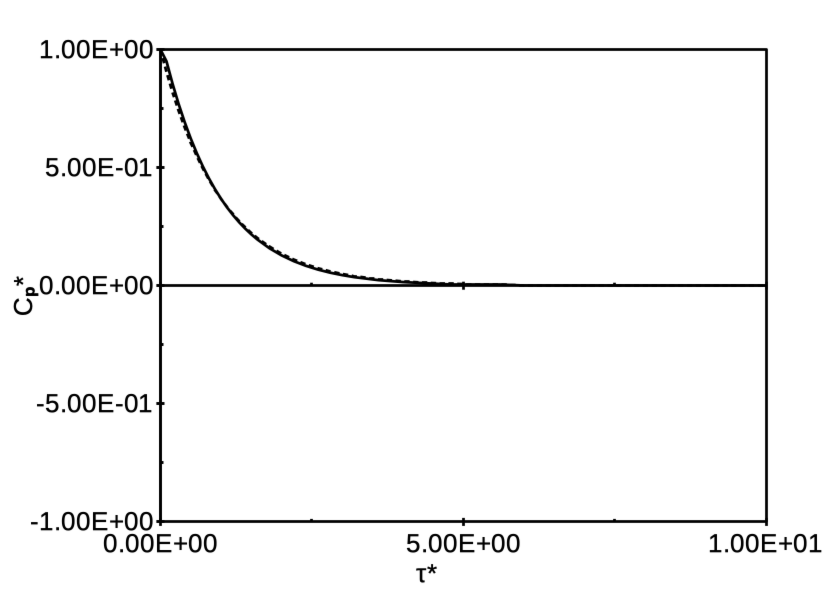}}{a} 
  \subsubfloat{\includegraphics[width=0.32\columnwidth]{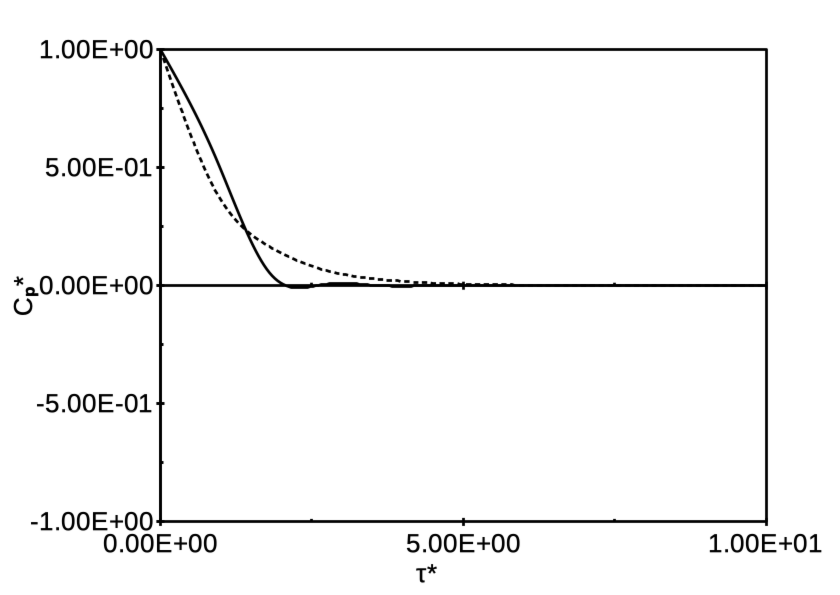}}{b}
  \subsubfloat{\includegraphics[width=0.32\columnwidth]{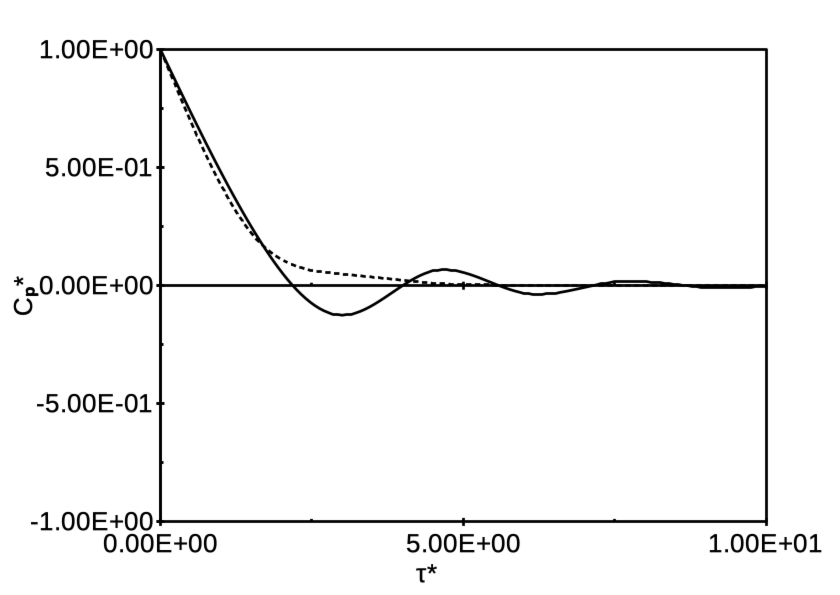}}{c}
  \qquad
  \subsubfloat{\includegraphics[width=0.32\columnwidth]{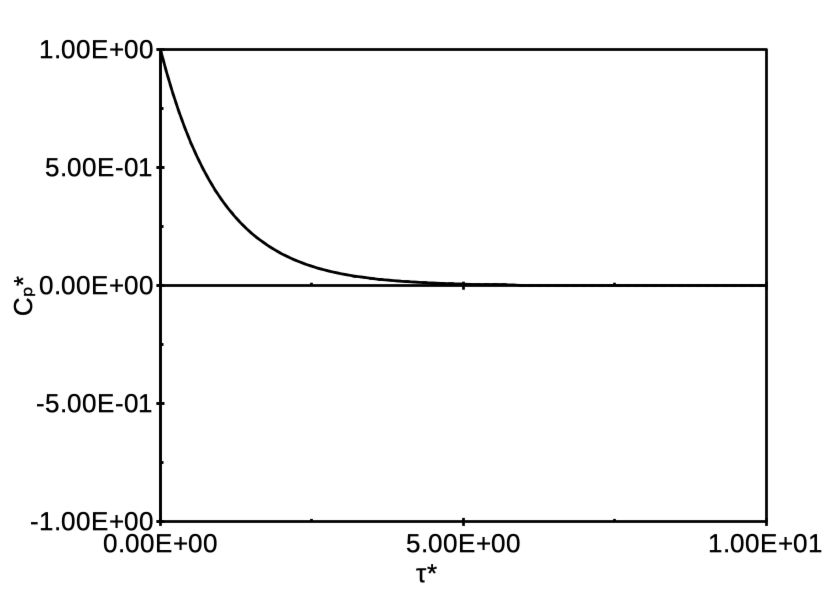}}{d} 
  \subsubfloat{\includegraphics[width=0.32\columnwidth]{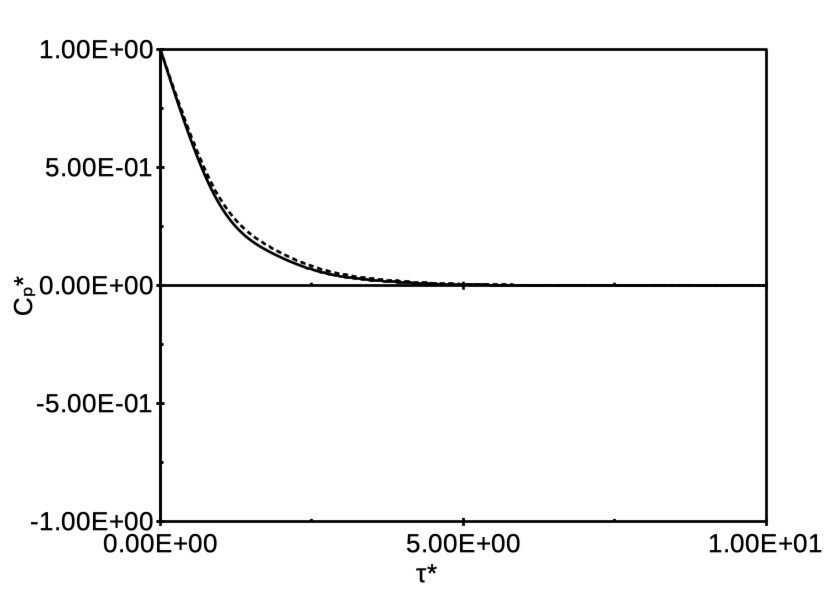}}{e}
  \subsubfloat{\includegraphics[width=0.32\columnwidth]{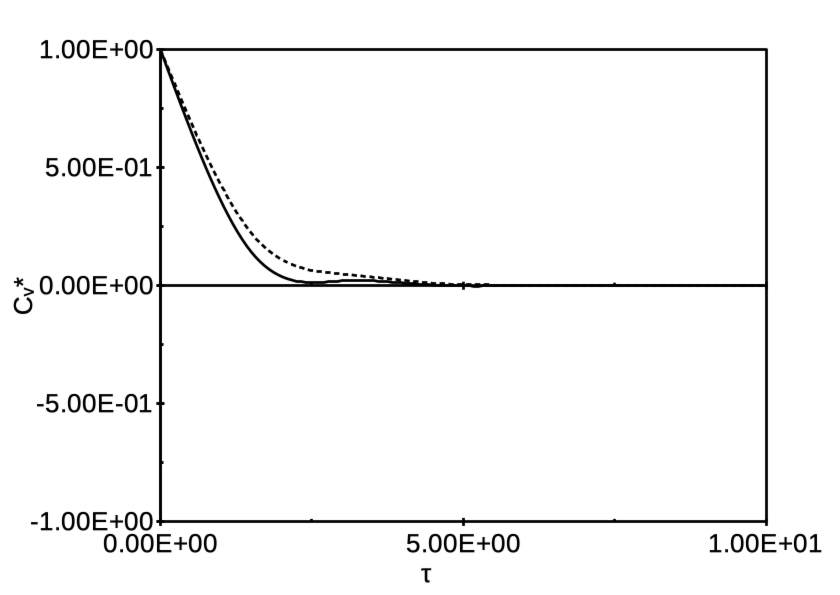}}{f}
  \qquad
  \subsubfloat{\includegraphics[width=0.32\columnwidth]{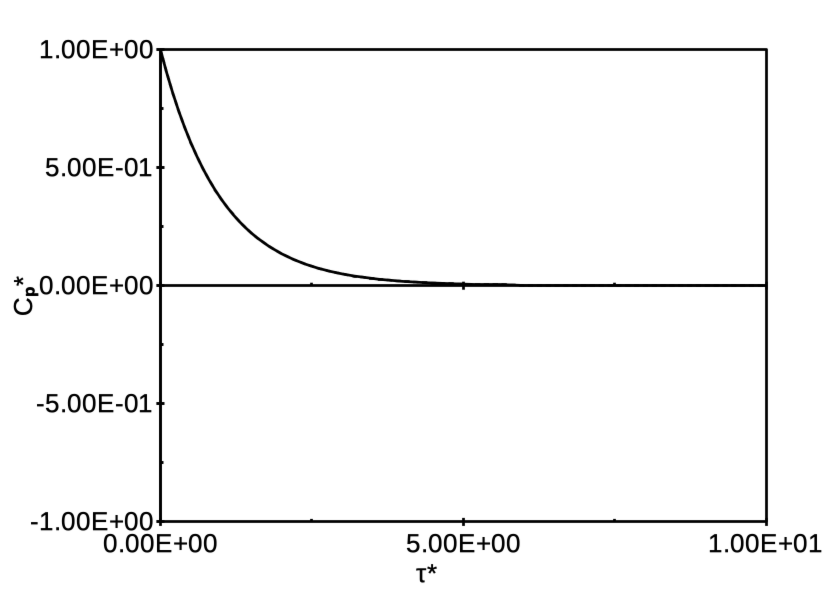}}{g} 
  \subsubfloat{\includegraphics[width=0.32\columnwidth]{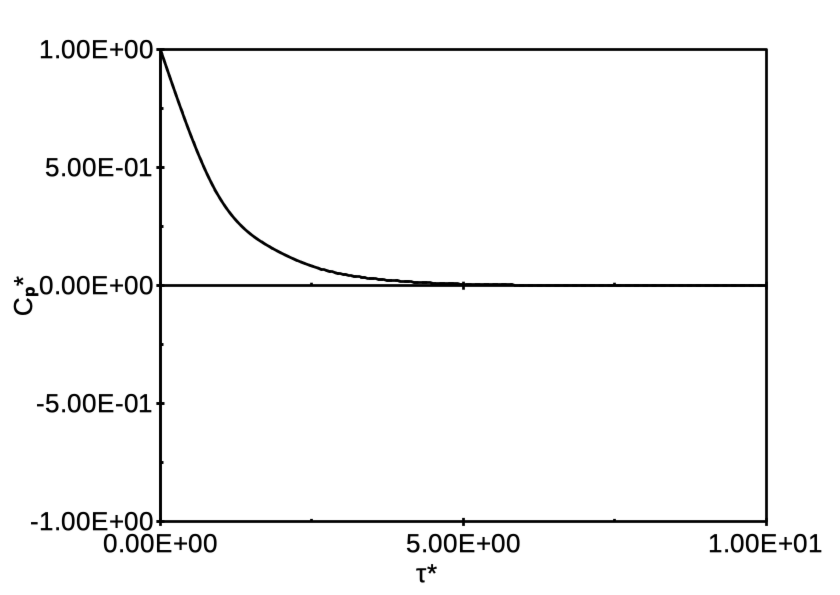}}{h}
  \subsubfloat{\includegraphics[width=0.32\columnwidth]{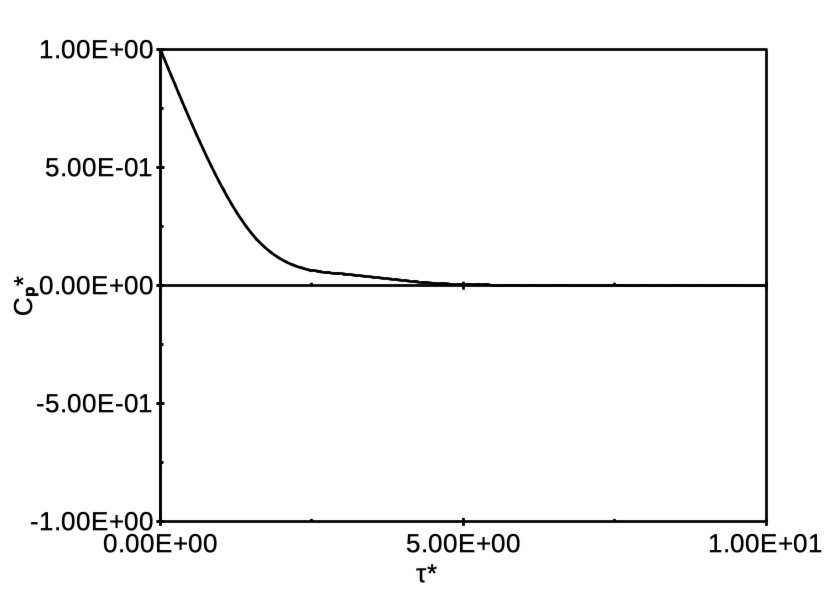}}{i}
  \qquad
  \subsubfloat{\includegraphics[width=0.32\columnwidth]{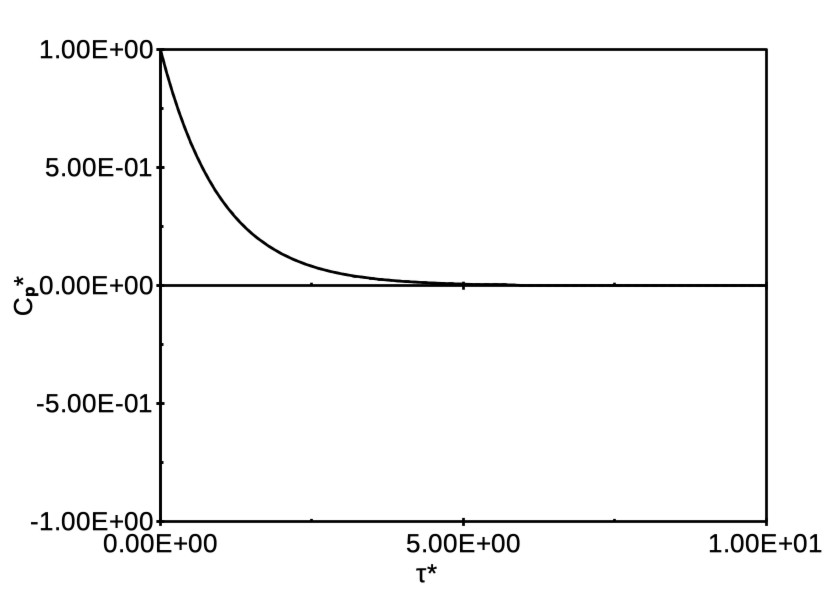}}{j} 
  \subsubfloat{\includegraphics[width=0.32\columnwidth]{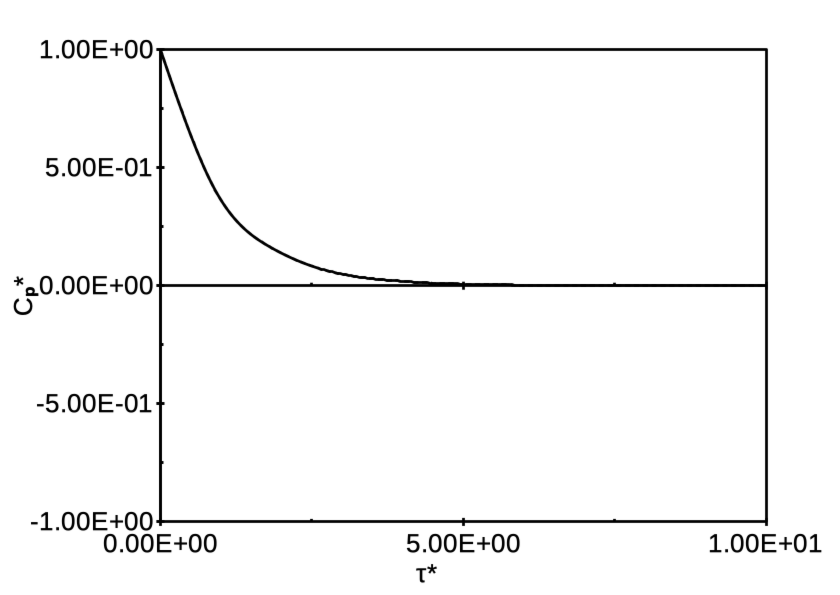}}{k}
  \subsubfloat{\includegraphics[width=0.32\columnwidth]{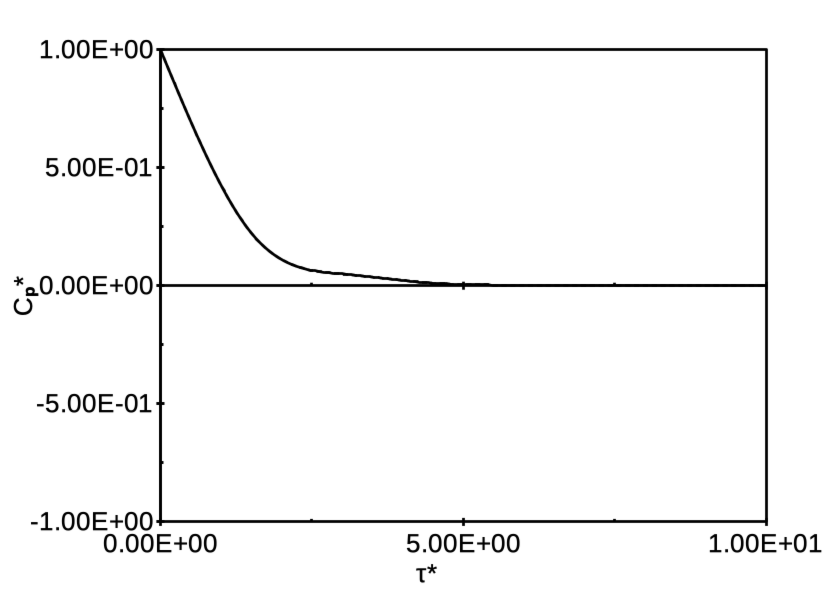}}{l}

\caption{Simulated (solid) and theoretically expected (dashed) momentum autocorrelation function of a free particle for $\lambda05$--VV (a,b,c), GJF (d,e,f), LI (g,h,i), and BAOAB (j,k,l) integrators at time steps $\Delta^*$=0.1 (a,d,g,j), 0.9 (b,e,h,k), and 1.5 (c,f,i,l).} \label{fig:autocorr}

\end{figure}

Another validation is to reproduce the correct autocorrelation properties of position and momentum of a given particles in absence of external potential. By using the Green-Kubo relations~\cite{green1954markoff} for a free particle undergoing Langevin dynamics, the following expression holds:
\begin{eqnarray}
  C_{p}^*\left( \tau^* \right) &=&
  \left.
  \left<
    \textbf{p}^{*} \left( t_1 \right)
    \textbf{p}^{*} \left( t \right)
  \right>
  \right/ \left( 3 N \right)
  =
  e^{- \tau^*} \label{eq:Green-Kubo-2}
\end{eqnarray}
with
\begin{eqnarray}
   \int_{0}^{\infty} C_{p}^*\left( \tau^* \right) d\tau^* &=& I^* = 1 \label{eq:Green-Kubo-1}
\end{eqnarray}
Using the  method described in ~\cite{frenkel2001understanding}, the simulated value $I^*_{\mathrm{sim}}$ (here calculated via the trapezoidal rule) and  $C_{p(\mathrm{sim})}^*\left( \tau^* \right)$ determines the integrator accuracy ~(Fig.~\ref{fig:autocorr}-~\ref{fig:autocorr-eps}):
\begin{eqnarray}
  \Bigm\lvert I^*_{\mathrm{sim}} - 1 \Bigm\rvert
  &=& \epsilon'
  \label{eq:autocorr-eps1}
\\
  \frac{1}{t_2^* - t_1^*}
  \int_{0}^{t_2^* - t_1^*}
  \Bigm\lvert
  \frac
  {C_{p(\mathrm{sim})}^*\left( \tau^* \right)
 - C_{p}^*\left( \tau^* \right)}
  {C_{p}^*\left( \tau^* \right)}
  \Bigm\rvert
  d\tau^*
  &=&
  \epsilon''
  \label{eq:autocorr-eps2}
\end{eqnarray}
A similar accuracy metrics was suggested in the study of IBD and LTID ~\cite{beard2000inertial} for the harmonic potential case. We use it for the free particle case as well as the same harmonic one (in the following section). Criteria like~\eqref{eq:autocorr-eps2} have been applied only for the  range such that $\lvert C_{p}^*\left( \tau^* \right) \rvert > 0.01$, ignoring smaller values with spurious digits which are more sensitive to implementation details than to the integrator structure itself.

\subsection{Harmonic potential}
The harmonic potential case is a basic model for microscopical interactions and correspondingly, to validate the quality of auto-correlations ~\cite{beard2000inertial}. In addition, several real systems can be approximated at first order by such interactions in the case of small departures from equilibrium positions. The corresponding Langevin equation is given by:
\begin{eqnarray}
  \dot{\textbf{p}}^*
    &=&
    - \kappa^* \textbf{q}^*
    - \textbf{p}^*
    + \textbf{f}^*\left(t^*\right) \label{eq:langevin-harm}
\\
  \kappa^*
  &\stackrel{\mathrm{def}}{=}&
  \kappa
  \frac{\Delta_0}{\gamma_0},
  \label{kappa-def}
\end{eqnarray}
where $\kappa$ is Hooke's constant. Expected autocorrelation functions can be derived using the Wiener--Khinchin theorem~\cite{coffey2012langevin}. We use units with dimensions again for better reference:
\begin{eqnarray}
  C_{q}\left( \tau \right) =
  \left.
  \left<
    \textbf{q} \left( t_1 \right)
    \textbf{q} \left( t \right)
  \right>
  \right/ \left( 3 N \right)
  &=&
  \frac
  {\beta_0 \mykB T}
  {\pi m_0}
  \int_{-\infty}^{\infty}
    \frac
    {e^{-i \omega \tau} d\omega}
    {\left( \omega_0^2 - \omega^2 \right)^2 + \omega^2 \beta_0^2}
  \label{eq:harm-autocorr-q}
\\
  C_{p}\left( \tau \right) &=&
  - m_0^2
  \frac
  {d^2}
  {d\tau^2}
  C_{q}\left( \tau \right),
  \label{eq:harm-autocorr-p}
\end{eqnarray}
where $\beta_0 = \left. 1 \right/ \Delta_0$ and $\omega_0 = \sqrt{\left. \kappa \right/ m_0}$. The autocorrelation functions can be derived for the underdamped ~($\kappa^* > 0.25$), critical~($\kappa^* = 0.25$), and overdamped~($0 < \kappa^* < 0.25$) cases ~ \cite{beard2000inertial,gronbech2013simple} and read:
\begin{eqnarray}
  C_{q}\left( \tau \right)
  \biggr\rvert_{\kappa^* > 1/4}
  &=&
    \frac
    {\mykB T}
    {m_0 \omega_0^2}
    e^{-\beta_0 \tau / 2}
    \left(
      \cos \omega_1 \tau +
      \frac{\beta_0}{2\omega_1}
      \sin \omega_1 \tau
    \right)
  \label{eq:harm-autocorr-q-sol-underdamped}
\\
  C_{p}\left( \tau \right)
  \biggr\rvert_{\kappa^* > 1/4}
  &=&
    \mykB T m_0
    e^{-\beta_0 \tau / 2}
    \left(
      \cos \omega_1 \tau -
      \frac{\beta_0}{2\omega_1}
      \sin \omega_1 \tau
    \right)
  \label{eq:harm-autocorr-p-sol-underdamped}
\\
  C_{q}\left( \tau \right)
  \biggr\rvert_{\kappa^* = 1/4}
  &=&
    \frac
    {2 \mykB T}
    {m_0 \beta_0^2}
    e^{-\beta_0 \tau / 2}
    \left(
      2 + \beta_0 \tau
    \right)
  \label{eq:harm-autocorr-q-sol-critical}
\\
  C_{p}\left( \tau \right)
  \biggr\rvert_{\kappa^* = 1/4}
  &=&
    \mykB T m_0
    e^{-\beta_0 \tau / 2}
    \left(
      1 - \frac{\beta_0 \tau}{2}
    \right)
  \label{eq:harm-autocorr-p-sol-critical}
\\
  C_{q}\left( \tau \right)
  \biggr\rvert_{\kappa^* < 1/4}
  &=&
  \frac
  {\mykB T}
  {2 \omega_0^2 \beta_1 m_0}
  \left(
    - e^{-\beta_{+}\tau} \beta_{-}
    + e^{-\beta_{-}\tau} \beta_{+}
  \right)
  \label{eq:harm-autocorr-q-sol-overdamp}
\\
  C_{p}\left( \tau \right)
  \biggr\rvert_{\kappa^* < 1/4}
  &=&
  \frac
  {\mykB T m_0}
  {2 \beta_1}
  \left(
    + e^{-\beta_{+}\tau} \beta_{+}
    - e^{-\beta_{-}\tau} \beta_{-}
  \right)
  \label{eq:harm-autocorr-p-sol-overdamp}
\end{eqnarray}
where the following definitions have been used:
\begin{eqnarray}
  \omega_1^2
  &\stackrel{\mathrm{def}}{=}&
  \omega_0^2 - \left. \beta_0^2 \right/ 4
  \label{eq:omega1-def}
\\
  \beta_1^2
  &\stackrel{\mathrm{def}}{=}&
  -\omega_0^2 + \left. \beta_0^2 \right/ 4
  \label{eq:beta1-def}
\\
  \beta_{\pm}
  &\stackrel{\mathrm{def}}{=}&
  \sqrt{\beta_0
          \left(
            \frac{\beta_0}{2} \pm \beta_1
          \right)
        - \omega_0^2
       }
  \label{eq:betapm-def}
\end{eqnarray}

The corresponding integrals read:
\begin{eqnarray}
  \int_{0}^{\infty} C_{q}\left( \tau \right) d\tau
  &\stackrel{\mathrm{def}}{=}&
  \zeta_q
  \label{eq:harm-int-def-q}
\\
  \int_{0}^{\infty} C_{p}\left( \tau \right) d\tau
  &\stackrel{\mathrm{def}}{=}&
  \zeta_p
  \label{eq:harm-int-def-p}
\\
  \zeta_q
  \biggr\rvert_{\kappa^* > 1/4}
  &=&
  \frac
  {\mykB T \gamma_0}
  {\kappa^2}
  \label{eq:corr-q-underdamp}
\\
  \zeta_p
  \biggr\rvert_{\kappa^* > 1/4}
  &=& 0
  \label{eq:corr-p-underdamp}
\\
  \zeta_q
  \biggr\rvert_{\kappa^* = 1/4}
  &=&
  \frac
  {16 \mykB T}
  {m_0 \beta_0^3}
  \label{eq:corr-q-crit}
\\
  \zeta_p
  \biggr\rvert_{\kappa^* = 1/4}
  &=& 0
  \label{eq:corr-p-crit}
\\
  \zeta_q
  \biggr\rvert_{\kappa^* < 1/4}
  &=&
  \frac
  {\mykB T}
  {2 m_0 \beta_1 \omega_0^2}
  \left(
    \frac{\beta_{+}}{\beta_{-}}
  - \frac{\beta_{-}}{\beta_{+}}
  \right)
  \label{eq:corr-q-overdamp}
\\
  \zeta_p
  \biggr\rvert_{\kappa^* < 1/4}
  &=& 0
  \label{eq:corr-p-overdamp}
\end{eqnarray}
and dimensionless form: $\zeta_q
  \biggr\rvert_{\kappa^* > 1/4} = 1 \left. \right/ \kappa^2$, $\zeta_q
  \biggr\rvert_{\kappa^* = 1/4} = 16$, and
\begin{eqnarray}
  \zeta_q
  \biggr\rvert_{\kappa^* < 1/4} =
  \frac
  {1}
  {2 \kappa^* \sqrt{1/4 - \kappa^*}}
  \left[
    \frac
    {\beta_{+}^* \left( \tau^* \right)}
    {\beta_{-}^* \left( \tau^* \right)}
  - \frac
    {\beta_{-}^* \left( \tau^* \right)}
    {\beta_{+}^* \left( \tau^* \right)}
  \right]
  \label{eq:corr-q-overdamp-dimless}
\end{eqnarray}
where $\beta_{\pm}^*\left( \tau^* \right)$ are defined in eqs.~\eqref{eq:t_dim}~\eqref{eq:beta1-def}, and \eqref{eq:betapm-def}.

\begin{figure}
  \centering

  \subsubfloat{\includegraphics[width=0.32\columnwidth]{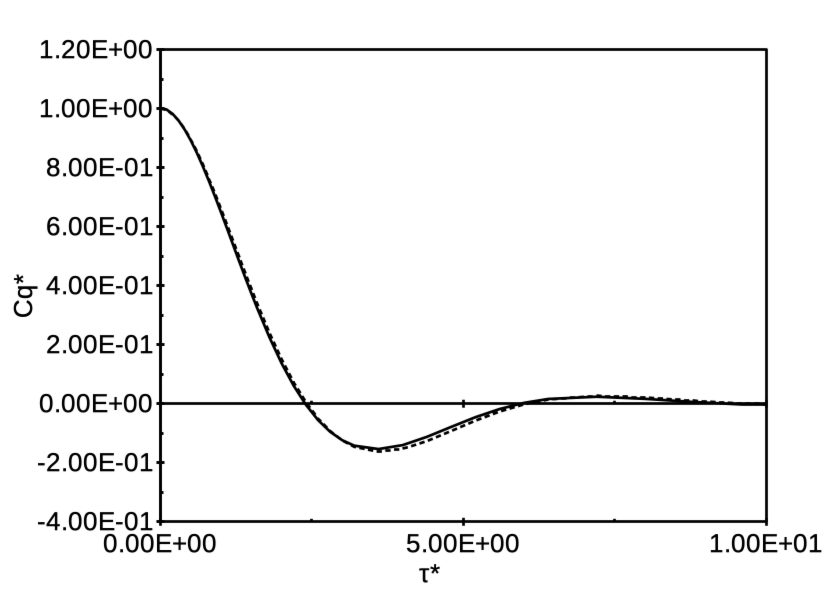}}{a} 
  \qquad
  \subsubfloat{\includegraphics[width=0.32\columnwidth]{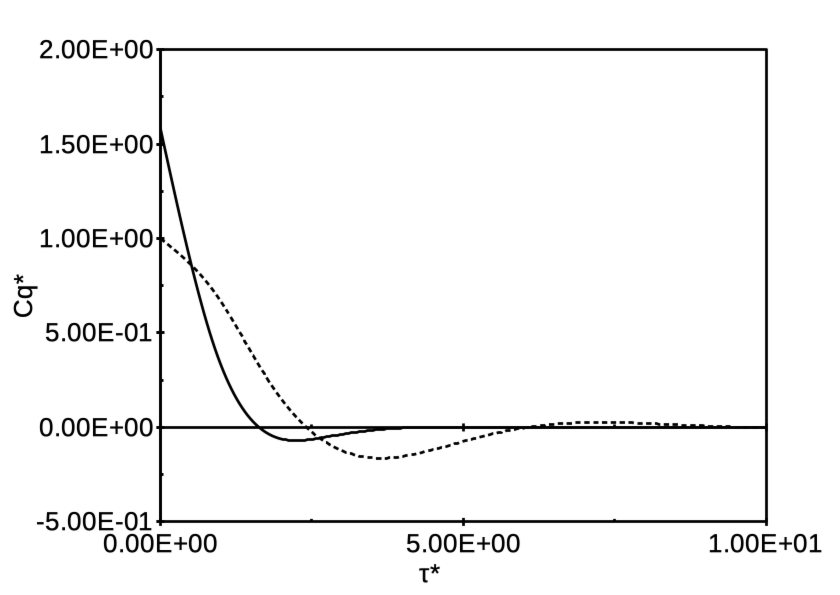}}{b}
  \qquad
  \subsubfloat{\includegraphics[width=0.32\columnwidth]{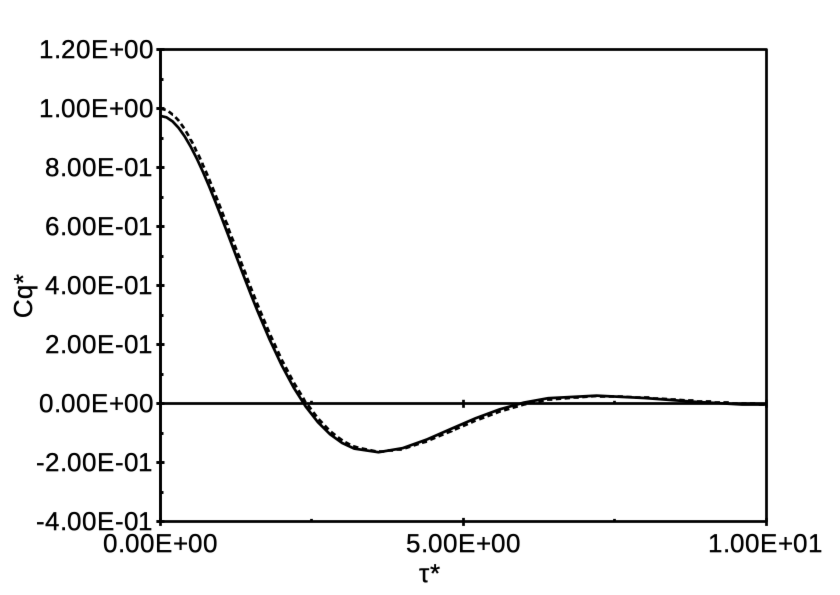}}{c} 
  \subsubfloat{\includegraphics[width=0.32\columnwidth]{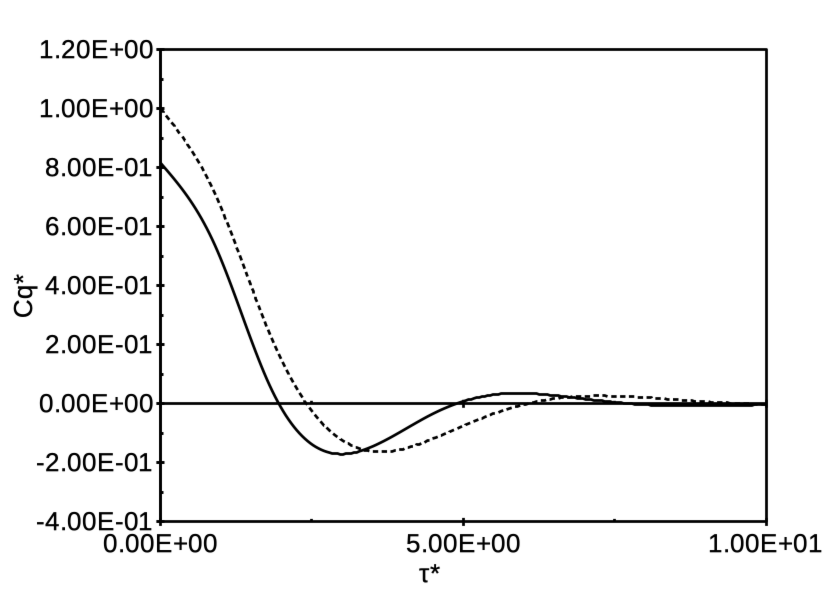}}{d}
  \subsubfloat{\includegraphics[width=0.32\columnwidth]{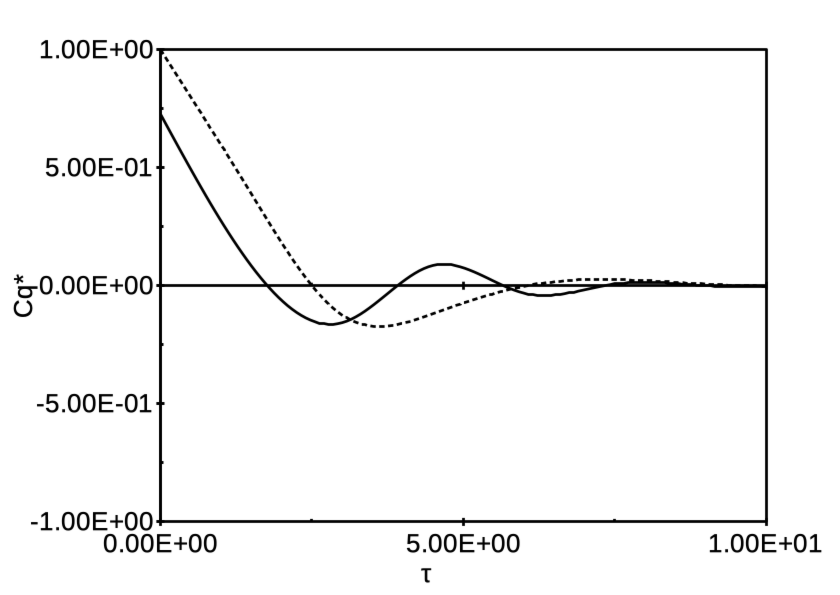}}{e}
  \qquad
  \subsubfloat{\includegraphics[width=0.32\columnwidth]{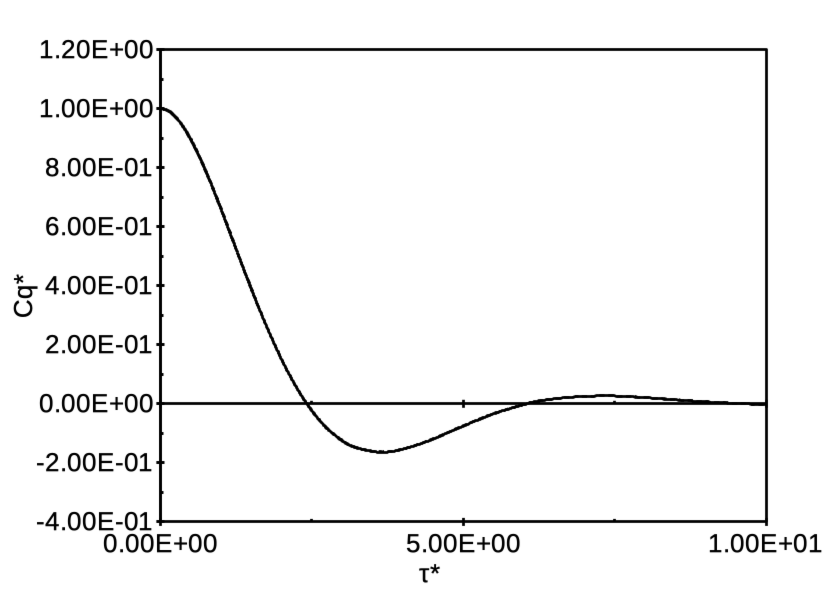}}{f} 
  \subsubfloat{\includegraphics[width=0.32\columnwidth]{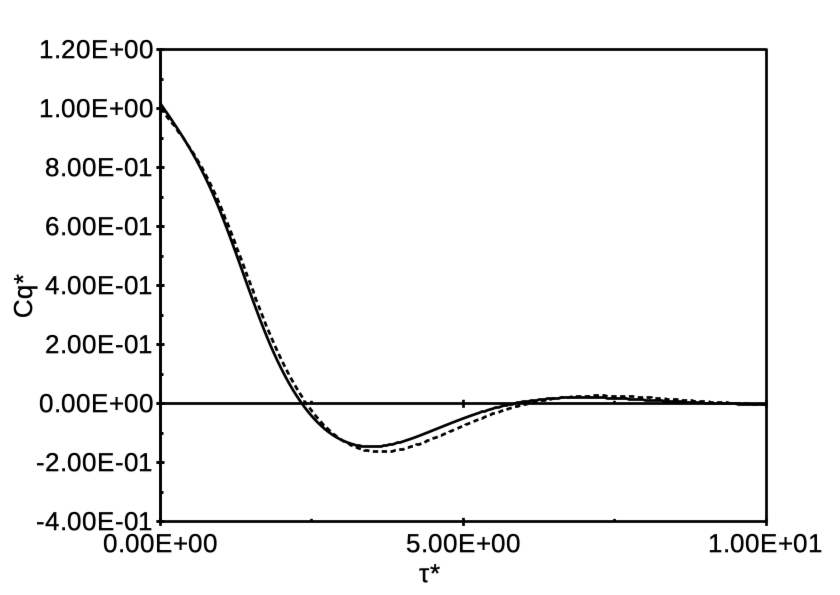}}{g}
  \subsubfloat{\includegraphics[width=0.32\columnwidth]{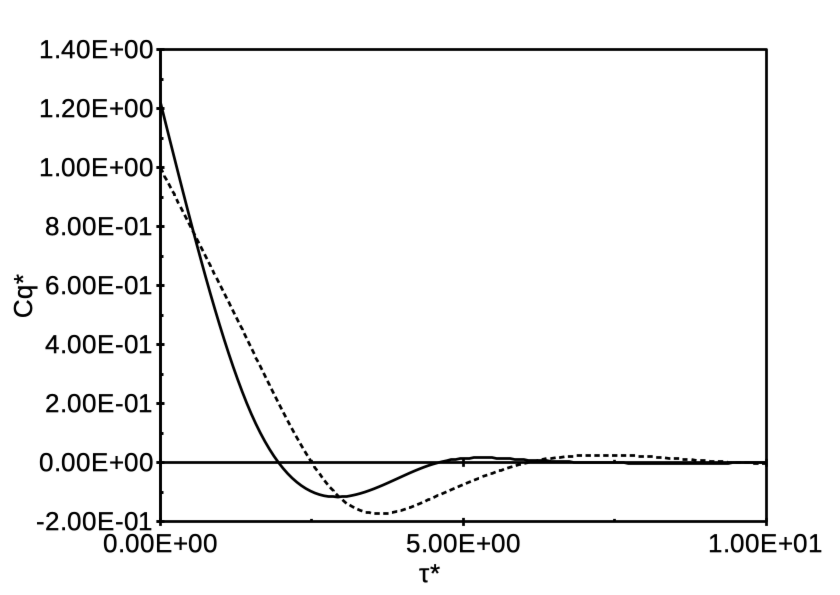}}{h}
  \qquad
  \subsubfloat{\includegraphics[width=0.32\columnwidth]{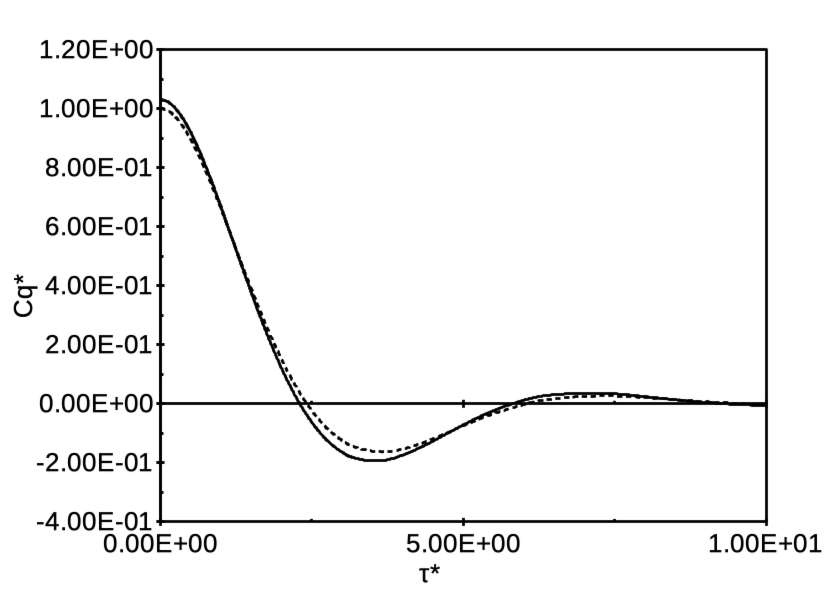}}{i} 
  \subsubfloat{\includegraphics[width=0.32\columnwidth]{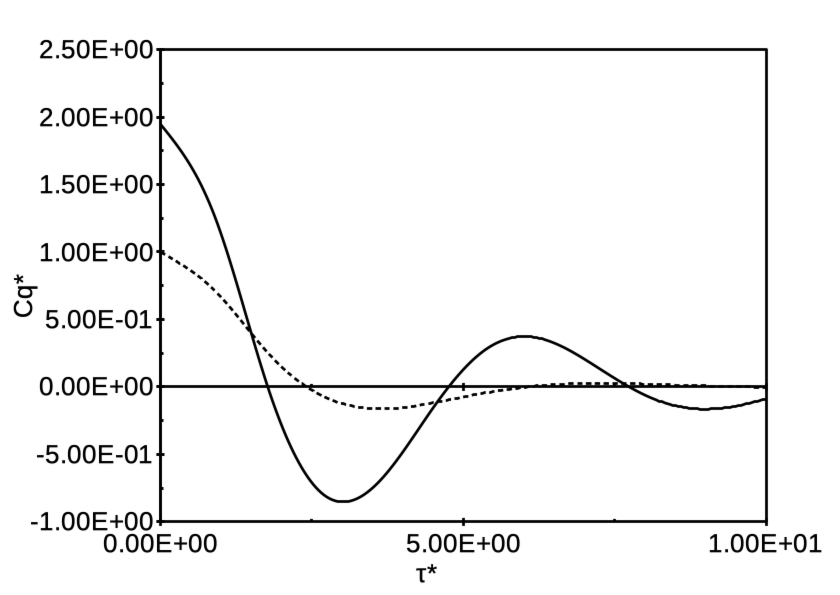}}{j}
  \subsubfloat{\includegraphics[width=0.32\columnwidth]{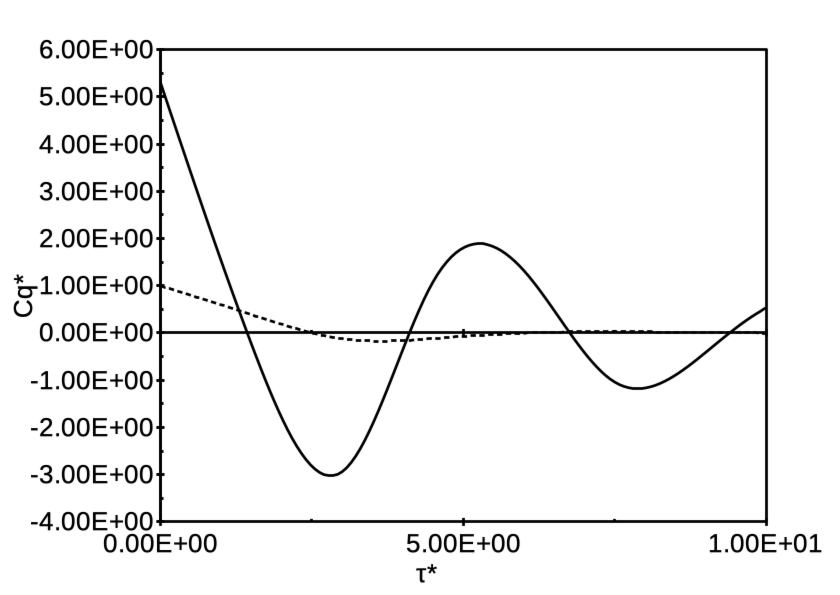}}{k}

\caption{Simulated (solid) and theoretical (dashed) position autocorrelation function of a particle in the harmonic potential (underdamped case $\kappa^*$=1) for $\lambda05$--VV (a,b), GJF (c,d,e), LI (f,g,h), and BAOAB (i,j,k) integrators at time steps $\Delta^*$=0.1 (a,c,f,i), 0.9 (b,d,g,j), and 1.5 (e,h,k).} \label{fig:autocorr-harm-pos}

\end{figure}

\begin{figure}
  \centering

  \subsubfloat{\includegraphics[width=0.32\columnwidth]{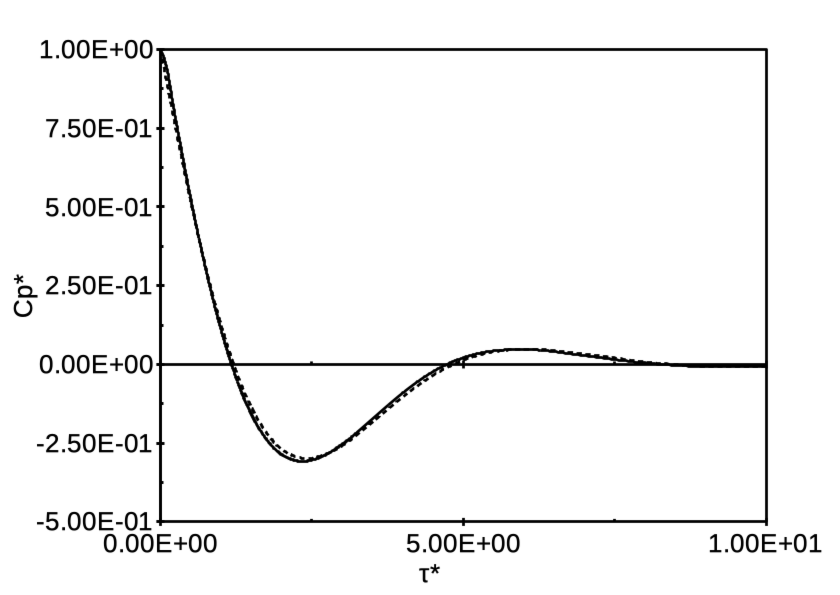}}{a} 
  \qquad
  \subsubfloat{\includegraphics[width=0.32\columnwidth]{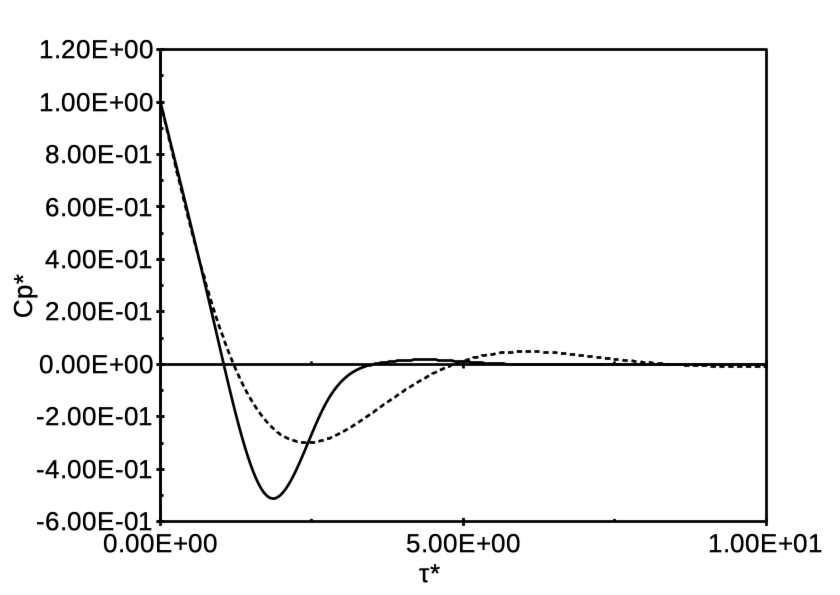}}{b}
  \qquad
  \subsubfloat{\includegraphics[width=0.32\columnwidth]{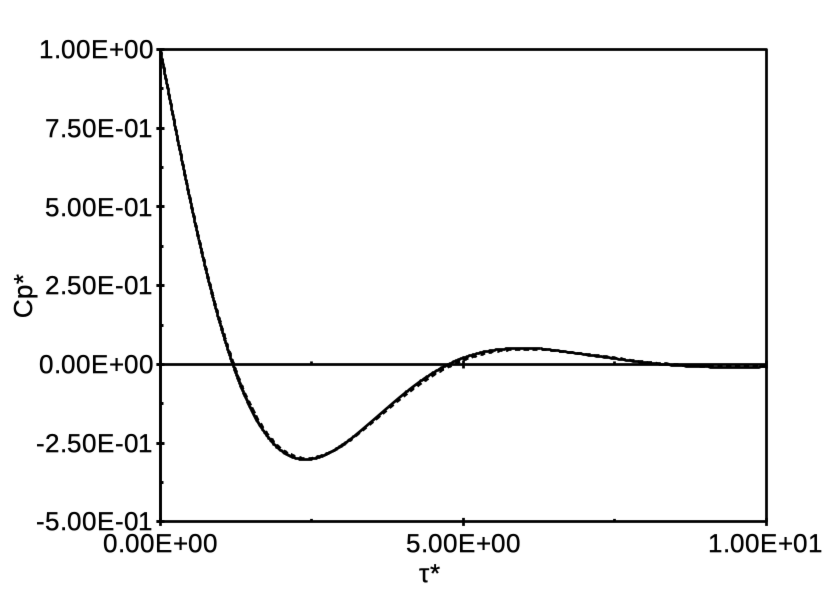}}{c} 
  \subsubfloat{\includegraphics[width=0.32\columnwidth]{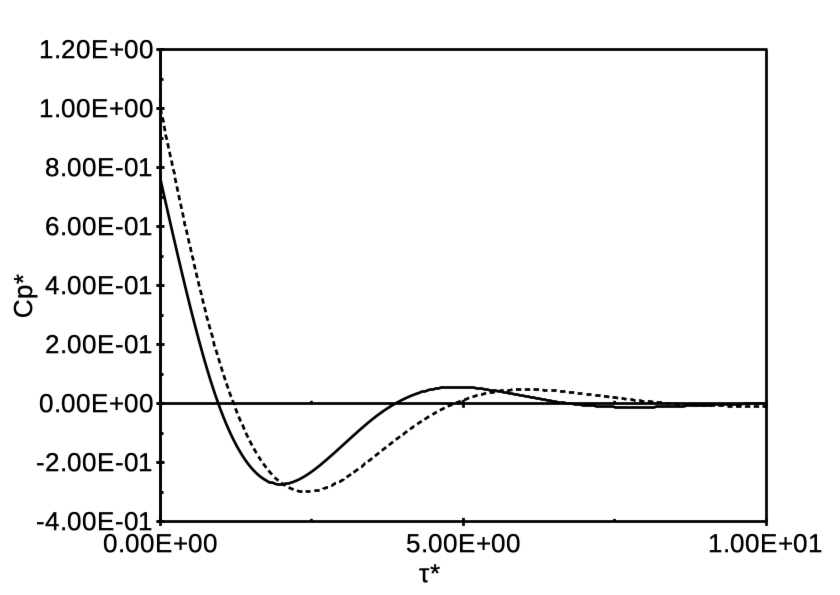}}{d}
  \subsubfloat{\includegraphics[width=0.32\columnwidth]{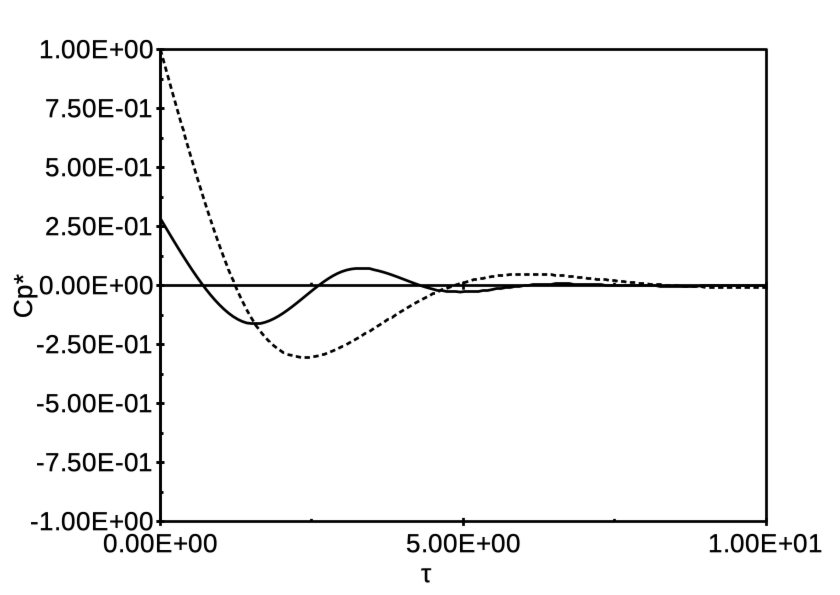}}{e}
  \qquad
  \subsubfloat{\includegraphics[width=0.32\columnwidth]{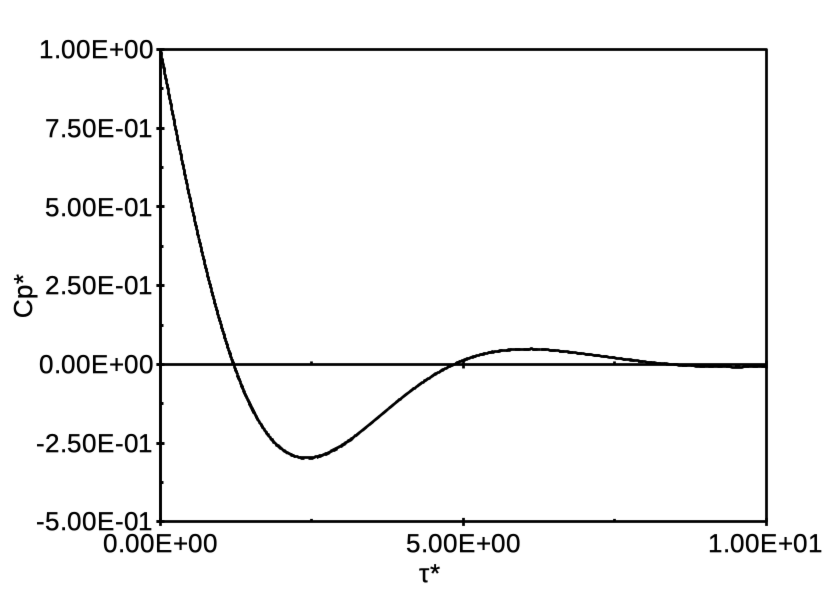}}{f} 
  \subsubfloat{\includegraphics[width=0.32\columnwidth]{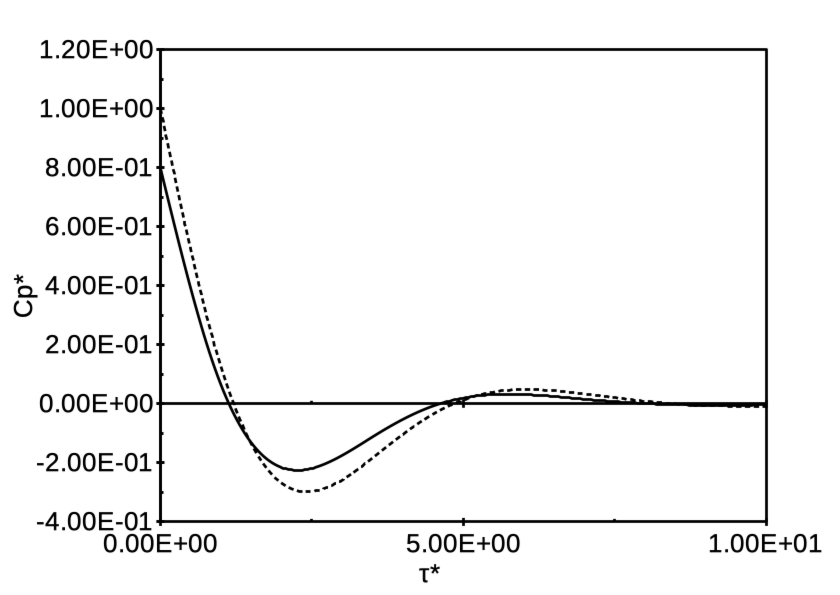}}{g}
  \subsubfloat{\includegraphics[width=0.32\columnwidth]{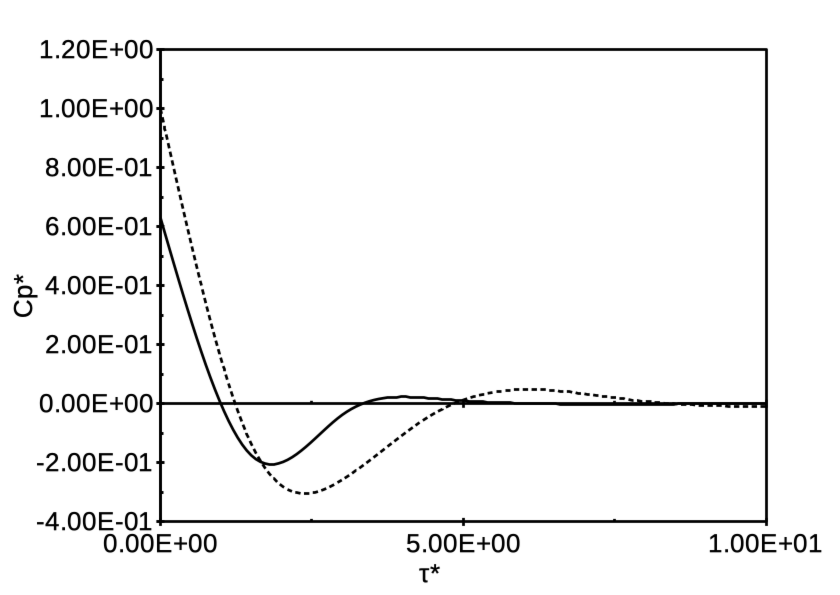}}{h}
  \qquad
  \subsubfloat{\includegraphics[width=0.32\columnwidth]{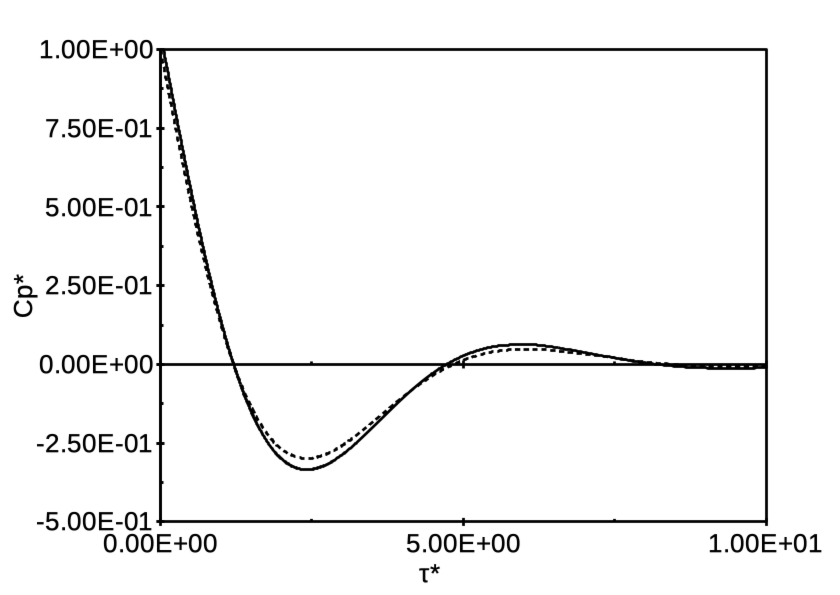}}{i} 
  \subsubfloat{\includegraphics[width=0.32\columnwidth]{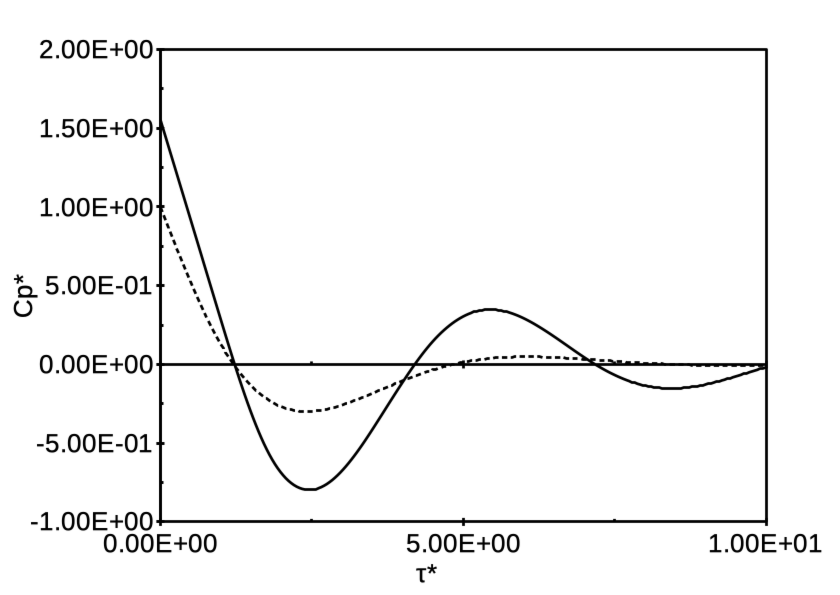}}{j}
  \subsubfloat{\includegraphics[width=0.32\columnwidth]{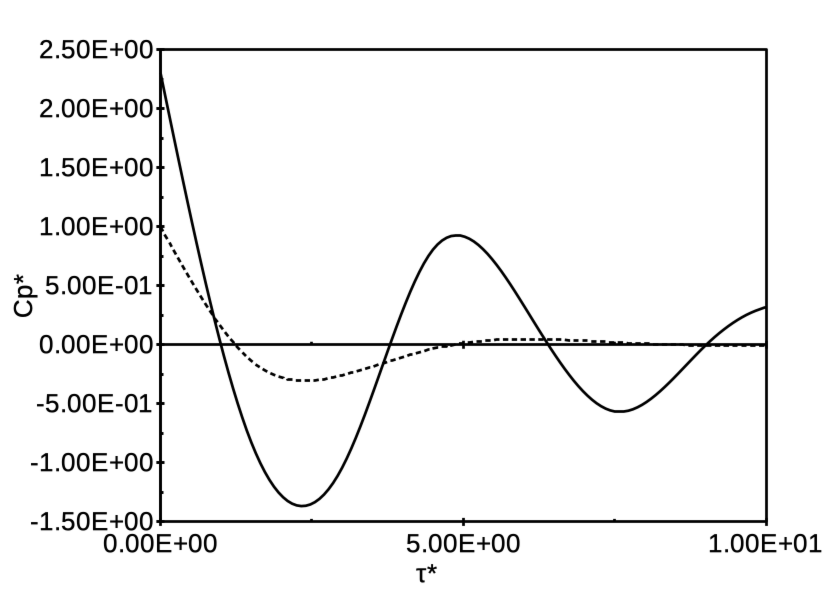}}{k}

\caption{Simulated (solid) and theoretical (dashed) momentum autocorrelation function for a particle in the harmonic potential (underdamped case $\kappa^*$=1) for $\lambda05$--VV (a,b), GJF (c,d,e), LI (f,g,h), and BAOAB (i,j,k)  at time steps $\Delta^*$=0.1 (a,c,f,i), 0.9 (b,d,g,j), and 1.5 (e,h,k).} \label{fig:autocorr-harm-momentum}

\end{figure}

The simulated values $\zeta^*_{(\mathrm{sim})q/p}$ and dependencies $C_{q/p(\mathrm{sim})}^*\left( \tau^* \right)$~(Fig.~\ref{fig:autocorr-harm-pos}-Fig.~\ref{fig:autocorr-harm-momentum}) determine the given integrator precision~(Fig.~\ref{fig:autocorr-harm-eps}):
\begin{eqnarray}
  \Bigm\lvert
    \frac
    {\zeta^*_{(\mathrm{sim})q/p} - \zeta^*_{q/p}}
    {\zeta^*_{q/p}}
  \Bigm\rvert
  &=& \epsilon'
  \label{eq:autocorr-harm-eps1}
\\
  \frac{1}{t_2^* - t_1^*}
  \int_{0}^{t_2^* - t_1^*}
  \Bigm\lvert
  \frac
  {C_{q/p(\mathrm{sim})}^*\left( \tau^* \right)
 - C_{q/p}^*\left( \tau^* \right)}
  {C_{q/p}^*\left( \tau^* \right)}
  \Bigm\rvert
  d\tau^*
  &=&
  \epsilon''
  \label{eq:autocorr-harm-eps2}
\end{eqnarray}
\subsection{Constant force}

\begin{figure}
  \centering

  \subsubfloat{\includegraphics[width=0.32\columnwidth]{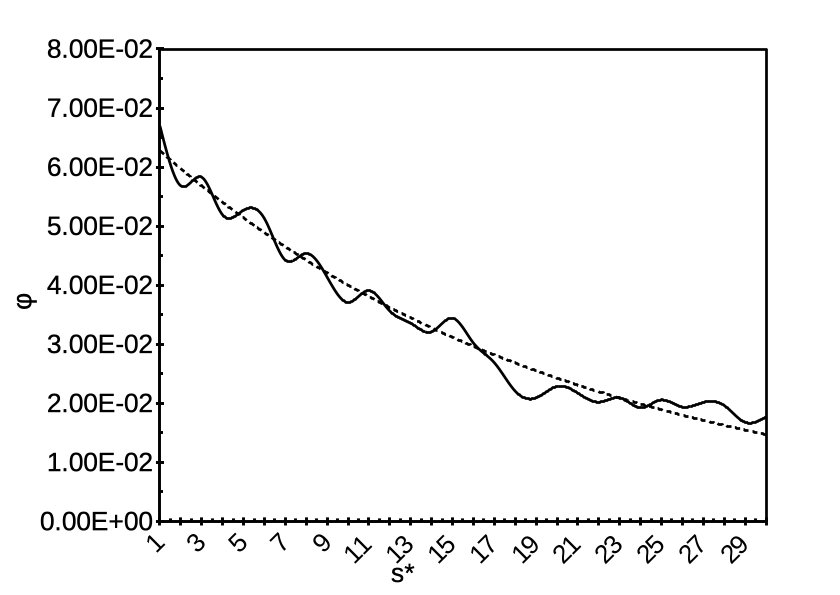}}{a} 
  \subsubfloat{\includegraphics[width=0.32\columnwidth]{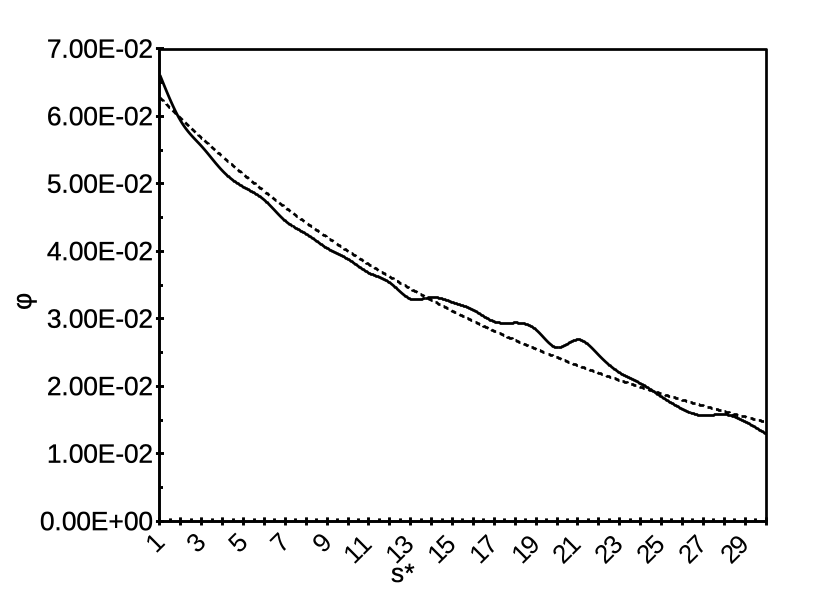}}{b}
  \subsubfloat{\includegraphics[width=0.32\columnwidth]{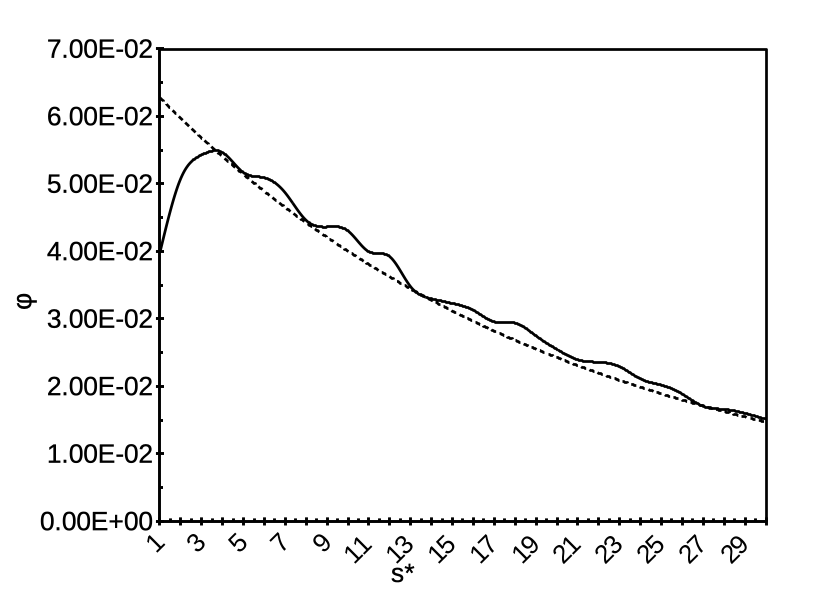}}{c}
  \qquad
  \subsubfloat{\includegraphics[width=0.32\columnwidth]{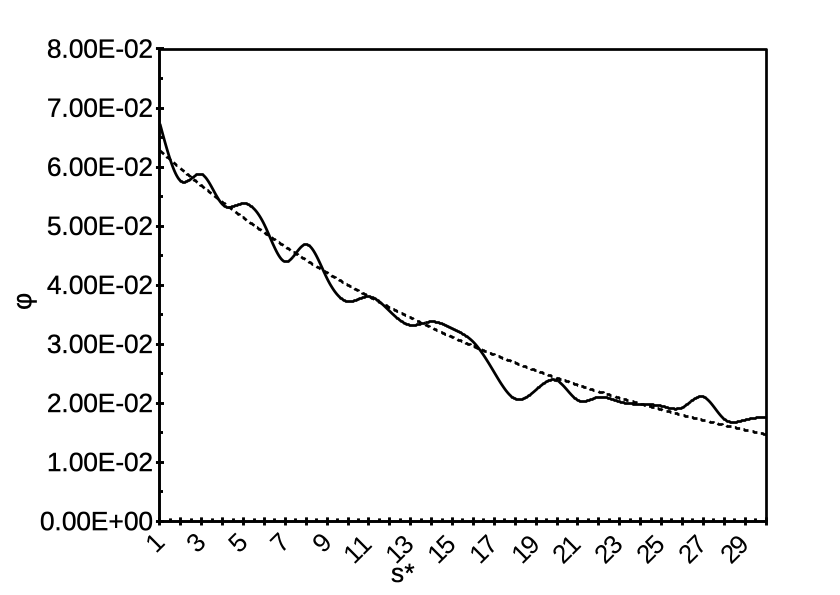}}{d} 
  \subsubfloat{\includegraphics[width=0.32\columnwidth]{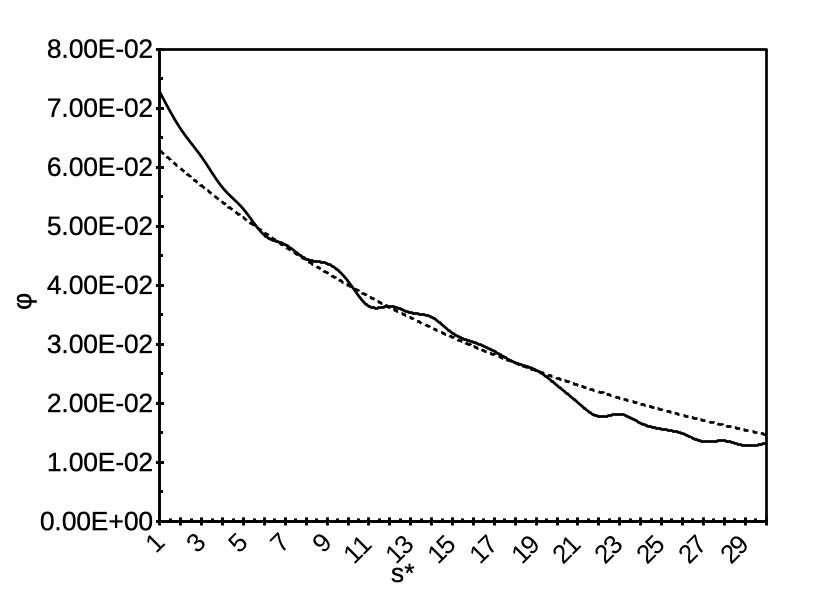}}{e}
  \subsubfloat{\includegraphics[width=0.32\columnwidth]{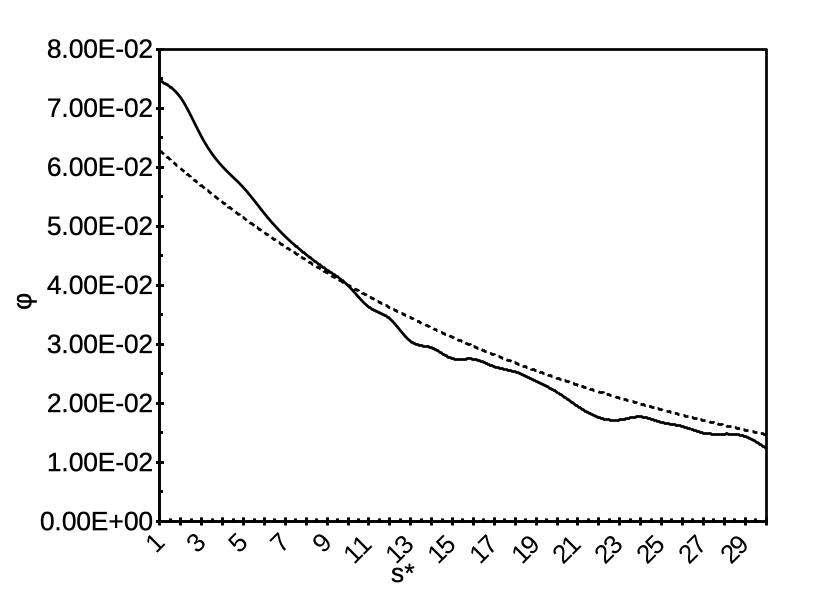}}{f}
  \qquad
  \subsubfloat{\includegraphics[width=0.32\columnwidth]{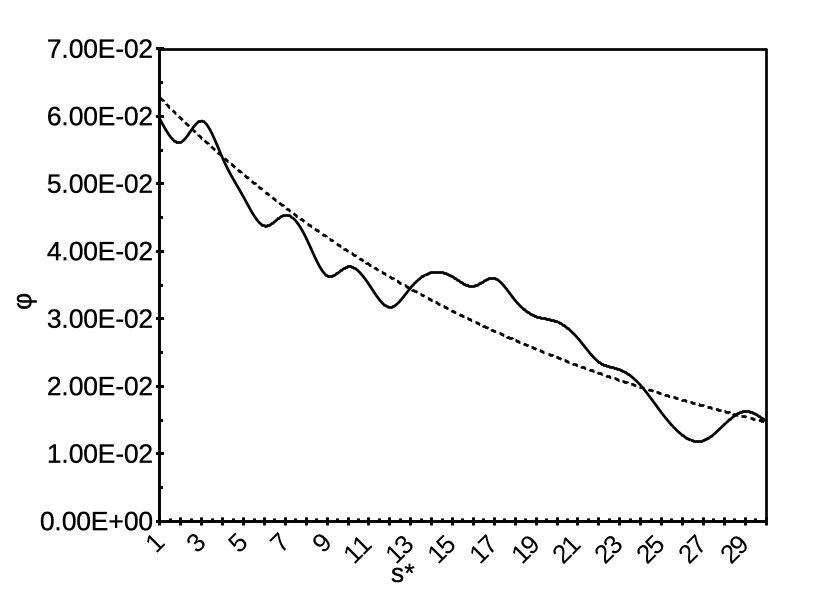}}{g} 
  \subsubfloat{\includegraphics[width=0.32\columnwidth]{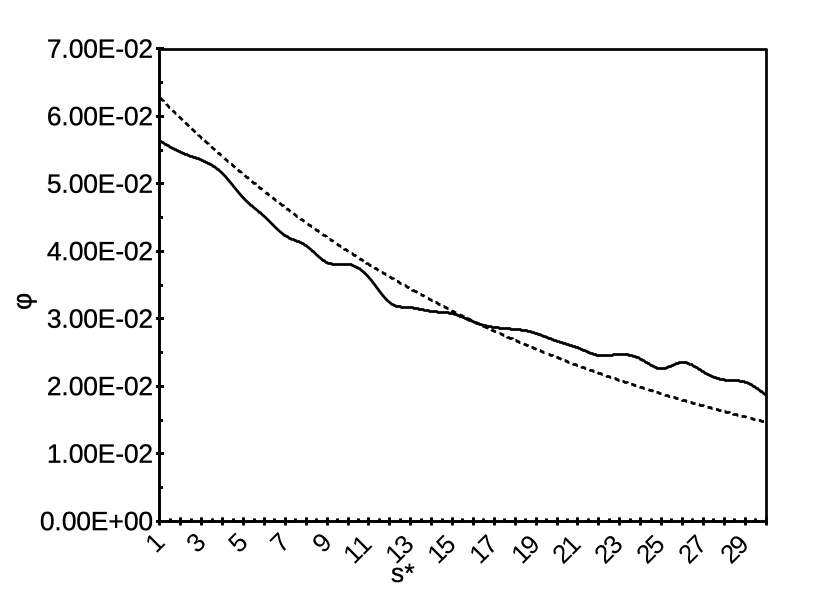}}{h}
  \subsubfloat{\includegraphics[width=0.32\columnwidth]{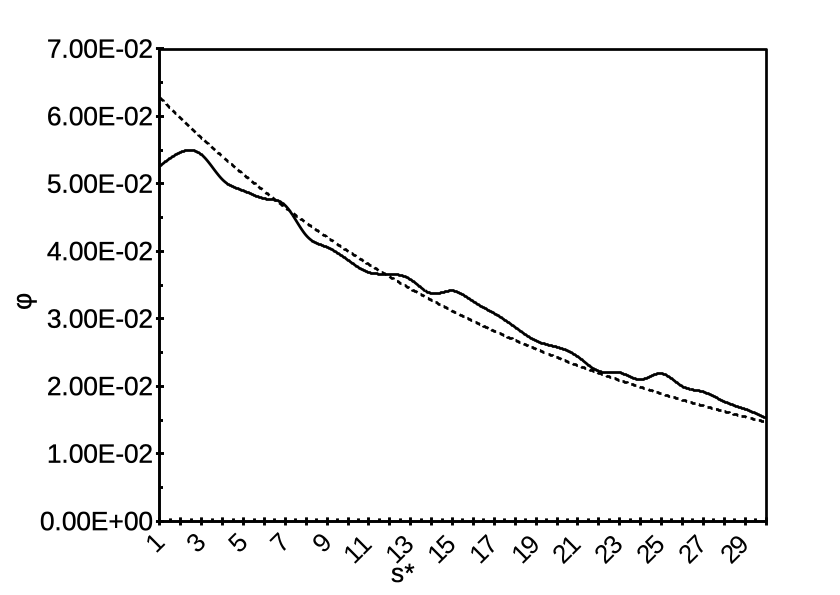}}{i}
  \qquad
  \subsubfloat{\includegraphics[width=0.32\columnwidth]{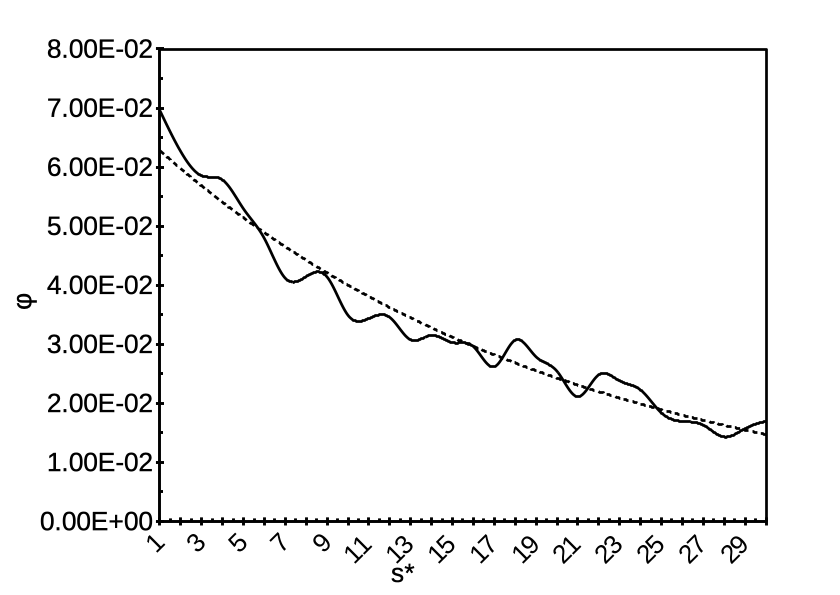}}{j} 
  \subsubfloat{\includegraphics[width=0.32\columnwidth]{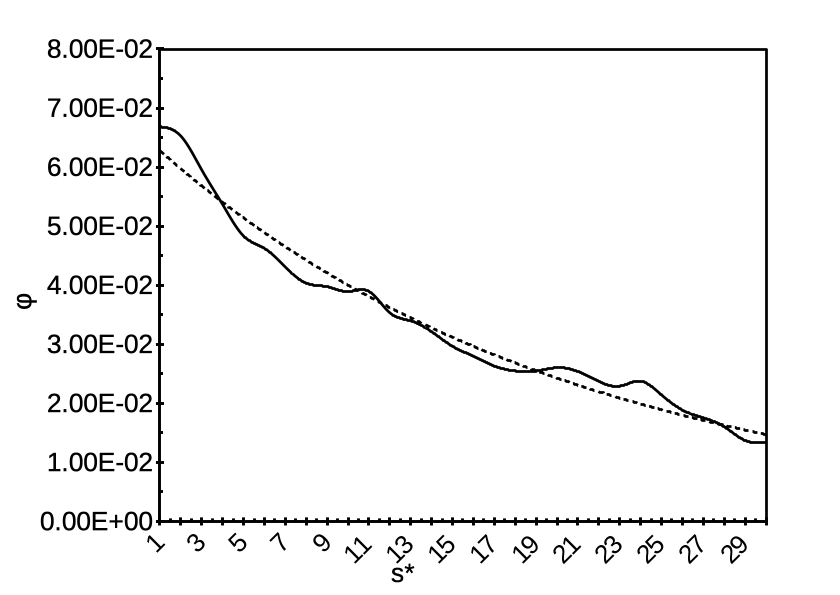}}{k}
  \subsubfloat{\includegraphics[width=0.32\columnwidth]{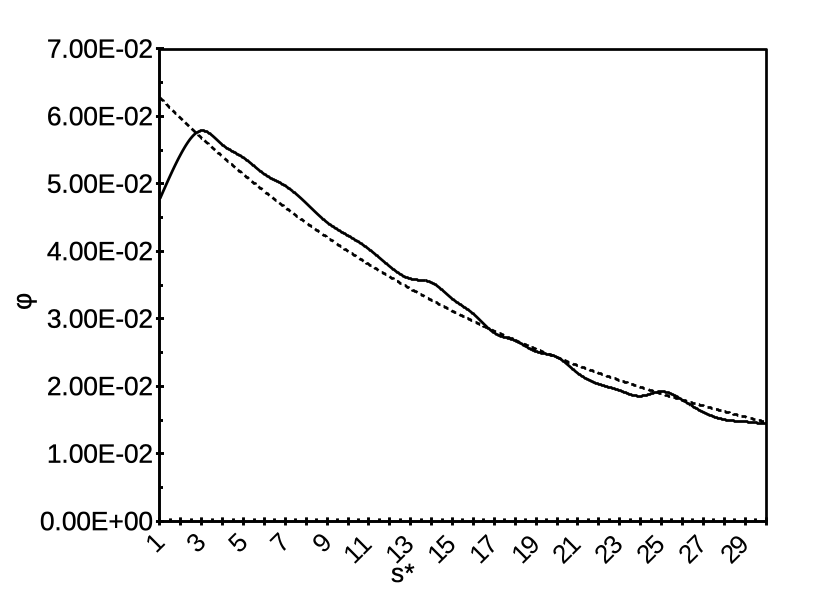}}{l}
  \qquad

\caption{Simulated (solid) and theoretical  (dashed) particles number density distribution in space with a constant external force applied for $\lambda05$--VV (a,b,c), GJF (d,e,f), LI (g,h,i), and BAOAB (j,k,l)  for time steps $\Delta^*$=0.1 (a,d,g,j), 0.9 (b,e,h,k), and 1.5 (c,f,i,l). } \label{fig:boltzmann}

\end{figure}

To consider the effect of external forces, especially in the spatially uniform field as the simplest and common case (e.g. gravity, electrostatic field), we validated  $N=300$ particles in a box of edge length $L^* = 2500$ and walls represented by a repulsive Lennard-Jones potential:
\begin{eqnarray}
    U^{\mathrm{LJ*}}_i
    =
    4E^{*}
    \left[
      \left(
      \frac
      {\sigma^{*}}
      {|\pmb{\rho}_i|}
      \right)^{12}
      -
      \left(
      \frac
      {\sigma^{*}}
      {|\pmb{\rho}_i|}
      \right)^{6}
    \right]
    \biggr\rvert_{\rho_i^{*} < \rho_0^{*}}
    \label{eq:lj_potential} \\
    U^{\mathrm{LJ*}}_i = 0
    \biggr\rvert_{\rho_i^{*} \geq \rho_0^{*}}
\end{eqnarray}
The cut-off distance $\rho_0^{*}=\sigma^{*} 2^{1/6}$ was selected to make a repulsive-only potential. The repulsion potential energy magnitude $E^{*}=1$ and zero-energy distance $\sigma^{*}=40$ has been selected for better performance and stability. The resulting particles have a number density that distributes in space according to the Boltzmann profile ~(Fig.~\ref{fig:boltzmann}):
\begin{eqnarray}
  \varphi \left( s^{*} \right) =
    \frac
    {N \lvert \textbf{F}^{*}_{\mathrm{ext}} \rvert}
    {  L^{*2}
       \left(
         1 - e^{-\lvert \textbf{F}_{\mathrm{ext}}^* \rvert L^{*}}
       \right)
    }
    e^{-\lvert \textbf{F}_{\mathrm{ext}}^* \rvert s^{*}},
  \label{eq:boltzmann}
\end{eqnarray}
where $s^{*}$ is the dimensionless coordinate along the force direction and $\textbf{F}^{*}_{\mathrm{ext}}=3 \cdot 10^{-4}$. The area that is distorted by the stiff repulsive potential has been excluded. The corresponding accuracy parameter is defined as~(Fig.~\ref{fig:boltzmann-eps}):
\begin{eqnarray}
  \frac{1}{L^*}
  \int_{0}^{L^*}
  \Bigm\lvert
  \frac
  {\varphi_{\mathrm{sim}} \left( s^{*} \right)
 - \varphi \left( s^{*} \right)}
  {\varphi \left( s^{*} \right)}
  \Bigm\rvert
  ds^{*}
  =
  \epsilon''
  \label{eq:boltzmann-eps2}
\end{eqnarray}

\section{Discussion} \label{results_dis}

Previous studies (e.g. ~\cite{Schlick2010,gronbech2013simple}) have addressed the accuracy and stability of popular LD integrators in the low-- and high--viscosity regimes. At described in the previous section, the current study supplements and extends such known results where the integrators have now been studied for large time steps. The viscosity and temperature were defined implicitly in the units~\eqref{eq:t_dim},~\eqref{eq:p_dim}, and \eqref{eq:bbk-temporal-limit}, while the reduced Langevin equation does not have these parameters at all~\eqref{eq:langevin-dimless}. Accordingly, the time step of an integrator already accounts for these parameters and additionally, the parameters $\textbf{F}^{*}_{\mathrm{ext}}$ and $\kappa^*$ are retained in dimensionless form.

The accuracy of an integrator in the context of balance between viscosity and conservative forces can be validated by altering any of these parameters. Following common practice (e.g.~\cite{beard2000inertial}), we focused on $\kappa^*$ corresponding to underdamped, critically damped and overdamped cases~\cite{beard2000inertial,gronbech2013simple} to investigate the large time steps regime as well as for analysing the effects of viscosity on algorithmic accuracy.

The time step cannot be increased indefinitely for the vBBK and $\lambda05$--VV integrators as it is revealed in the harmonic potential tests (Fig.~\ref{fig:autocorr-harm-eps}).
The unlimited time step for the remaining integrators depends on implementation and force types, e.g. to avoid particle crossing the wall, etc. However, the possibility to use large time steps exists at the level of the integrator structure because all these integrators have a solved closed-form of the Langevin equation, i.e. a damping factor exponential functions and/or terminal velocities. This is not the case for the BBK and $\lambda05$--VV schemes which are more ``fundamental'' Verlet--like integrators. It is crucial to mention that BAOAB combines features of both (including the limit of $t$-symmetry) without stability drawbacks. It is not a surprise that BAOAB method is claimed to be a very effective and robust one in~\cite{leimkuhler2013robust} as verified by independent researchers.

\subsection{Analysis in detail}

\subsubsection{Thermal properties}
\begin{figure}
  \centering

  \subsubfloat{\includegraphics[width=0.9\columnwidth]{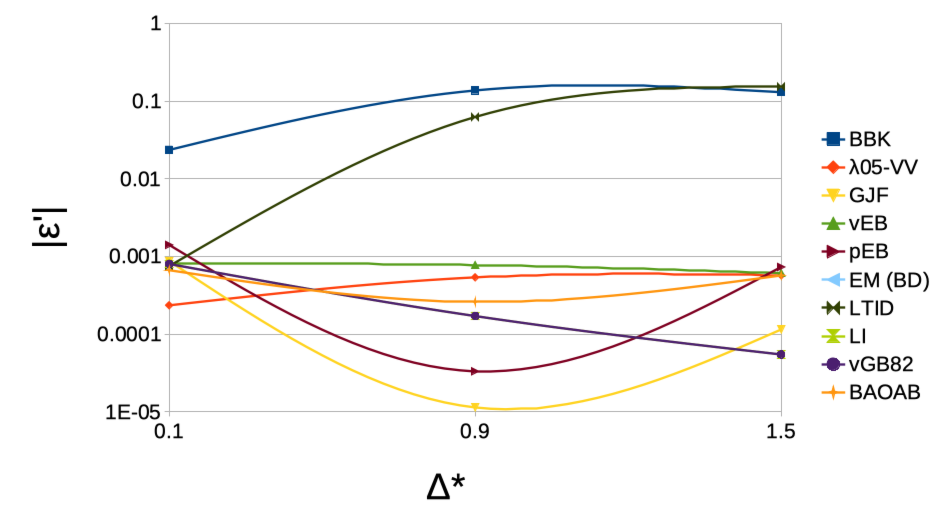}}{a}
  \subsubfloat{\includegraphics[width=0.9\columnwidth]{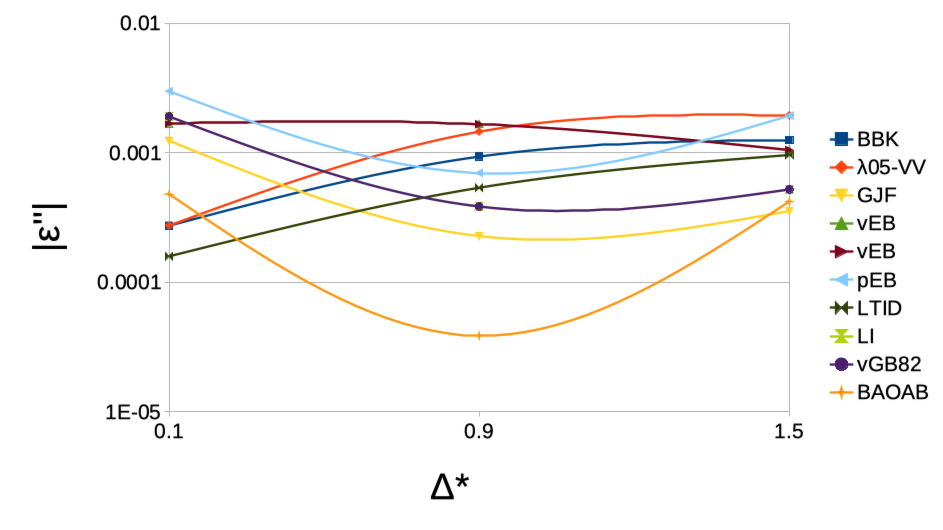}}{b}


\caption{The linear regression coefficients $\epsilon'$ (a) and $\epsilon''$ (b) represent an integrator precision in terms of a general temperature shift and its stationarity over time accordingly. The vGB82 and LI integrator dependencies are identical. Here and hereinafter, the simulation standard error bars are shown.} \label{fig:Tdrift}

\end{figure}

The average temperature shift $\lvert \epsilon' \rvert$ from the thermodynamically target is minor (less than 5\%) for all the integrators (Fig.~\ref{fig:Tdrift}a) at the time step $\Delta^{*}=0.1$. Vice versa, the parameter $\lvert \epsilon' \rvert$ is large in the cases of LTID and vBBK up to stable value starting from $\Delta^{*} \approx 0.9$. All other integrators exhibit a much better $10^{-5}< \lvert \epsilon' \rvert <10^{-3}$. The LTID integrator accuracy belongs to this range at small time steps. The good values of  $\lvert \epsilon' \rvert$ for temperature are found for GJF, vGB82, LI and pEB with prominent results for GJF fall in the intermediate time step region. The thermal stationarity precision parameter $\lvert \epsilon'' \rvert$ varies in the range of $10^{-3}..10^{-5}$ with the best values in the case of BAOAB (Fig.~\ref{fig:Tdrift}b).

\begin{figure}
  \centering

  \subsubfloat{\includegraphics[width=0.32\columnwidth]{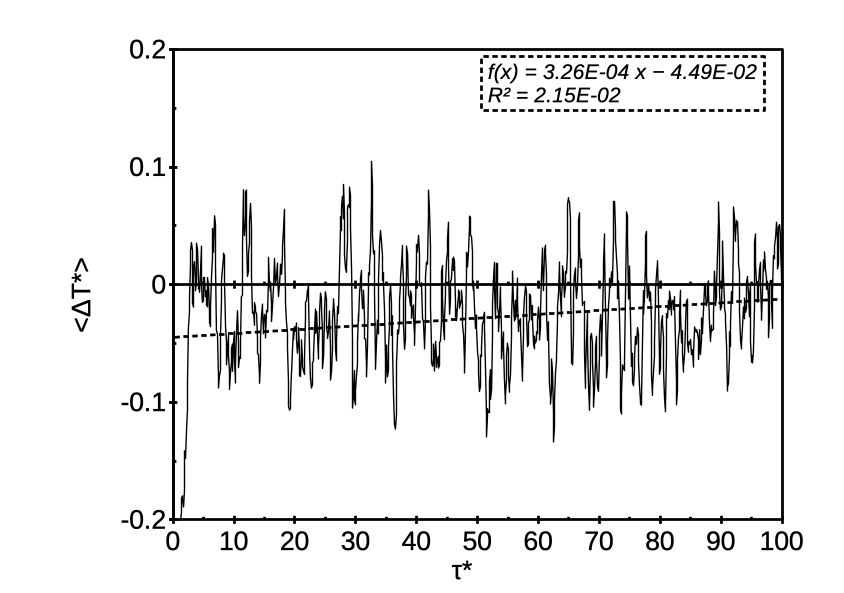}}{a} 
  \subsubfloat{\includegraphics[width=0.32\columnwidth]{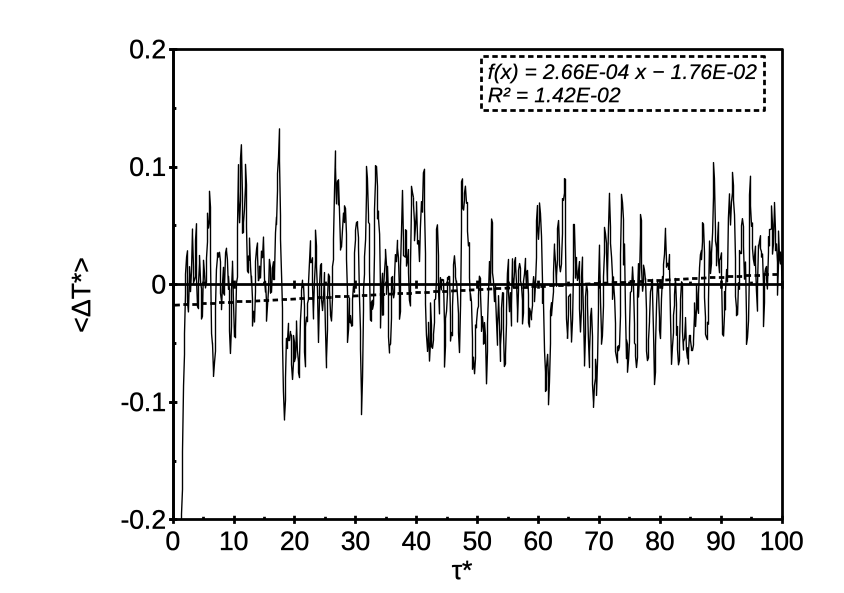}}{b}
  \subsubfloat{\includegraphics[width=0.32\columnwidth]{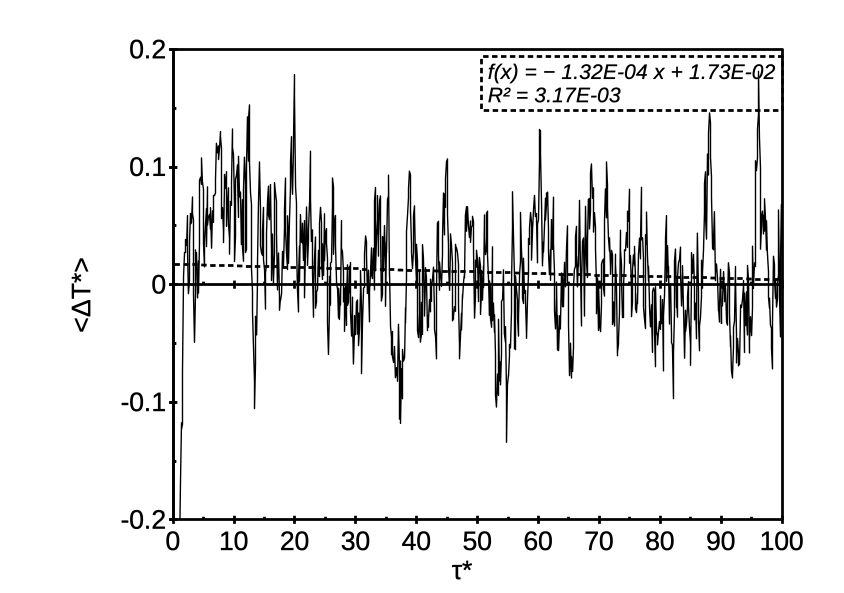}}{c}
  \qquad
  \subsubfloat{\includegraphics[width=0.32\columnwidth]{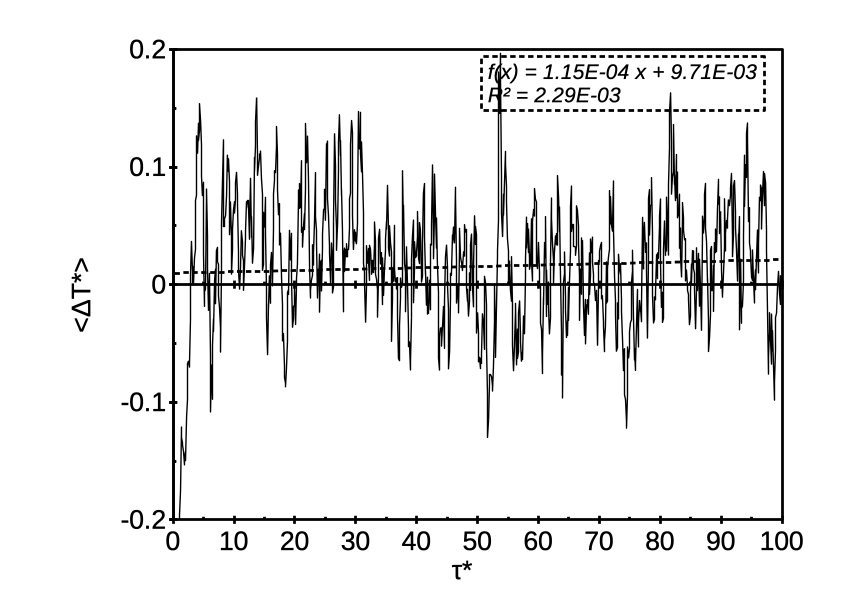}}{d} 
  \subsubfloat{\includegraphics[width=0.32\columnwidth]{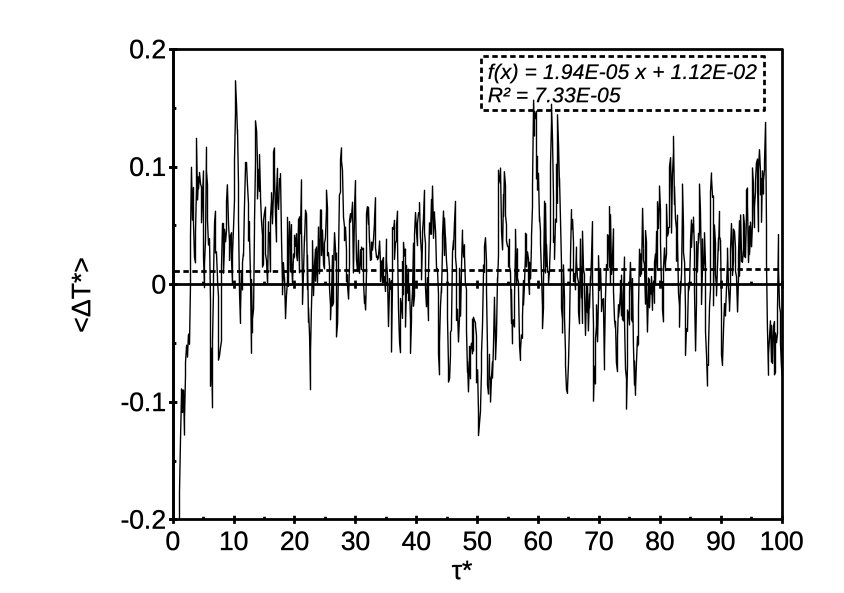}}{e}
  \subsubfloat{\includegraphics[width=0.32\columnwidth]{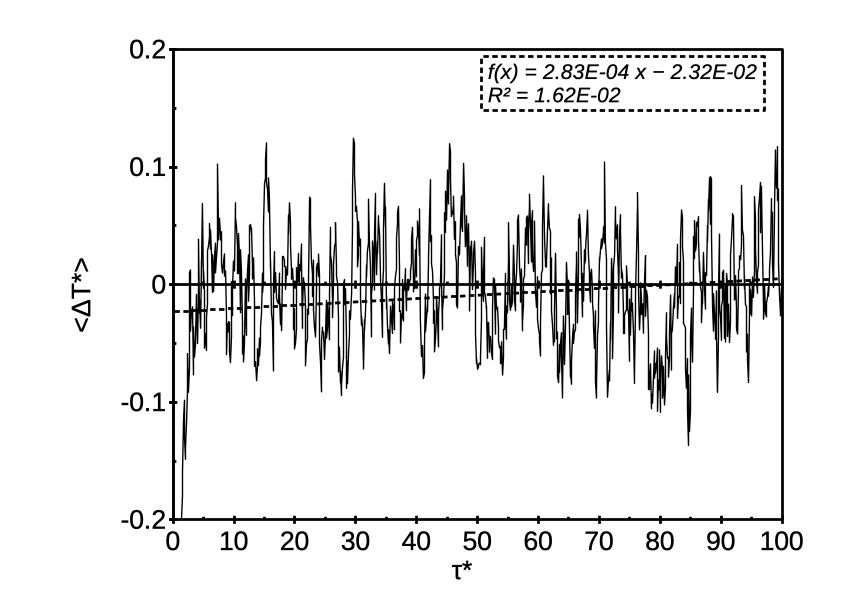}}{f}
  \qquad
  \subsubfloat{\includegraphics[width=0.32\columnwidth]{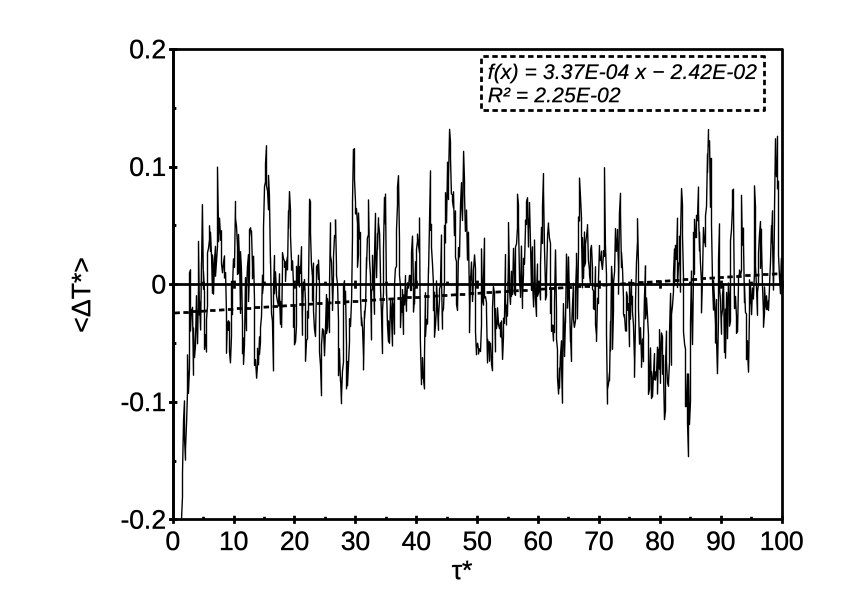}}{g} 
  \subsubfloat{\includegraphics[width=0.32\columnwidth]{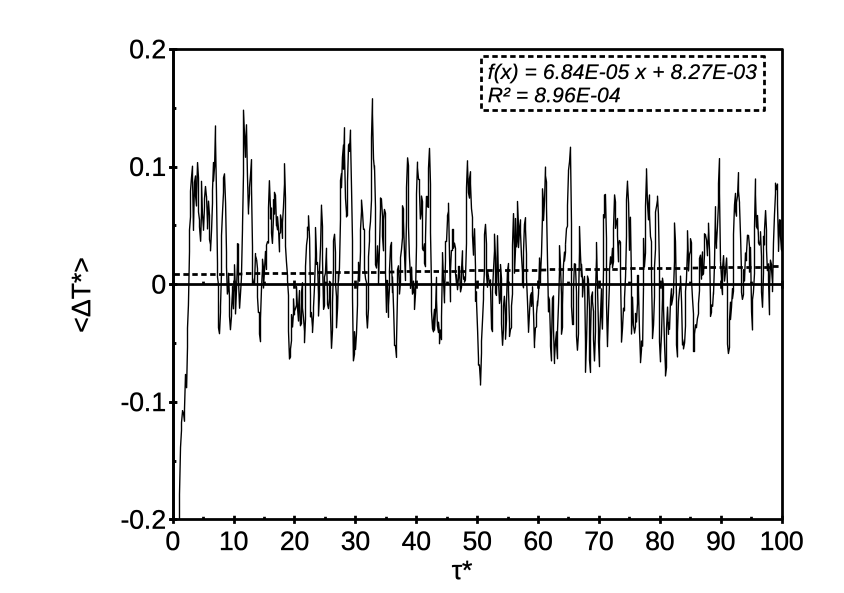}}{h}
  \subsubfloat{\includegraphics[width=0.32\columnwidth]{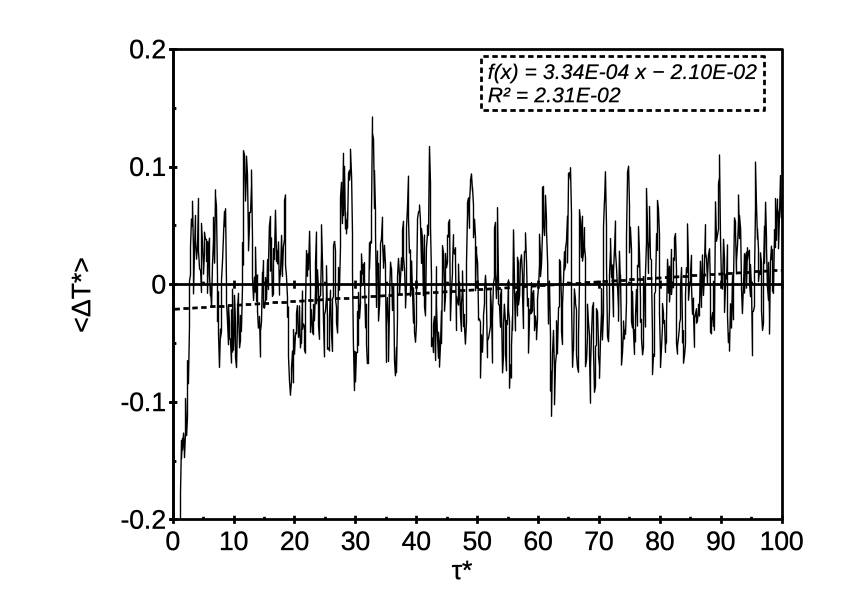}}{i}

\caption{The deviation of a in silico temperature of the model ferrofluid against the expected unitary value over the course of time for vBBK (a), $\lambda05$--VV (b), LTID (c), vEB (d), pEB (e), LI (f), vGB82 (g), BAOAB (h), and GJF (i) integrators at the time step $\Delta^*$=0.1.} \label{fig:Tdrift-dd}

\end{figure}

For ferrofluid simulation, the best accuracy in terms of thermal stability is achieved by BAOAB and vEB  (Fig.~\ref{fig:Tdrift-dd}h,d) and thermal stationarity is best appreciated for BAOAB and pEB (Fig.~\ref{fig:Tdrift-dd}h,e).

LI and vGB82 have identical quality for zero external force by design~\cite{skeel2002impulse}. Hence, their results overlap in this and in the following sections except for the case of external force.

\subsubsection{Diffusion}
\begin{figure}
  \centering

  \subsubfloat{\includegraphics[width=0.32\columnwidth]{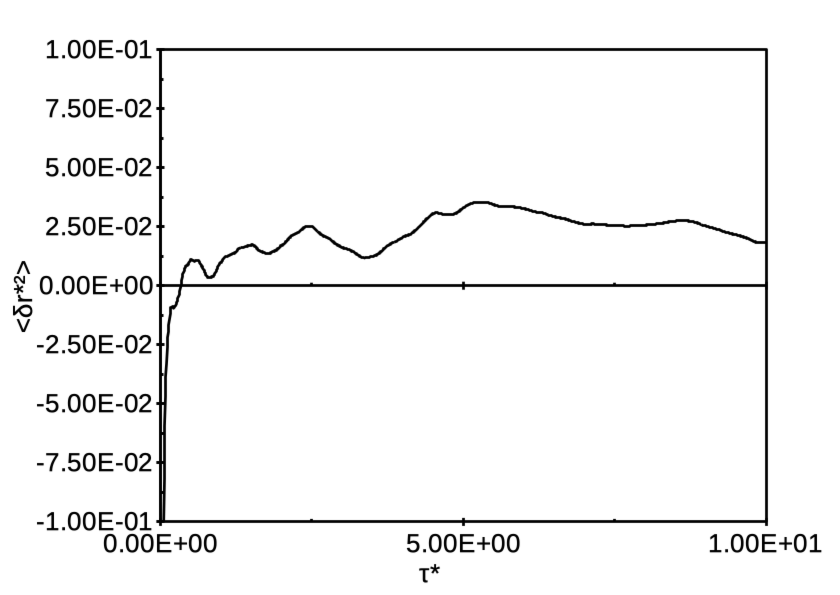}}{a} 
  \subsubfloat{\includegraphics[width=0.32\columnwidth]{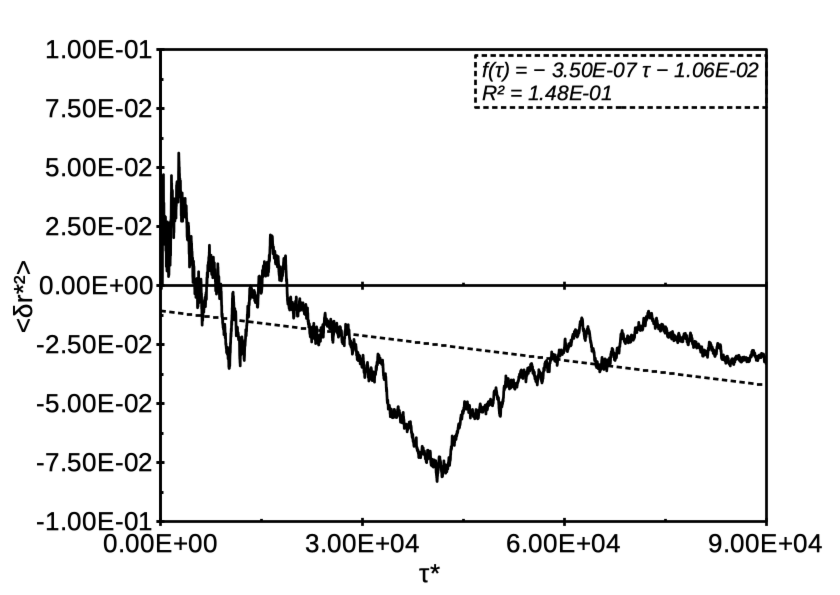}}{b}
  \subsubfloat{\includegraphics[width=0.32\columnwidth]{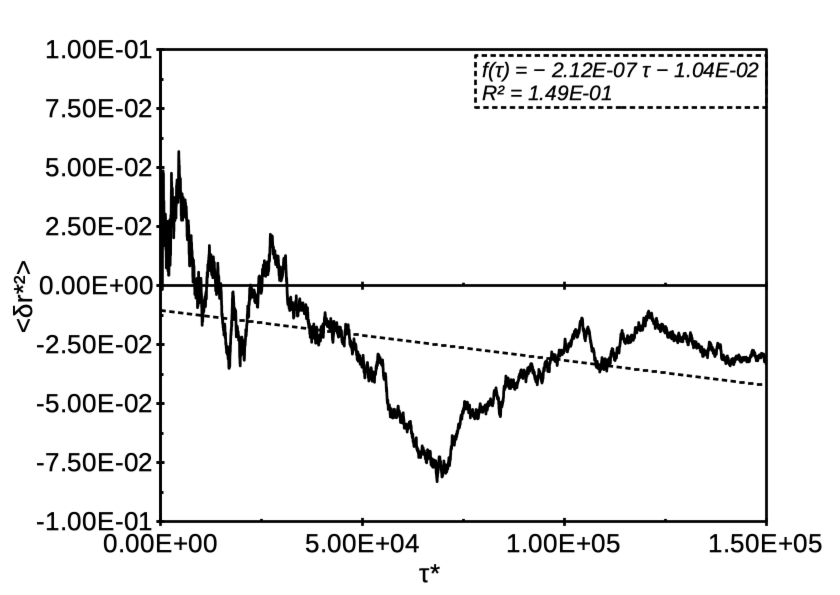}}{c}
  \qquad
  \subsubfloat{\includegraphics[width=0.32\columnwidth]{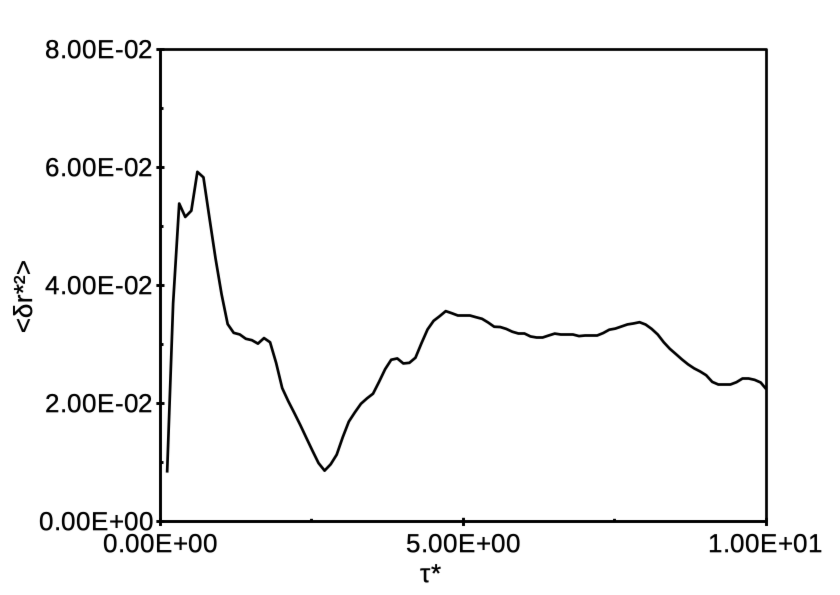}}{d} 
  \subsubfloat{\includegraphics[width=0.32\columnwidth]{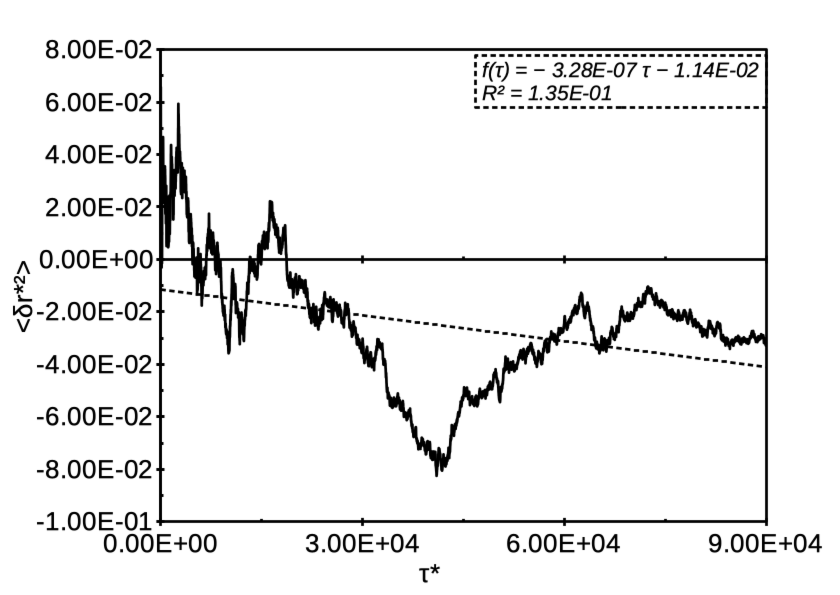}}{e}
  \subsubfloat{\includegraphics[width=0.32\columnwidth]{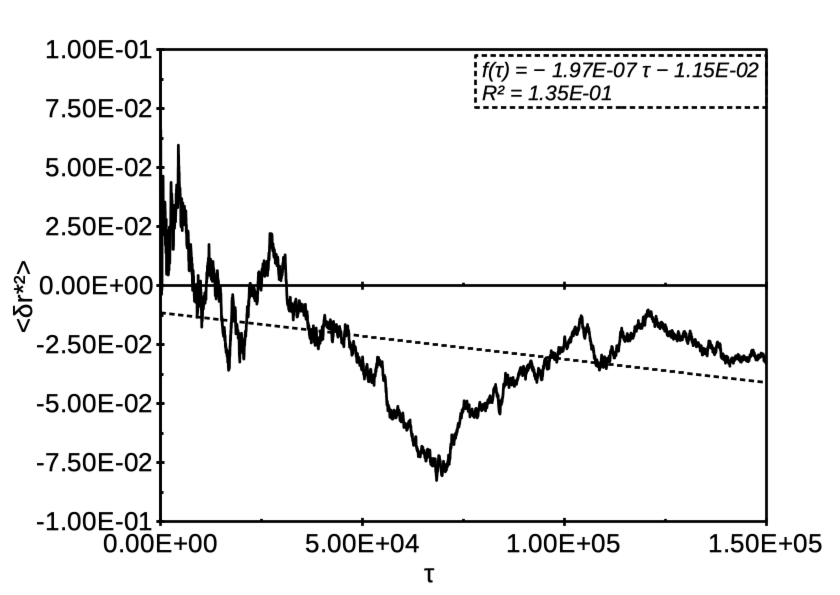}}{f}
  \qquad
  \subsubfloat{\includegraphics[width=0.32\columnwidth]{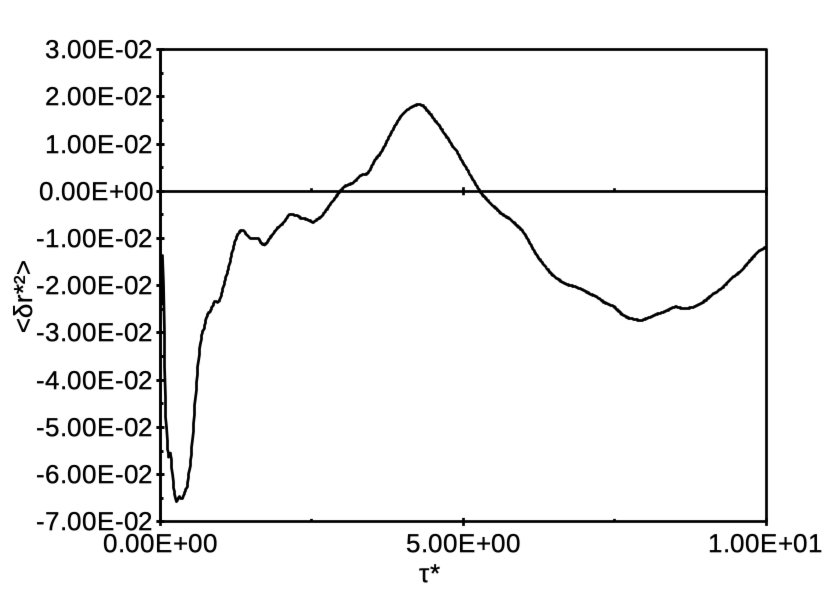}}{g} 
  \subsubfloat{\includegraphics[width=0.32\columnwidth]{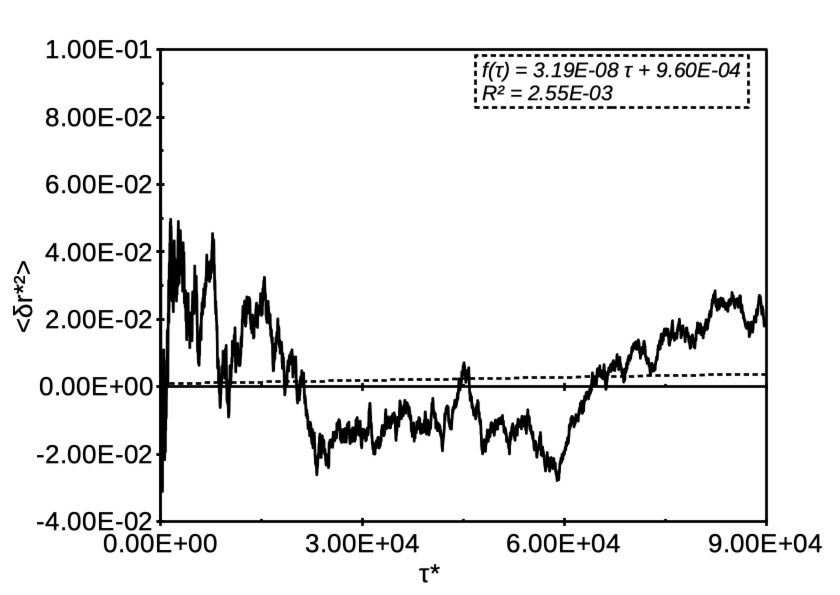}}{h}
  \subsubfloat{\includegraphics[width=0.32\columnwidth]{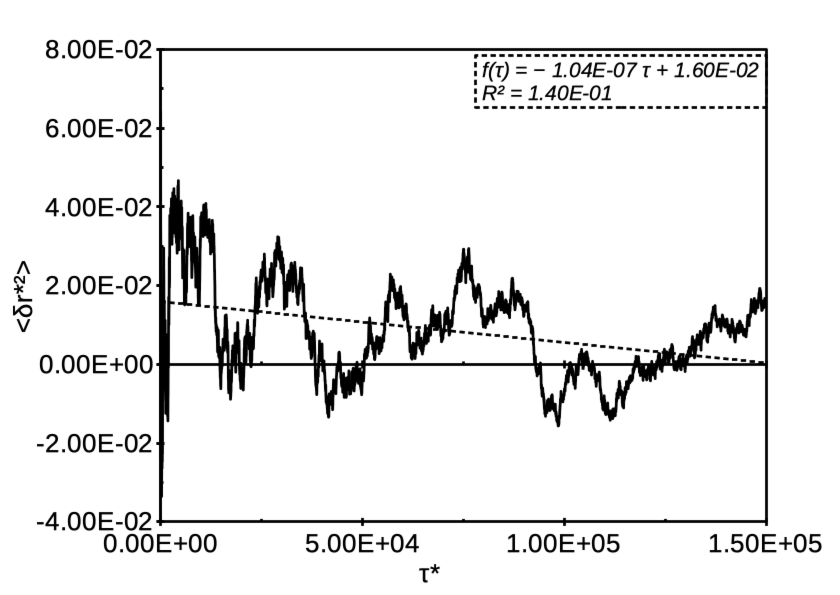}}{i}
  \qquad
  \subsubfloat{\includegraphics[width=0.32\columnwidth]{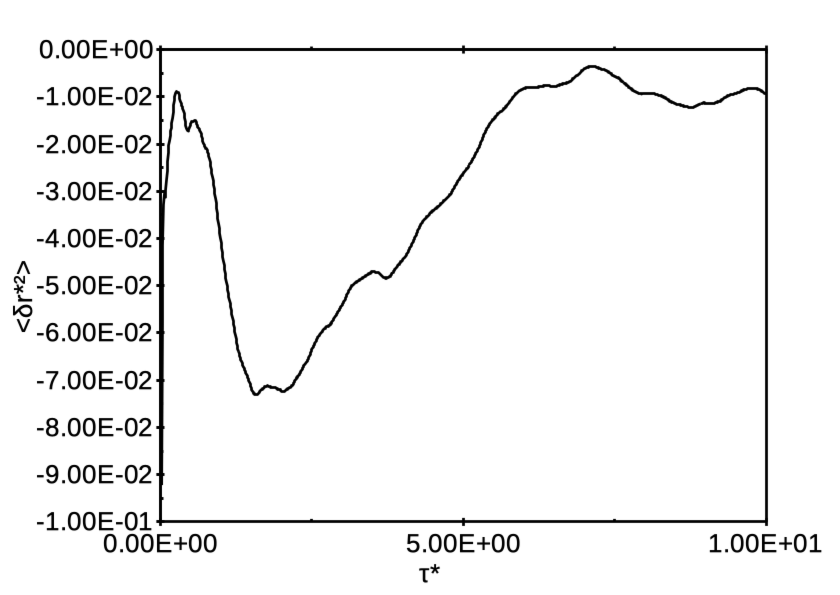}}{j} 
  \subsubfloat{\includegraphics[width=0.32\columnwidth]{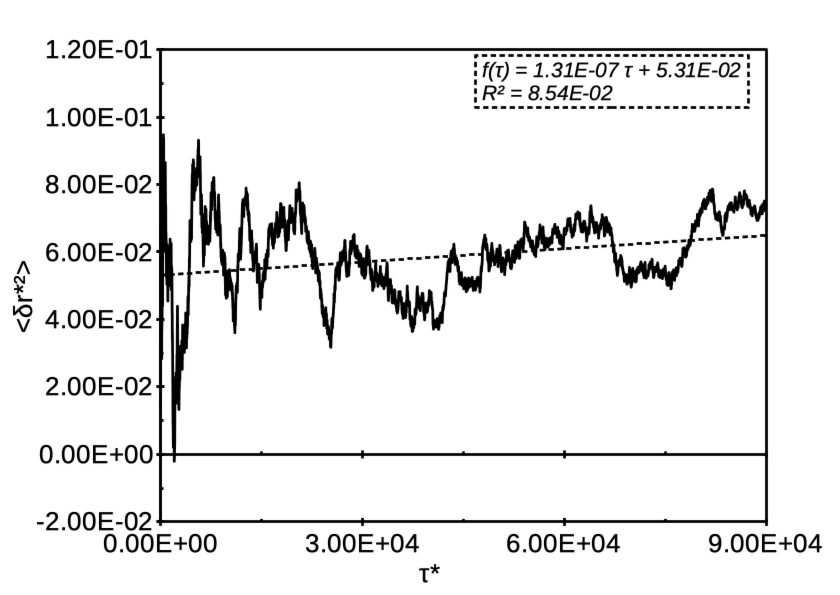}}{k}
  \subsubfloat{\includegraphics[width=0.32\columnwidth]{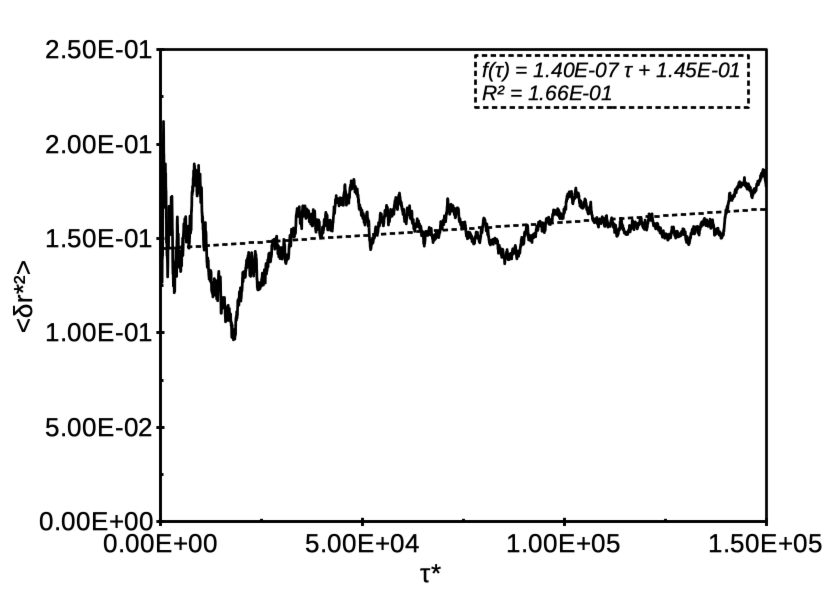}}{l}
  \qquad

\caption{The relative deviation of a positional variance over the course of time for $\lambda05$--VV (a,b,c), GJF (d,e,f), LI (g,h,i), and BAOAB (j,k,l) integrators at time steps $\Delta^*$=0.01 (a,d,g,j), 0.9 (b,e,h,k), and 1.5 (c,f,i,l).} \label{fig:diffusion}

\end{figure}

The relative deviation of a positional variance $\left( \delta r* \right)^2$~\eqref{eq:diff-dr2} temporal dependence within the initial time frames exhibits complex structure for all the integrators with ranges up to $10^{-1}$ arbitrary units (Fig.~\ref{fig:diffusion}a,d,g,j). The GJF, vGB82, and LI propagators exhibit asymptotic convergence $\left( \delta r* \right)^2 \xrightarrow[\Delta^{*} \to 0]{} 0$ as expected theoretically. There is probably no such property in the case of BAOAB. The calculation accuracy does not allow to claim this conclusion for other integrators. GJF and $\lambda05$--VV  have the scaling feature of the dependence $\left( \delta r* \right)^2 \left( \tau^{*} \right)$, which preserve its form at $\Delta^{*} \rightarrow \mu \Delta^{*}$ for any positive value of $\mu$~(Fig.~\ref{fig:diffusion}b,c and Fig.~\ref{fig:diffusion}e,f).

\begin{figure}
  \centering

  \subsubfloat{\includegraphics[width=0.9\columnwidth]{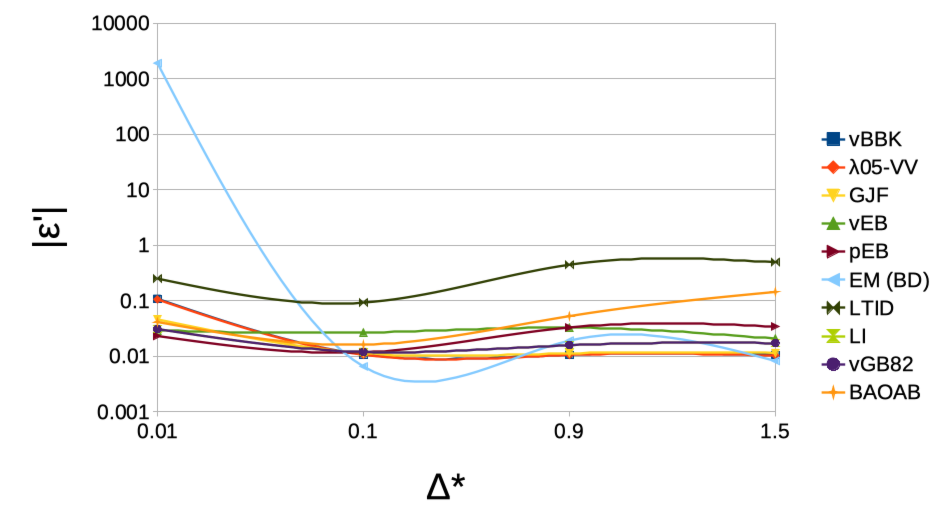}}{a}
  \subsubfloat{\includegraphics[width=0.9\columnwidth]{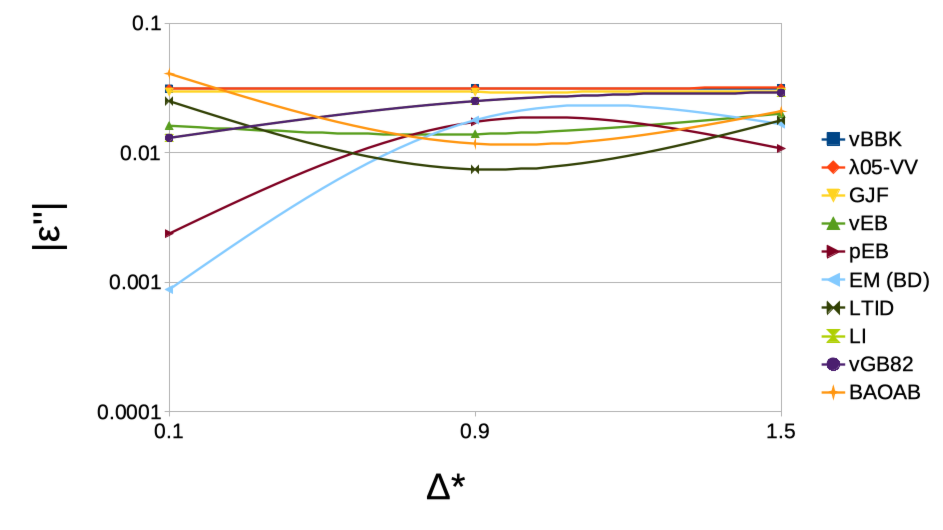}}{b}

\caption{The particles diffusion: the linear regression coefficients $\epsilon'$~(a) and $\epsilon''$~(b) representing integrators' precision. The vGB82 and LI integrator dependencies are identical.} \label{fig:diffusion-eps}

\end{figure}

The accuracy parameter $\lvert \epsilon' \rvert$ defines a basic deviation of the position variance from the expected value. The accuracy parameter $\lvert \epsilon'' \rvert$ defines stationarity of the deviation, ensuring that its magnitude does not grow with time. The $\lvert \epsilon' \rvert$ varies in the range $10^{-3}..10^{-1}$ for all  integrators at  time steps within their stability limits~(Fig.~\ref{fig:diffusion-eps}a). One notable exception is EM at $\Delta^{*}=0.01$ for the case of the particle kick experiment. The case of the thermalized start is well-studied in previous comparative studies. As already mentioned, the motivation of the current research is to consider new use cases. Particularly, the zero temperature 'kick' simulation is a good validation for each integrator with potentially experimental implications. The best accuracy in terms of $\lvert \epsilon' \rvert$ is achieved by the LI integrator.

The $\lvert \epsilon'' \rvert$ varies in the range $10^{-3}..10^{-2}$ for all integrators at all time steps within their stability limits (Fig.~\ref{fig:diffusion-eps}b). For small time steps, the best accuracy corresponds to EM and pEB whereas for larger time steps the best accuracy is given by LTID. The pEB precision parameter $\lvert \epsilon'' \rvert$ has the lowest value starting from $\Delta^* \approx 1.4$.

\subsubsection{Autocorrelation properties}
The momentum autocorrelation dependencies fits theoretically expected values for all the integrators at small time steps (Fig.~\ref{fig:autocorr}) while for large time steps, BBK, $\lambda05$--VV, and GJF show larger deviations.

\begin{figure}
  \centering

  \subsubfloat{\includegraphics[width=0.9\columnwidth]{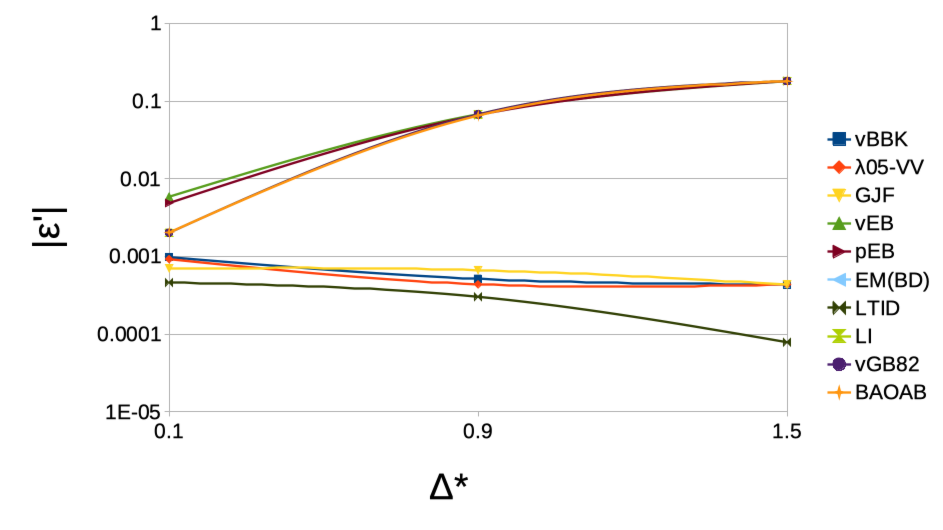}}{a}
  \subsubfloat{\includegraphics[width=0.9\columnwidth]{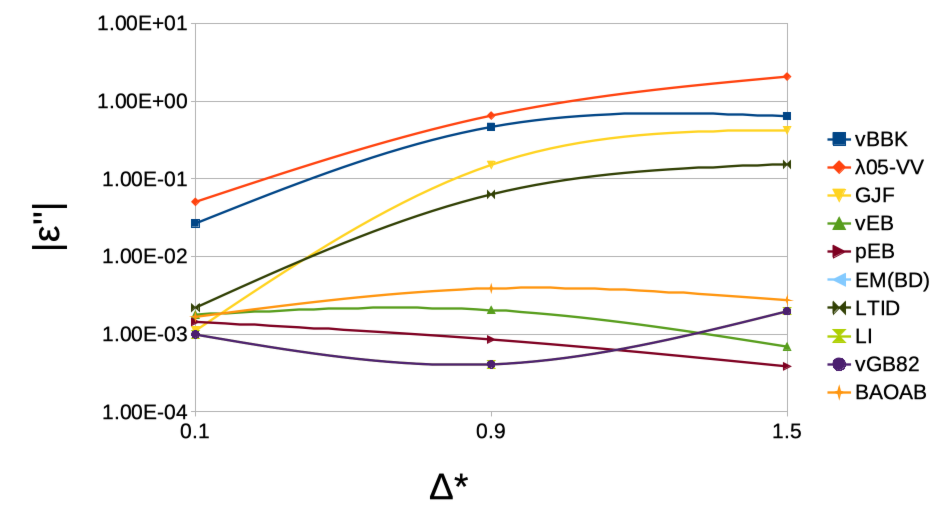}}{b}

\caption{Momentum autocorrelation function of a free particle: the linear regression coefficients $\epsilon'$~(a) and $\epsilon''$~(b) representing integrators' precision. The vGB82 and LI integrator dependencies are identical.} \label{fig:autocorr-eps}

\end{figure}

The ``integral'' precision parameter $\epsilon'$~\eqref{eq:autocorr-eps1} has very different magnitudes for the different integrators:  $10^{-4}$ in the cases of GJF, LTID, vBBK, and $\lambda05$--VV up to $10^{-1}$ in the cases of BAOAB, pEB, vEB, LI, and vGB82 at large time steps (Fig.~\ref{fig:autocorr-eps}a). The difference is less marked for small time steps and the best accuracy is achieved by LTID. 
The worst results for BAOAB, pEB, vEB, LI, and vGB82  are somehow surprising because qualitatively the temporal dependence of the autocorrelation function looks fitting. The explanation is the discrepancy between real improper integral~\eqref{eq:Green-Kubo-1} (unlimited temporal range) and simulated one with the range of $\tau^{*}$ which is restricted by the simulation duration. Also, the discrepancy can be accounted for the integrator intrinsic accuracy.

The precision parameter $\epsilon''$~\eqref{eq:autocorr-eps2} does not show the issue of the improper integration because it tracks the relative deviation of the momentum autocorrelation from the expected curve for each value of  $\tau^{*}$ ~\eqref{eq:autocorr-eps2}. We then removed numerical artifacts by discarding deviations smaller than $1\%$. Corresponding dependencies on the time step (Fig.~\ref{fig:autocorr-eps}b) vary in the range $10^{-4}..1$ with much better results for LI, vGB82, pEB, vEB, and BAOAB.

\subsubsection{Harmonic potential}

Both position and momentum autocorrelation time dependencies fits theoretically the expected forms for all  integrators at small time steps (Fig.~\ref{fig:autocorr-harm-pos}-~\ref{fig:autocorr-harm-momentum}). All the integrators  diverge from the expected curves at large time steps leading to significant mismatch and extents. This divergence does not depend on the integrator's time step stability limit as it was in the case of a free particle.

The positional ``integral'' accuracy parameter $\epsilon'$~\eqref{eq:autocorr-harm-eps1} in the underdamped case has generally better values for EM and LI  (Fig.~\ref{fig:autocorr-harm-eps}a) for the time step $\Delta^{*}=0.9$. At small time steps vEB performs better. At large time steps vGB82 show the best performances. The same precision parameter for momentum shows that $\lambda05$--VV and vBBK perform the best at the time step $\Delta^{*}=0.9$ (Fig.~\ref{fig:autocorr-harm-eps}c). For bigger time steps BAOAB and GJF propagators are more precise in this aspect.

In the critically damped case, the positional precision parameter $\epsilon'$ exhibits good performances in the range $10^{-4}..10^{-2}$ for all  integrators at all time steps except for LTID, BAOAB, and GJF with values falling around $10^{-1}$ (Fig.~\ref{fig:autocorr-harm-eps}e). The same accuracy parameter for momentum has approximately the same values for all integrators at small time steps except for vEB with a value about $10^{-4}$ (Fig.~\ref{fig:autocorr-harm-eps}g). At large time steps, the leading accuracy is found for $\lambda05$--VV and vBBK.

In the overdamped case, the positional precision parameter $\epsilon'$ has similar values for the most of integrators except for BAOAB, LTID, and GJF showing a lower accuracy (Fig.~\ref{fig:autocorr-harm-eps}i). The same accuracy parameter for momentum has similar values at  small time steps for all  integrators and best performances for BAOAB and GJF are found at large time steps (Fig.~\ref{fig:autocorr-harm-eps}k).

All the accuracy parameters in this subsection were related to theoretically expected integral of the position and momentum autocorrelation functions. In order to test correctness of the detailed structure of these autocorrelation dependencies, another precision parameter $\epsilon''$~\eqref{eq:autocorr-harm-eps2} has been used. This parameter has better values for vGB82 and LI in all the use cases discussed above (Fig.~\ref{fig:autocorr-harm-eps}b,d,f,h,j,l) with the only exception at small time steps for the overdamped  autocorrelation case (Fig.~\ref{fig:autocorr-harm-eps}j,l) where vEB has higher accuracy.

\subsubsection{Constant force}

\begin{figure}
  \centering

  \subsubfloat{\includegraphics[width=0.9\columnwidth]{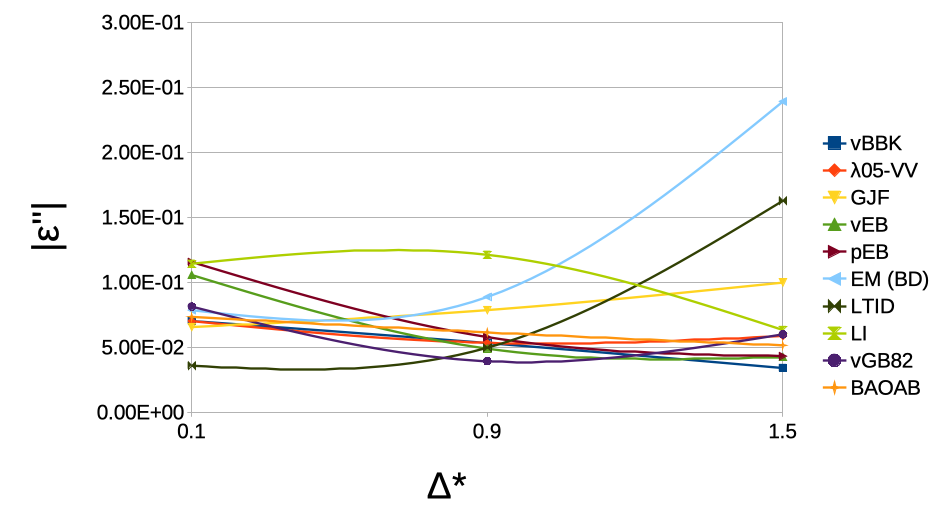}}

\caption{Accuracy parameter $\epsilon''$ for the particles number density distribution with a constant external force applied. It reflects an average relative deviation against the expected Boltzmann profile.} \label{fig:boltzmann-eps}

\end{figure}

The deviation of the particles density spatial distribution from the expected Boltzmann distribution has been evaluated for all the selected integrators~(Fig.~\ref{fig:boltzmann-eps}). Comparison of this deviation can be revealed using the precision parameter~$\epsilon''$~\eqref{eq:boltzmann-eps2} (Fig.~\ref{fig:boltzmann-eps}). The smaller time steps corresponds to a better precision for LTID integrator. The vGB82 integrator shows the best precision at the intermediate time steps. The larger time steps correspond to  better results for vBBK integrator.

\subsection{Benchmarks}

Naturally, there are multiple ways to benchmark the selected integrators. The situation becomes  more complex where there is strong dependency on the details of an integrator implementation and its validation test level (the coding/scripting level of details). This is why it is important to use the common (preferably open-source and well-maintained one) software platform and, also,  identical test scripts for each one without tuning and modification of tests. The present cross-disciplinary tests have been implemented as Python tests using the open-source ESPResSo package~\cite{TheESPResSoprojectCopyrightC2016}. The present set of use cases, selected according to authors opinion, are enough foundational and, also, the compare directly to the more complex molecular dynamics applications as done in a previous comparative report~\cite{leimkuhler2013robust}. We do not claim the objective and comprehensive comparison of LD integrators in the present report, rather our tests for selected use cases and time step range extent the level of confidence on the most popular integrators.

Summarizing, the validation results can be based on the accuracy factors $\epsilon'$ and $\epsilon''$ defined above. It is very common to use a logarithmic scale for such  comparison (e.g.~\cite{beard2000inertial}). Hence, we use a sum of integers $-\log_{10} \lvert \epsilon' \rvert$ and $-\log_{10} \lvert \epsilon'' \rvert$ for all the validation cases to evaluate each integrator effectiveness. If $\lvert \epsilon' \rvert>1$ or $\vert \epsilon'' \rvert>1$, or the integrator is unstable, or either an integrator is not designed for a particular use case, one is counted as -1 in the total score. Each of the following theoretical (and usability) features gives +3 score for each integrator: arbitrary small time steps possible, arbitrary large time steps possible, and $t$--symmetrical integrator type.

\begin{figure}
  \centering

  \subsubfloat{\includegraphics[width=0.9\columnwidth]{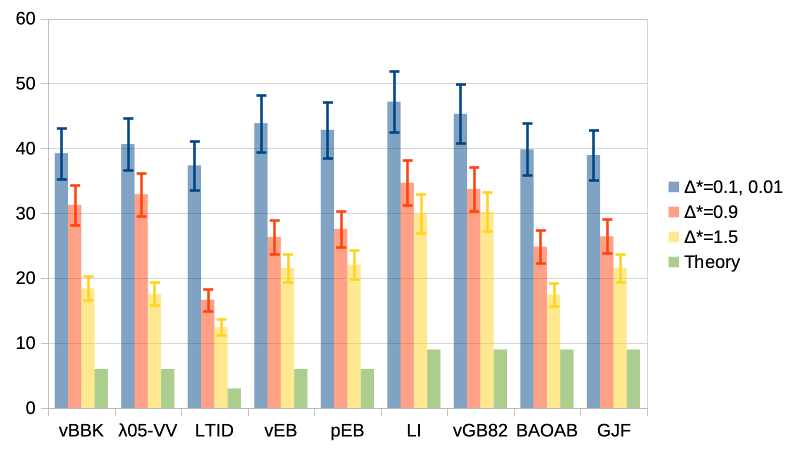}} 

\caption{Cumulative benchmarks for all integrators where higher values correspond to better accuracy. Cases of different representative dimensionless timesteps $\Delta^{*}$ are shown. The benchmark score is calculated as $-\log_{10}$ of relative errors of expected values in all the selected test cases. Theoretical benchmarking represents an absence of restrictions towards smaller or bigger timesteps and $t$--symmetrical integrator type.} \label{fig:benchmarking}

\end{figure}

The final cumulative comparison is a sum of all scores for all the validation tests at each time step~(Fig.~\ref{fig:benchmarking}). The best results at the small time steps were determined for LI, vGB82, vEB, and pEB integrators. For intermediate time steps, the most precise integrators are LI, vGB82, $\lambda05$--VV, and vBBK. The big time steps correspond to better results of vGB82 and LI integrators. The better score of theoretical features has been identified for BAOAB and GJF integrators. 

Among the considered integrators it is important that some of them are derived with minimal approximations or assumptions -- only time quantization (Table~\ref{table:integrators-def}): vBBK, vGB82, LI, and BAOAB. They also have a feature of $t$--symmetry in the deterministic (non-stochastic) limit. There are many precision issues for vBBK reported in the literature, see for example ~\cite{gronbech2013simple,cances2007theoretical}.
The BAOAB integrator deserves special attention --- both theoretically and practically. In fact, it was assessed as a leading integrator for certain complex molecular dynamics use cases~\cite{leimkuhler2013robust}. Meanwhile, other top integrators here are widely used as well for many use cases, such as $\lambda05$--VV \cite{Limbach2006,Martys1999} that was assessed as a {\em de facto} standard for rotational motion simulation~\cite{Martys1999} and, often, is referred to as the default velocity Verlet integrator without special naming~\cite{TheESPResSoprojectCopyrightC2016}. 

It is required to select a probability distribution of the discrete random $\xi_k$ used in all the above methods. Because of the model nature of LD, its selection is somehow arbitrary. The methods without the upper stability limits in the time step can provide physically meaningful results even after a single simulation step (e.g. classical Brownian diffusion random spatial  step ~\cite{einstein1905movement}). Random displacements of the Brownian particle is known to match the a normal distribution as it was confirmed in experiments~\cite{perrin2013brownian}. Conversely, the methods relying on small time steps require a certain number of steps to achieve the statistically significant averages as long as the process is stationary. A probability distribution of $\xi_k$ can be arbitrarily chosen according to the central limit theorem for these propagators~\cite{shardlow2003splitting,Pottier2014}. 

One limitation of the present results is related to the validation approach rather than to its fundamental value. For instance, the EM integrator is based on solid theoretical ground and is designed to provide positional trajectories without associated momentum (and referred as such, more specifically, as the EM/BD propagator~\cite{Schlick2010}), while we considered the corresponding momentum tests as failed. This is why it was omitted in the final benchmarks. The majority of integrators are based on approximations~(Tab.~\ref{table:integrators-def}) requiring to employ certain time steps range and those are consistent. Hence, our comparison is not conceptually critical, rather it corresponds to practical usability for given use cases. The ability to vary the time step is practically useful. For instance, increasing the time step to a maximum value is crucial for DNA plasmid juxtaposition~\cite{Beard2000}, simple diffusion (e.g. via a membrane), and other mesoscale studies. 

\section{Conclusions}
We have considered different use cases for the validation of popular integrators for Langevin and Brownian dynamics. Verification of quantitative matching of expected and numerical results was supplemented by usability testing, e.g. by inspecting integrator stability at large time steps. The final benchmarks showed the prominent quality and effectiveness of the ``Langevin Impulse'' integrator suggested by Skeel and Izaguirre in 2002. As a common basis for the implementation of all integrators, the open-source platform ``ESPResSo'' was used for simplicity of reference and reuse of testing scripts.

\section{Acknowledgements}
The research was supported by the scholarship IAC-RM UA-SC 01-2022 from the National Research Council (Italy, CNR). We wish to thank the ESPResSo team for their open-source software maintenance which was of significant value for the present research. We thank the open-source community in general for supplying multiple digital tools used in this work.

\section*{References}

\bibliography{Manuscript}

\pagebreak

\pagebreak

\pagebreak

\pagebreak

\pagebreak

\pagebreak

\pagebreak

\pagebreak

\pagebreak

\begin{figure}
  \centering

  \subsubfloat{\includegraphics[width=0.49\columnwidth]{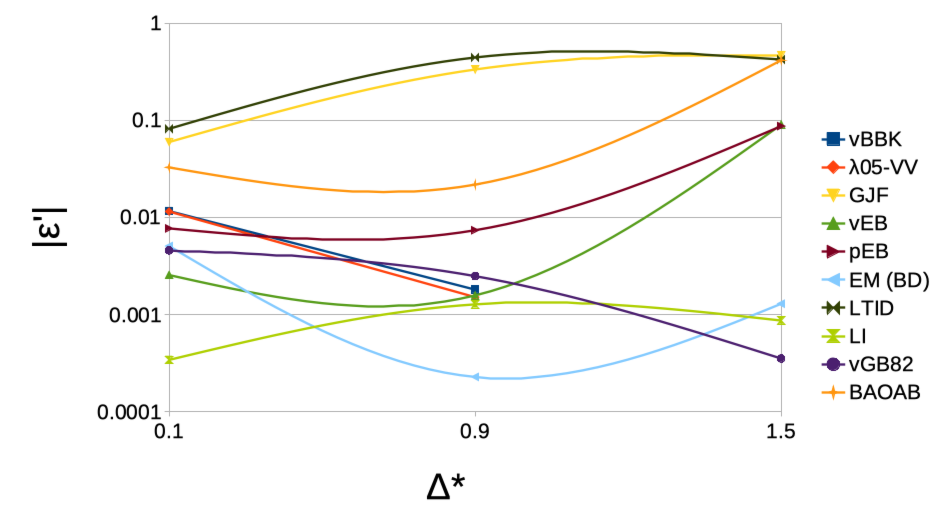}}{a} 
  \subsubfloat{\includegraphics[width=0.49\columnwidth]{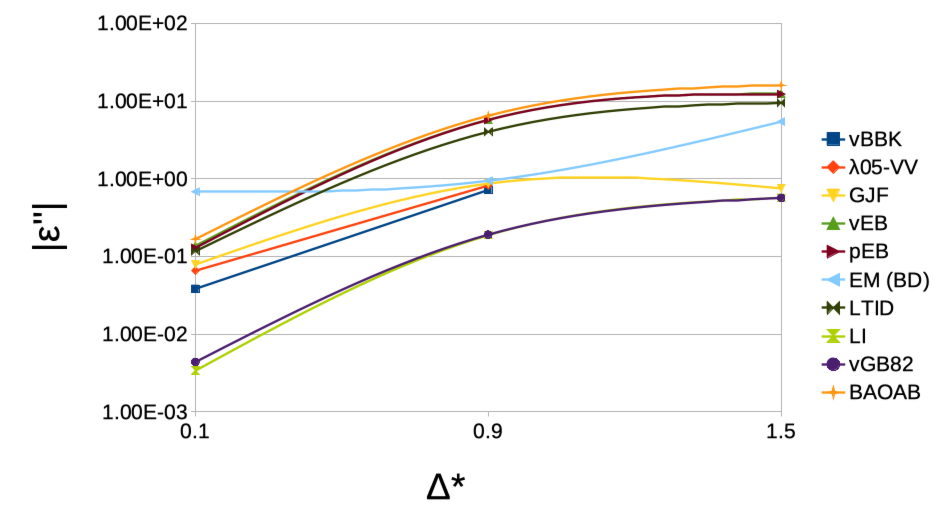}}{b}
  \subsubfloat{\includegraphics[width=0.49\columnwidth]{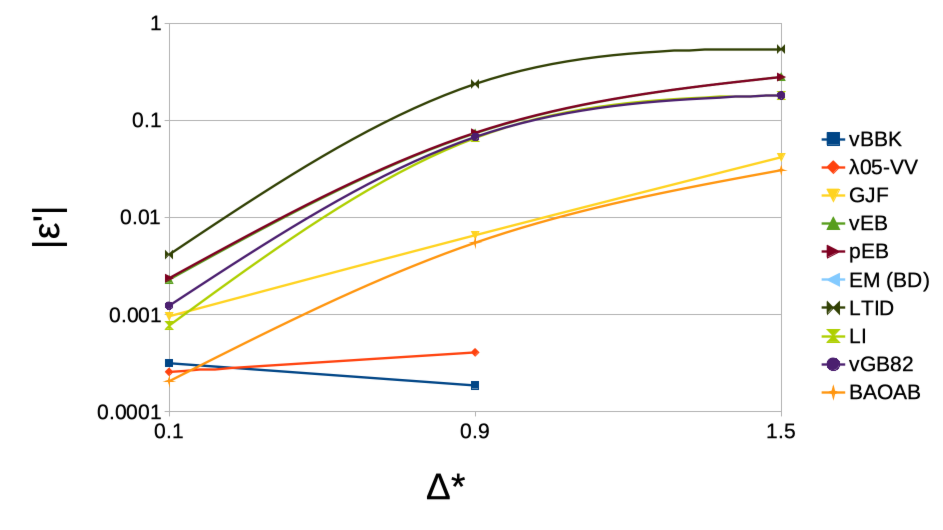}}{c} 
  \subsubfloat{\includegraphics[width=0.49\columnwidth]{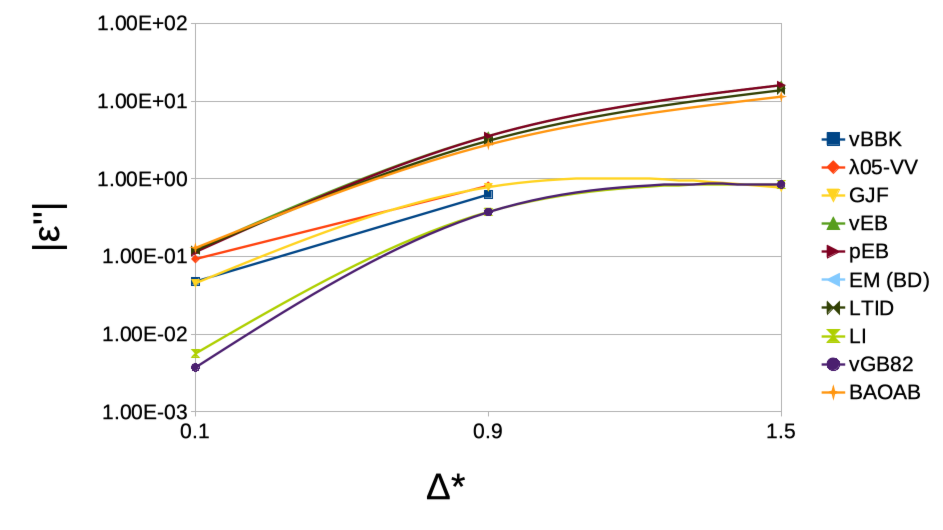}}{d}
  \subsubfloat{\includegraphics[width=0.49\columnwidth]{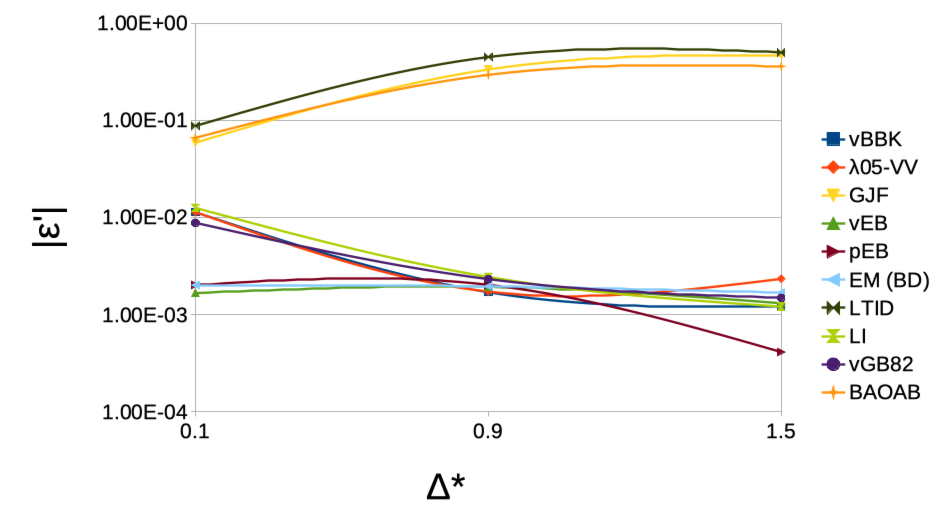}}{e} 
  \subsubfloat{\includegraphics[width=0.49\columnwidth]{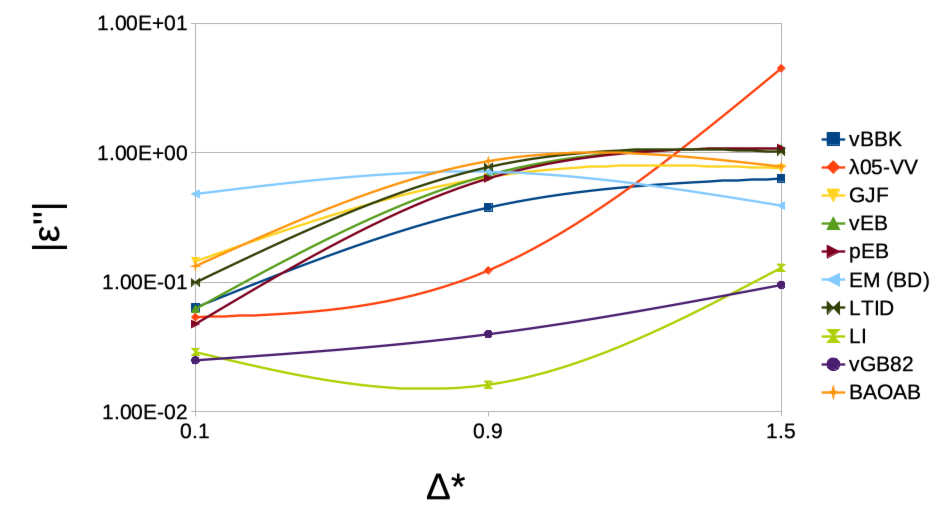}}{f}

\caption{Accuracy parameters $\epsilon'$~(a,c,e,g,i,k) and $\epsilon''$~(b,d,f,h,j,l) vs time step for the position (a,b,e,f,i,j) and momentum (c,d,g,h,k,l) autocorrelation functions for a particle in the harmonic potential in the underdamped $\kappa^*$=1 (a,b,c,d), critically damped $\kappa^*$=0.25 (e,f,g,h), and overdamped $\kappa^*$=0.1 (i,j,k,l) regimes. The EM momentum-related dependencies are absent by the integrator design. \emph{(cont.)}} \label{fig:autocorr-harm-eps}

\end{figure}

\begin{figure}
  \ContinuedFloat
  \centering
  \subsubfloat{\includegraphics[width=0.49\columnwidth]{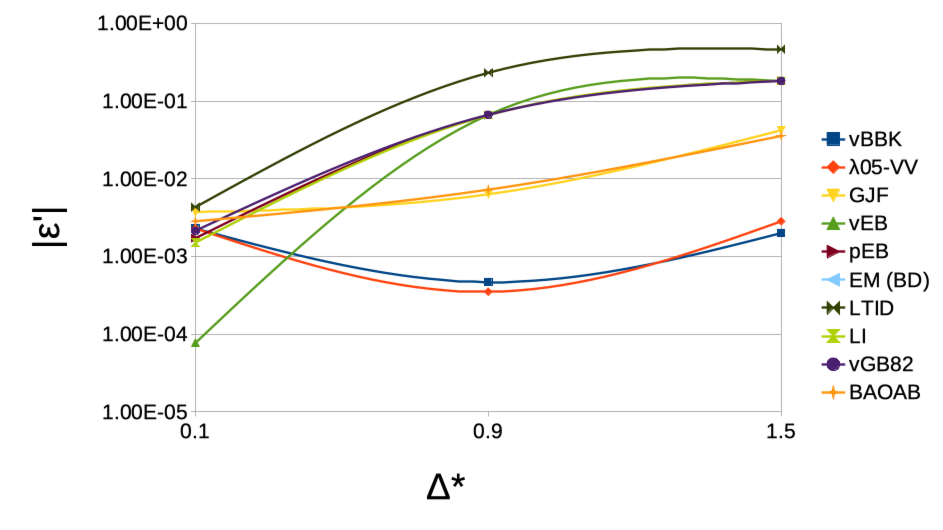}}{g} 
  \subsubfloat{\includegraphics[width=0.49\columnwidth]{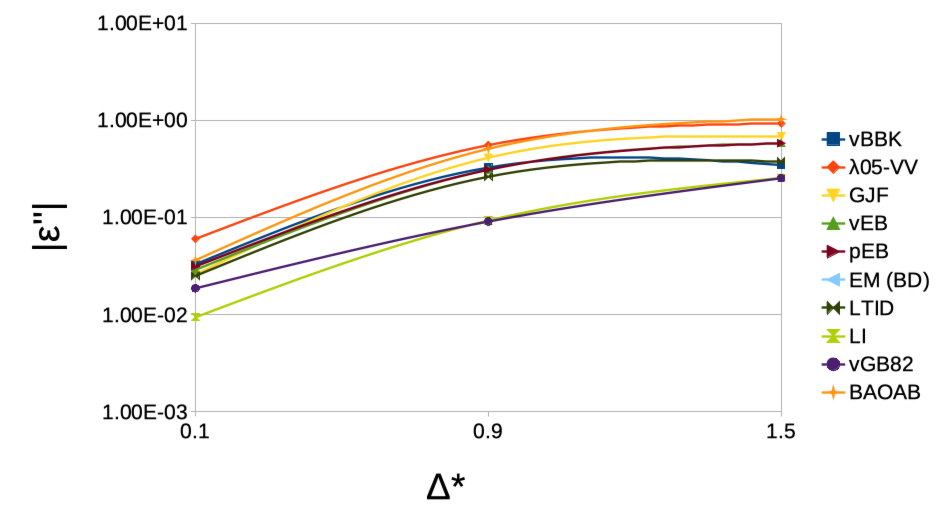}}{h}
  \subsubfloat{\includegraphics[width=0.49\columnwidth]{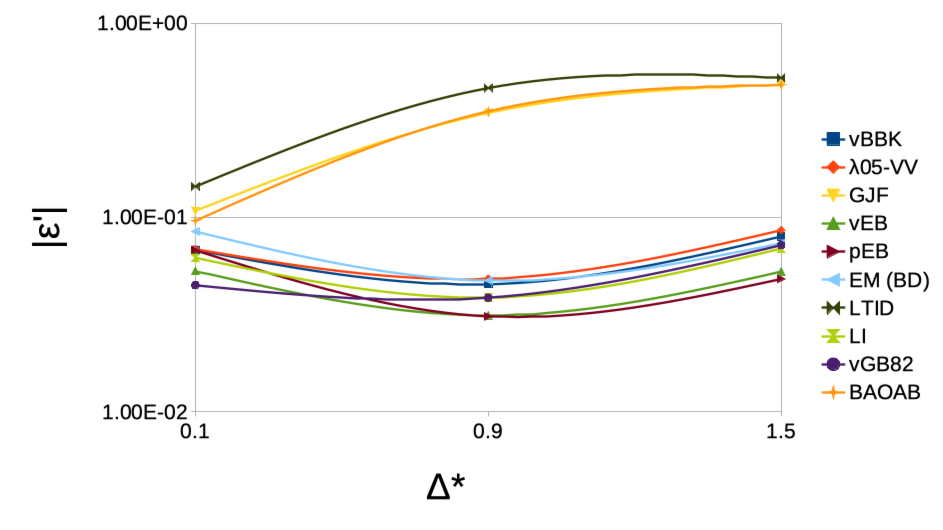}}{i} 
  \subsubfloat{\includegraphics[width=0.49\columnwidth]{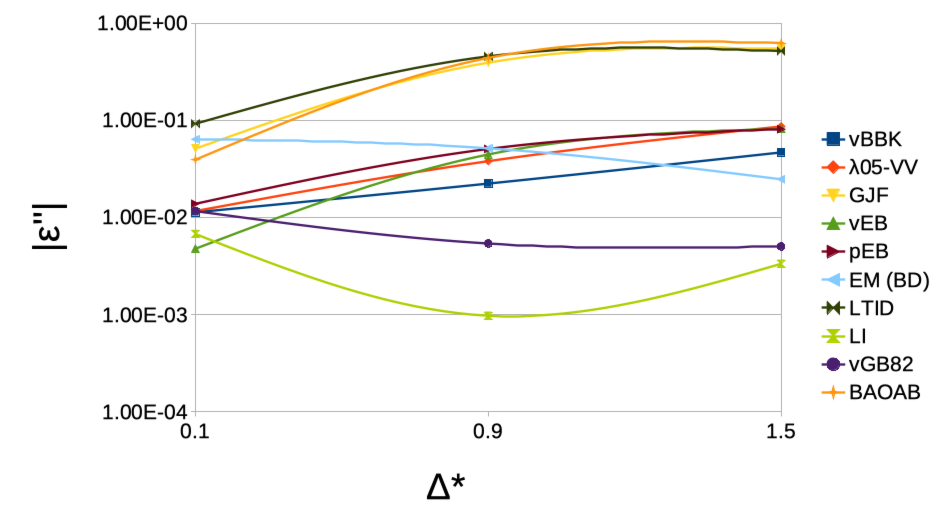}}{j}
  \subsubfloat{\includegraphics[width=0.49\columnwidth]{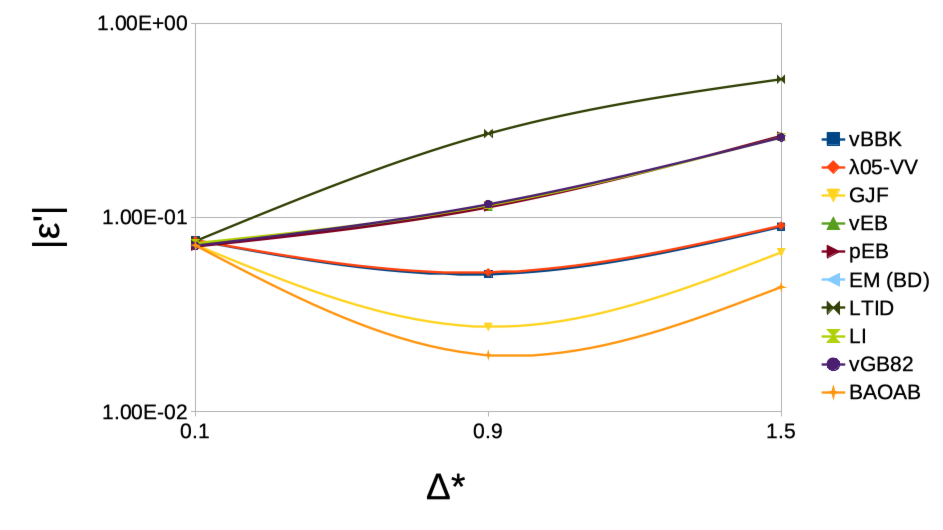}}{k} 
  \subsubfloat{\includegraphics[width=0.49\columnwidth]{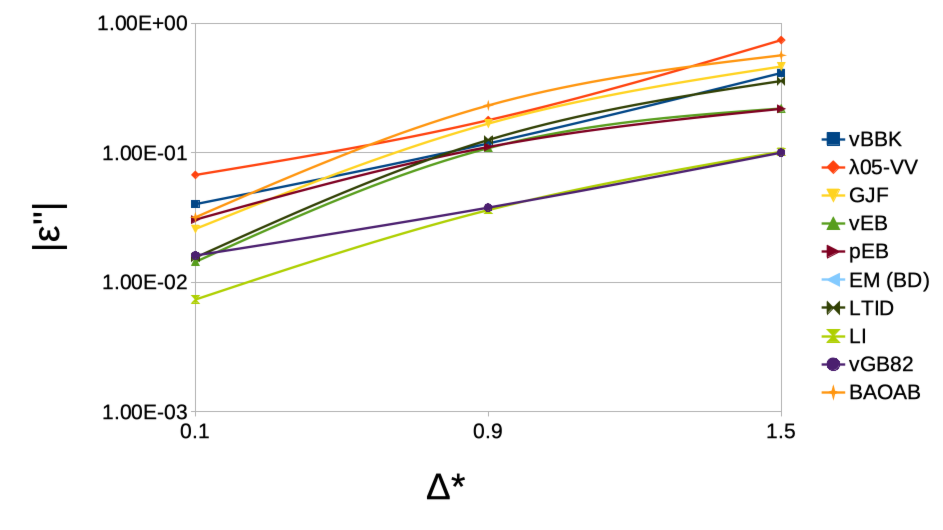}}{l}
\caption{Accuracy parameters $\epsilon'$~(a,c,e,g,i,k) and $\epsilon''$~(b,d,f,h,j,l) vs  time step for the position (a,b,e,f,i,j) and momentum (c,d,g,h,k,l) autocorrelation functions and for a particle in the harmonic potential in the underdamped $\kappa^*$=1 (a,b,c,d), critically damped $\kappa^*$=0.25 (e,f,g,h), and overdamped $\kappa^*$=0.1 (i,j,k,l) cases. The EM momentum-related dependencies are absent by the integrator design.} 

\end{figure}

\pagebreak

\pagebreak

\pagebreak

\end{document}